\documentclass[modern]{aastex701}

\newcommand{\teff}{T$_\mathrm{eff}$}
\newcommand{\logg}{$\log{\mathrm{g}}$}
\newcommand{\kzz}{$\log\kappa_\mathrm{zz}$}
\newcommand{\fsed}{$f_{sed}$}
\newcommand{\rjup}{R$_\mathrm{Jup}$}
\newcommand{\meth}{CH$_4$}
\newcommand{\wat}{H$_2$O}
\newcommand{\chhhh}{CH$_4$}
\newcommand{\coo}{CO$_2$}
\newcommand{\phhh}{PH$_3$}

\newcommand{\chisquare}{$\chi^2$}
\newcommand{\ldl}{$\lambda/\Delta\lambda$}
\renewcommand{\micron}{$\mathrm{\mu}$m}

\usepackage{subcaption}

\accepted{2025 Nov 11}

\submitjournal{AJ}

\shorttitle{RUBIES Brown Dwarfs}
\shortauthors{Morrissey et al.}
\begin{document}

\title{Discovery of Seven Cold and Distant Brown Dwarfs with JWST RUBIES}

\correspondingauthor{Adam J.\ Burgasser}

\author[orcid=0009-0006-0192-9630,sname=Morrissey]{Sara J. Morrissey}
\affiliation{Department of Physics, UC San Diego, La Jolla, CA 92093, USA}
\affiliation{Department of Physics and Astronomy, University of Notre Dame, Notre Dame, Indiana 46556, USA}
\email{smorriss@nd.edu}

\author[orcid=0000-0002-6523-9536,sname=Burgasser]{Adam J.\ Burgasser}
\affiliation{Department of Astronomy \& Astrophysics, UC San Diego, La Jolla, CA 92093, USA}
\email[show]{aburgasser@ucsd.edu}

\author[orcid=0000-0002-2380-9801,sname=de Graaff]{Anna de Graaff}
\affiliation{Max-Planck-Institut f\"ur Astronomie, K\"onigstuhl 17, D-69117, Heidelberg, Germany}
\affiliation{Harvard-Smithsonian Center for Astrophysics, 60 Garden Street, Cambridge, MA 02138, USA}
\email{degraaff@mpia.de}

\author[orcid=0000-0002-2446-8770,sname=McConachie]{Ian McConachie}
\affiliation{Department of Astronomy, University of Wisconsin-Madison, 475 N. Charter St., Madison, WI 53706 USA}
\email{ian.mcconachie@wisc.edu}

\author[orcid=0000-0003-2680-005X,sname=Nrammer]{Gabriel Brammer}
\affiliation{Cosmic Dawn Center (DAWN), Copenhagen, Denmark}
\affiliation{Niels Bohr Institute, University of Copenhagen, Jagtvej 128, Copenhagen, Denmark}
\email{gabriel.brammer@nbi.ku.dk}

\begin{abstract}

%% update with edits made for aas abstract
We report near-infrared spectral model fits to seven distant L- and T-type dwarfs observed with the JWST Near Infrared Spectrograph (NIRSpec) as part of the Red Unknowns: Bright Infrared Extragalactic Survey (RUBIES).
%We report the discovery of seven low-temperature (L- and T-type) 
%brown dwarfs in the 0.9--5.1~{\micron} low-resolution Prism spectra.
Comparison of 0.9--2.5~{\micron} near-infrared spectra of these sources to spectral standards indicates spectral types spanning L1 to T8 and spectrophotometric distances spanning 800--3,000~pc.
Fits to three grids of spectral models yield atmosphere parameters and spectrophotometric distances largely
consistent with our classifications, although fits to L dwarf spectra indicate missing components to the models. 
Three of our sources have vertical displacements from the Galactic plane exceeding 1~kpc, and have high probabilities of membership in the Galactic thick disk population.
Of these, the L dwarf 
RUBIES-BD-3 (RUBIES-EGS-3081) is well-matched to subdwarf standards, 
while the early T dwarf RUBIES-BD-5 (RUBIES-UDS-170428) is best fit by metal-poor atmosphere models; both \added{may} be a thick disk or halo brown dwarfs.
We critically examine the 1--5~{\micron} spectra of the current sample of 1--2~kpc mid- and late-T dwarfs, finding that temperature, surface gravity, metallicity, and vertical mixing efficiency can all contribute to observed variations in near-infrared spectral structure and the strength of the 4.2~{\micron} {\coo} band.  
This work aims to guide ongoing JWST, Euclid, and other space-based spectral surveys that are expected to uncover thousands of low-temperature stars and brown dwarfs throughout the Milky Way.
\end{abstract}



\section{Introduction\label{sec:intro}} 

Brown dwarfs are the lowest-temperature and lowest-mass ``stars'' in the Milky Way, with insufficient mass ($\lesssim$0.075~M$_\odot$) to sustain core
hydrogen fusion \citep{1962AJ.....67S.579K,1963ApJ...137.1121K,1963PThPh..30..460H}.
While comprising a significant fraction of star-like objects in the Galactic disk ($\sim$20\%; \citealt{Reyl__2018,Kirkpatrick_2021}),
these objects cool and dim over time, becoming exceedingly faint
($M_R$ $\gtrsim$ 18, $M_K$ $\gtrsim$ 10; \citealt{2017AJ....154..147D}).
As such, the majority of low-temperature stars and brown dwarfs---classified as late-M, L, T and Y dwarfs---are detected in the immediate solar neighborhood ($\lesssim$100~pc), and 
have predominantly solar elemental compositions.
Only a small fraction of local dwarfs are metal-poor members of the Galactic thick disk ($\sim$12\%) and halo populations ($\sim$0.5\%; \citealt{2008ApJ...673..864J,2020ApJ...898...77S}). While rare, these long-lived sources are essential for understanding how metallicity influences the formation, evolution, and hydrogen-burning mass threshold \added{of} brown dwarfs, and can potentially trace dynamical evolution and chemical enrichment throughout the history of the Milky Way \citep{2004ApJS..155..191B,2025ApJ...985...48H}.

The high infrared sensitivity of the current generation of space-based telescopes, including JWST \citep{2023PASP..135d8001R} and Euclid \citep{2011arXiv1110.3193L}, has made it possible to identify and study brown dwarfs at considerably larger distances.
%\citep{2016AJ....151...92R}.
Building on tools developed for deep imaging and spectroscopic surveys conducted with the Hubble Space Telescope \citep{2009ApJ...695.1591P,2011ApJ...739...83R,2016MNRAS.458..425V,2022ApJ...924..114A},
deep JWST surveys are now uncovering individual low-temperature stars and brown dwarfs beyond kiloparsec distances \citep{2023ApJ...942L..29N,2023ApJ...957L..27L,2023MNRAS.523.4534W,2024ApJ...962..177B,2024MNRAS.529.1067H,2024ApJ...964...66H,2024ApJ...975...31H,2025ApJ...980..230T,2025arXiv251002026T,2025arXiv251000111H}. 
Early Euclid studies are uncovering hundreds of cool dwarfs to similar scales \citep{2025arXiv250322497V,2025arXiv250322559M,2025ApJ...991...84D}.
Those surveys conducted at high galactic latitudes contain considerably higher fractions of thick disk and halo sources than in the local population, enabling more efficient and comprehensive studies of metal-poor low-mass stars and brown dwarfs.

In this article, we present the identification and spectral analysis of seven L and T dwarfs identified in deep spectroscopic data collected as part of the Red Unknowns: Bright Infrared Extragalactic Survey (RUBIES; \citealt{2025A&A...697A.189D}). 
In Section~\ref{sec:sample}, we describe the design and data associated with the RUBIES survey, and our method for identifying brown dwarfs from the spectral data.
In Section~\ref{sec:methods}, we present an empirical analysis of the spectra, including template classification and spectrophotometric distance estimates.
In Section~\ref{sec:results}, we fit the spectra to three sets of model grids
%using a Markov Chain Monte Carlo (MCMC) algorithm 
to obtain robust estimates of atmospheric parameters and their uncertainties, and evaluate 
the accuracy of the different grids over the broad range of specral types in our sample.
In Section~\ref{sec:discussion}, we quantify the probabilities of membership in the Galactic thin disk, thick disk, and halo populations, and critically assess the source of spectral variations among mid- and late-type T dwarfs in our sample.
We summarize our results in Section~\ref{sec:summary}.

\section{Sample} \label{sec:sample} 

\subsection{RUBIES Survey Data} \label{sec:data}
RUBIES is a spectroscopic survey conducted with the JWST Near Infrared Spectrograph (NIRSpec; \citealt{Birkmann_2022}).
The survey is focused on detecting and characterizing massive galaxies at high redshifts, drawing on sources selected from JWST Near Infrared Camera (NIRCam) imaging data obtained 
in the Extended Groth Strip (EGS) by the Cosmic Evolution Early Release Science Survey (CEERS, JWST-ERS-1345, PI S.\ Finkelstein), %\citealt{2017jwst.prop.1345F}), 
and in the Ultra Deep Survey (UDS) by the Public Release IMaging for Extragalactic Research Survey (PRIMER, JWST-GO-1837, PI J.\ Dunlap).
%; \citealt{2021jwst.prop.1837D}).
RUBIES prioritizes very red sources, and thus has a higher chance of observing brown dwarfs than surveys with differing color selection.
All RUBIES sources were acquired using the NIRSpec micro-shutter array (MSA; \citealt{2022A&A...661A..81F})
with a three-shutter slitlet, with total integrations of
48 minutes acquired in each of the low-resolution Prism/Clear
mode (0.6--5.3~{\micron}, {\ldl} = 30--300)
and the medium-resolution G395M/F290LP mode
(2.9--5.3~{\micron}, {\ldl} = 1000), using
a three-point nodding pattern.
Spectra were reduced using the JWST calibration pipeline version 1.12.5 \citep{bushouse_2023_10022973}
and {\tt msaexp} \citep{brammer_2023_8319596}
as described in detail in \citet{2025A&A...693A..60H} and \citet{2025A&A...697A.189D}.
Here, we focus on the NIRSpec/Prism data which provide the highest signal-to-noise (S/N) for our faint targets.

\subsection{Selection of Brown Dwarfs} \label{sec:selection}

%\begin{rotatetable}
\begin{deluxetable*}{lccccccc}
    \tabletypesize{\tiny}
    \tablecaption{Properties of the RUBIES Brown Dwarfs \label{tab:properties}}
    \tablehead{\\ \colhead{Name} & \colhead{BD-1} & \colhead{BD-2} & \colhead{BD-3} & \colhead{BD-4} & \colhead{BD-5} & \colhead{BD-6} & \colhead{BD-7} 
    }
    \startdata 
         \multicolumn{8}{c}{Survey Information} \\
         \hline
    RUBIES ID&  RUBIES-EGS-944743 & RUBIES-EGS-35616 & RUBIES-EGS-41280 & \added{RUBIES-}EGS63-30081 & RUBIES-UDS-170428 & RUBIES-UDS-170824 & RUBIES-UDS-140125 \\
    ASTRODEEP ID&  CEERS-43510 & CEERS-39492 & CEERS-47111 & CEERS-33796 & PRIMER-UDS-133345 & PRIMER-UDS-127017 & PRIMER-UDS-95456 \\
    Other ID &  o006\_s00089\tablenotemark{a} & o006\_s35616\tablenotemark{a} & o005\_s41280\tablenotemark{a} & \nodata & \nodata & \nodata & \nodata \\
    &  CEERS-EGS-BD-4\tablenotemark{b} & \nodata & \nodata & \nodata & \nodata & \nodata & \nodata \\
%    \hline
%         Spectral Type& T7$\pm$0.5 & L1$\pm$0.5 & d/sdL7$\pm$1 & sd:T0$\pm$3 & d/sdT1$\pm$2 & T8$\pm$1 & T5$\pm$0.5\\
         % SpT & T7$\pm$0.5 & L1$\pm$0.5 & d/sdL7$\pm$1 & T0$\pm$3 & T1$\pm$2 & T8$\pm$1 & T5$\pm$0.5\\
         % \hline
         R.A. (J2000) & 214$\fdg$9102828 & 214$\fdg$938816 & 214$\fdg$828434 & 214$\fdg$929872 & 34$\fdg$21396 & 34$\fdg$258098 & 34$\fdg$315054 \\
         Decl. (J2000) & +52$\fdg$860077 & +52$\fdg$873855 & +52$\fdg$810819 & +52$\fdg$856079 & $-$5$\fdg$10655 & $-$5$\fdg$106006 & $-$5$\fdg$149353 \\
%         Galactic $b$ & 59$\fdg$5 & 59$\fdg$5 & 59$\fdg$5 & 59$\fdg$5 & 60$\fdg$0 & 60$\fdg$0 & 60$\fdg$0 \\
         \hline
         \multicolumn{8}{c}{NIRCam Photometry} \\
         \hline
         % original photometry from RUBIES catalog - some flaws
         % F115W f$_\nu$ (nJy)& 638$\pm$32 & 1170$\pm$59 & 340$\pm$17 & 21$\pm$3 & 35$\pm$3 & 50$\pm$13 & 391$\pm$20  \\
         % F150W f$_\nu$ (nJy)& 364$\pm$41 & 1522$\pm$76 & 349$\pm$17 & 20$\pm$4 & 38$\pm$2 & 10$\pm$10 & 205$\pm$10 \\
         % F200W f$_\nu$ (nJy)& 260$\pm$32 & 1770$\pm$89 & 359$\pm$18 & 19$\pm$3 & 36$\pm$6 & 8$\pm$8 & 138$\pm$7 \\
         % F277W f$_\nu$ (nJy)& 211$\pm$16 & 1217$\pm$61 & 227$\pm$11 & 9$\pm$3 & 27.4$\pm$1.6 & 9$\pm$7 6 & 86$\pm$4 \\
         % F356W f$_\nu$ (nJy)& 410$\pm$21 & 1537$\pm$77 & 325$\pm$16 & 32$\pm$2 & 80$\pm$4 & 37$\pm$6 & 195$\pm$10 \\
         % F444W f$_\nu$ (nJy)& 1103$\pm$55 & 1181$\pm$59 & 253$\pm$13 & 31$\pm$3 & 87$\pm$4 & 122$\pm$9 & 430$\pm$22 \\
         % m$_{F115W}$ (AB)& 24.39$\pm$0.05 & 23.73$\pm$0.05 & 25.07$\pm$0.05 & 28.08$\pm$0.13 & 27.54$\pm$0.09 & 27.2$\pm$0.3 & 24.92$\pm$0.05 \\
         % m$_{F150W}$ (AB)& 25.00$\pm$0.12 & 23.44$\pm$0.05 & 25.04$\pm$0.05 & 28.13$\pm$0.23 & 27.46$\pm$0.07 & $\geq$28.9 & 25.62$\pm$0.05 \\
         % m$_{F200W}$ (AB)& 25.36$\pm$0.13 & 23.28$\pm$0.05 & 25.01$\pm$0.05 & 28.21$\pm$0.19 & 27.51$\pm$0.17 & $\geq$29.2 & 26.05$\pm$0.05 \\
         % m$_{F277W}$ (AB)& 25.59$\pm$0.08 & 23.69$\pm$0.05 & 25.51$\pm$0.05 & 29.04$\pm$0.32 & 27.80$\pm$0.06 & 29.0$\pm$0.8 & 26.57$\pm$0.05 \\
         % m$_{F356W}$ (AB)& 24.87$\pm$0.05 & 23.43$\pm$0.05 & 25.12$\pm$0.05 & 27.62$\pm$0.07 & 26.64$\pm$0.05 & 27.49$\pm$0.19 & 25.67$\pm$0.05 \\
         % m$_{F444W}$ (AB)& 23.79$\pm$0.05 & 23.72$\pm$0.05 & 25.39$\pm$0.05 & 27.66$\pm$0.09 & 26.55$\pm$0.05 & 26.18$\pm$0.08 & 24.82$\pm$0.05 \\
         % ASTRODEEP
         F115W f$_\nu$ (nJy)& 71.1$\pm$2 & 933$\pm$4 & 224$\pm$3 & 21$\pm$2 & 22$\pm$6 & 24$\pm$7 & 160$\pm$7  \\
         F150W f$_\nu$ (nJy)& 38$\pm$3 & 1308$\pm$6 & 226$\pm$3 & 21$\pm$4 & 28$\pm$5 & 8$\pm$6 & 127$\pm$6 \\
         F200W f$_\nu$ (nJy)& 27$\pm$2 & 1520$\pm$6 & 233$\pm$3 & 23$\pm$3 & 30$\pm$4 & 6$\pm$5 & 109$\pm$5 \\
         F277W f$_\nu$ (nJy)& 13.3$\pm$1.9 & 1080$\pm$5 & 154$\pm$2 & 16$\pm$3 & 24$\pm$3 & 2$\pm$4  & 69$\pm$3 \\
         F356W f$_\nu$ (nJy)& 40.0$\pm$1.5 & 1386$\pm$3 & 207$\pm$2 & 40$\pm$2 & 63$\pm$3 & 22$\pm$3 & 153$\pm$4 \\
         F444W f$_\nu$ (nJy)& 114$\pm$3 & 1134$\pm$5 & 182$\pm$3 & 38$\pm$3 & 79$\pm$4 & 83$\pm$5 & 371$\pm$5 \\
         m$_{F115W}$ (AB mag)& 26.77$\pm$0.06 & 23.97$\pm$0.05 & 25.53$\pm$0.05 & 28.11$\pm$0.13 & 28.06$\pm$0.28 & 27.9$\pm$0.3 & 25.89$\pm$0.07 \\
         m$_{F150W}$ (AB mag)& 27.45$\pm$0.10 & 23.61$\pm$0.05 & 25.52$\pm$0.05 & 28.10$\pm$0.21 & 27.78$\pm$0.19 & $\geq$28.3 & 26.14$\pm$0.07 \\
         m$_{F200W}$ (AB mag)& 27.81$\pm$0.11 & 23.45$\pm$0.05 & 25.48$\pm$0.05 & 27.99$\pm$0.16 & 27.71$\pm$0.15 & $\geq$28.5 & 26.31$\pm$0.07 \\
         m$_{F277W}$ (AB mag)& 28.59$\pm$0.16 & 23.82$\pm$0.05 & 25.93$\pm$0.05 & 28.39$\pm$0.21 & 27.94$\pm$0.15 & $\geq$28.8 & 26.80$\pm$0.07 \\
         m$_{F356W}$ (AB mag)& 27.39$\pm$0.06 & 23.55$\pm$0.05 & 25.61$\pm$0.05 & 27.39$\pm$0.07 & 26.90$\pm$0.07 & 28.04$\pm$0.18 & 25.94$\pm$0.06 \\
         m$_{F444W}$ (AB mag)& 26.26$\pm$0.06 & 23.76$\pm$0.05 & 25.75$\pm$0.05 & 27.44$\pm$0.10 & 26.66$\pm$0.07 & 26.61$\pm$0.08 & 24.98$\pm$0.05 \\
         \hline
         \multicolumn{8}{c}{NIRSpec/Prism Spectroscopy} \\
         \hline
         % 1.2 {\micron} [S/N]&4.8&53.9&13.9&3.2&5.7&1.9&16.9\\
         % 4.0 {\micron} [S/N]&7.8&21.7&7.2&1.2&8.1&5.0&11.5\\
         1.2 {\micron} S/N & 5 & 54 & 14 & 3.2 & 6 & 1.9 & 17 \\
         4.0 {\micron} S/N & 8 & 22 & 7 & 1.2 & 8 & 5 & 12 \\
         % \hline
         % RV Shift [km/s] & -1200 & -2350 & -3800 & +5500 & -2150 & 0 & -3500
    \enddata
    \tablenotetext{a}{As reported in \citet{2025ApJ...980..230T}.}
    \tablenotetext{b}{As reported in \citet{2024ApJ...964...66H}.}
\end{deluxetable*}
%\end{rotatetable}
%\clearpage

Our seven sources were selected among 
point sources identified in the CEERS/EGS and PRIMER/UDS NIRCam imaging \added{data}, with spectral data grades $\geq$2.5
(a quality assessment metric indicating the reliability of spectral features)
and automated redshift estimates of $z<0.1$. 
These criteria initially narrowed our selection to six sources in the EGS field and three sources in the UDS field. 
We visually inspected NIRCam images to ensure the sources were consistent with point sources,
and visually confirmed that the NIRSpec/Prism spectra were consistent with low-temperature dwarf morphologies as opposed to active galactic nuclei or other compact extragalactic sources. 
This evaluation allowed us to reject two duplicate EGS sources\footnote{ \added{The duplicate sources were the same spectra but processed by different versions of the {\tt msaexp} code. We used the most recent reductions for the analysis presented in this article.}}, resulting in a final sample of seven low-temperature stars and brown dwarfs, summarized in Table~\ref{tab:properties}.
We list \added{fluxes} in the NIRCam F115W, F150W, F200W, F277W, F356W, and F444W filters from the CEERS/EGS and PRIMER/UDS surveys as
compiled in the ASTRODEEP-JWST catalog \citep{2024A&A...691A.240M}.
For the corresponding magnitudes, we conservatively include a systematic uncertainty of 0.05~mag to account for zeropoint errors.
%PSF-matched NIRCam photometry from the CEERS/EGS and PRIMER/UDS data computed as described in \citet{2024MNRAS.533.1808W}.
Three of the sources---RUBIES-BD-1, -2, and -3---were \added{reported previously} by \citet{2024ApJ...964...66H} and \citet{2025ApJ...980..230T}, whose results we discuss below in comparison with our findings. 
%
% might have a problem here with our photometry?
%
% \textcolor{red}{We note that NIRCam flux densities reported here for RUBIES-BD-1 (aka 
% are approximately ten times greater than those reported in 
% \citet[][, source identification CEERS-EGS-BD-4]{2024ApJ...964...66H} 
% and \citet[][, source identification o006\_s00089, photometry drawn from \citealt{2024MNRAS.533.1808W}]{2025ApJ...980..230T}.
% [WHY?]}

\begin{figure}
    \centering
    RUBIES-BD-1\hspace{2.6in}RUBIES-BD-2 \\
    \includegraphics[width=.5\textwidth]{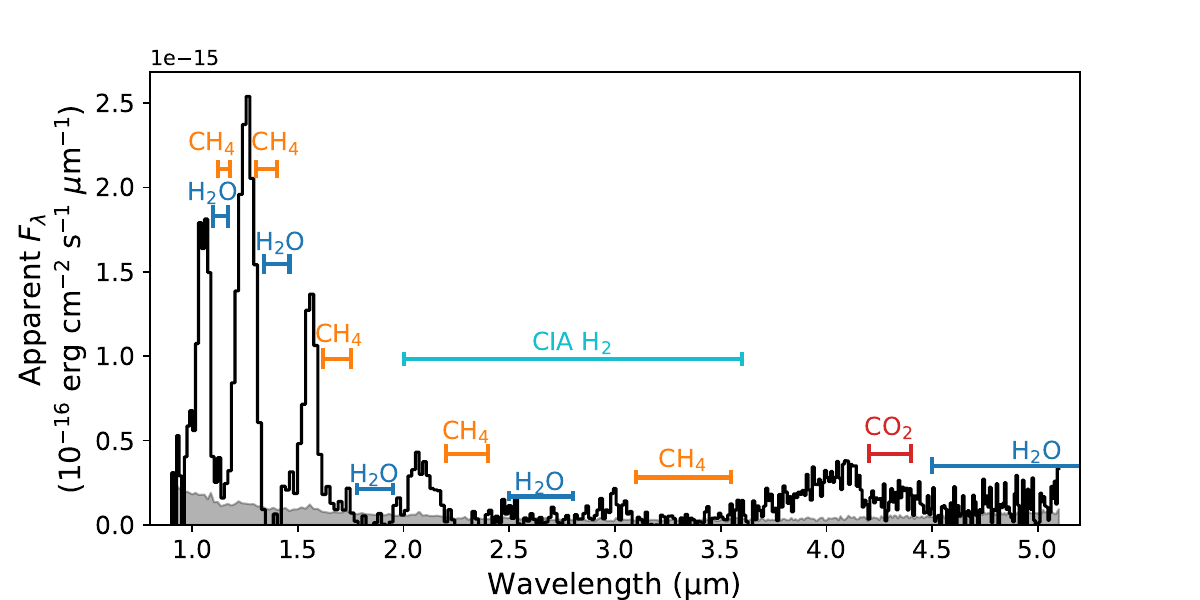}\hfill
    \includegraphics[width=.5\textwidth]{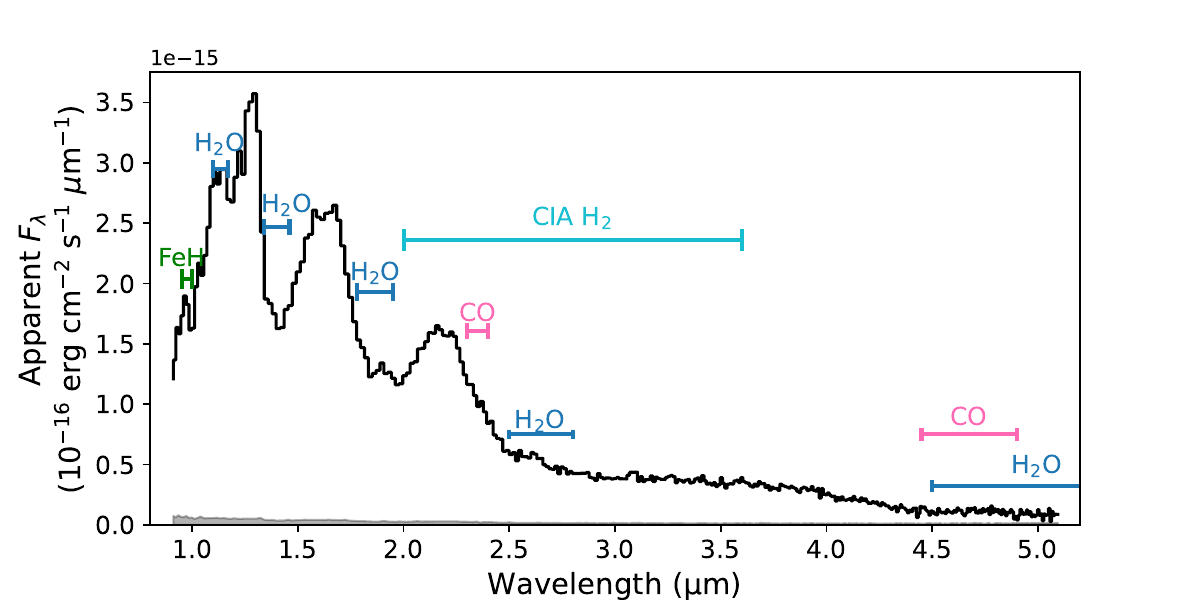}\hfill
    \\[\smallskipamount]
    RUBIES-BD-3\hspace{2.6in}RUBIES-BD-4 \\
    \includegraphics[width=.5\textwidth]{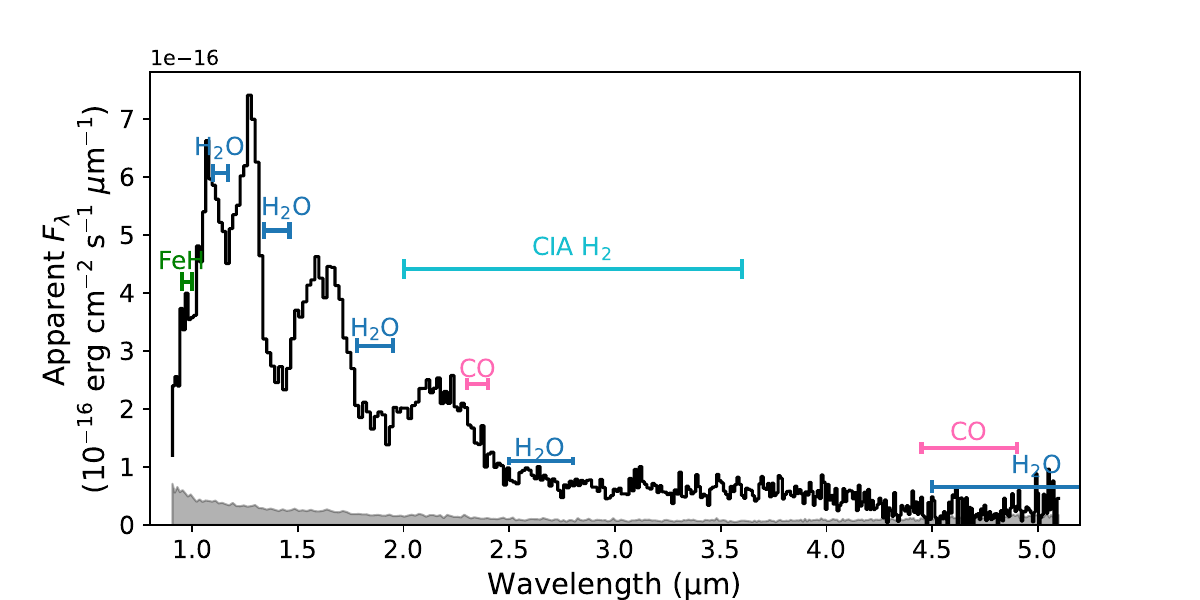}\hfill
    \includegraphics[width=.5\textwidth]{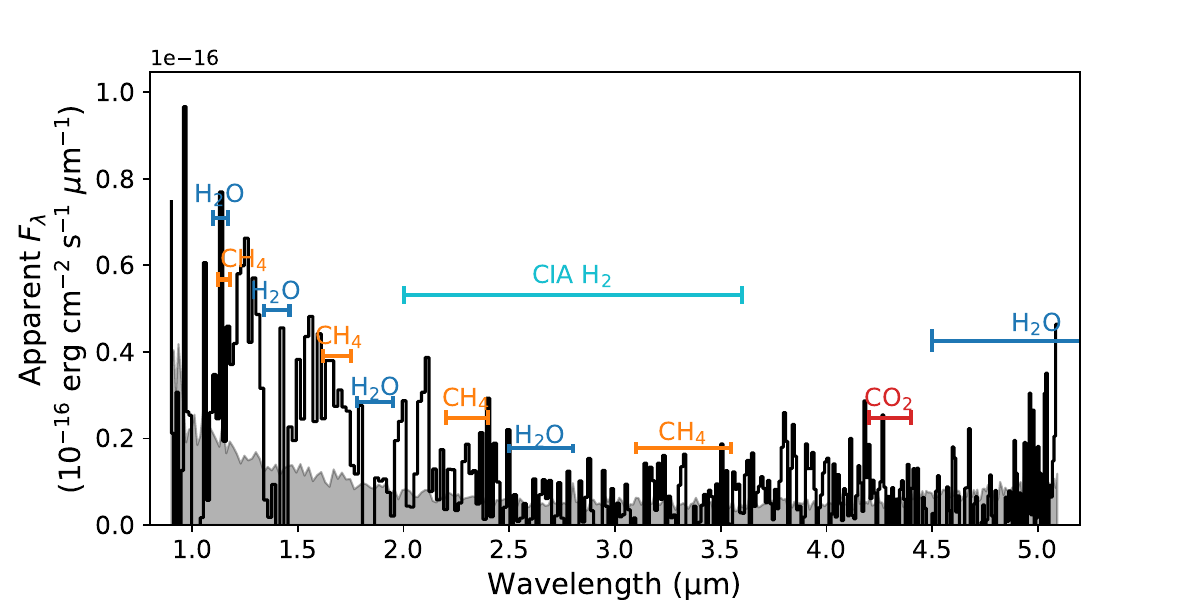}\hfill
    \\[\smallskipamount]
    RUBIES-BD-5\hspace{2.6in}RUBIES-BD-6 \\
    \includegraphics[width=.5\textwidth]{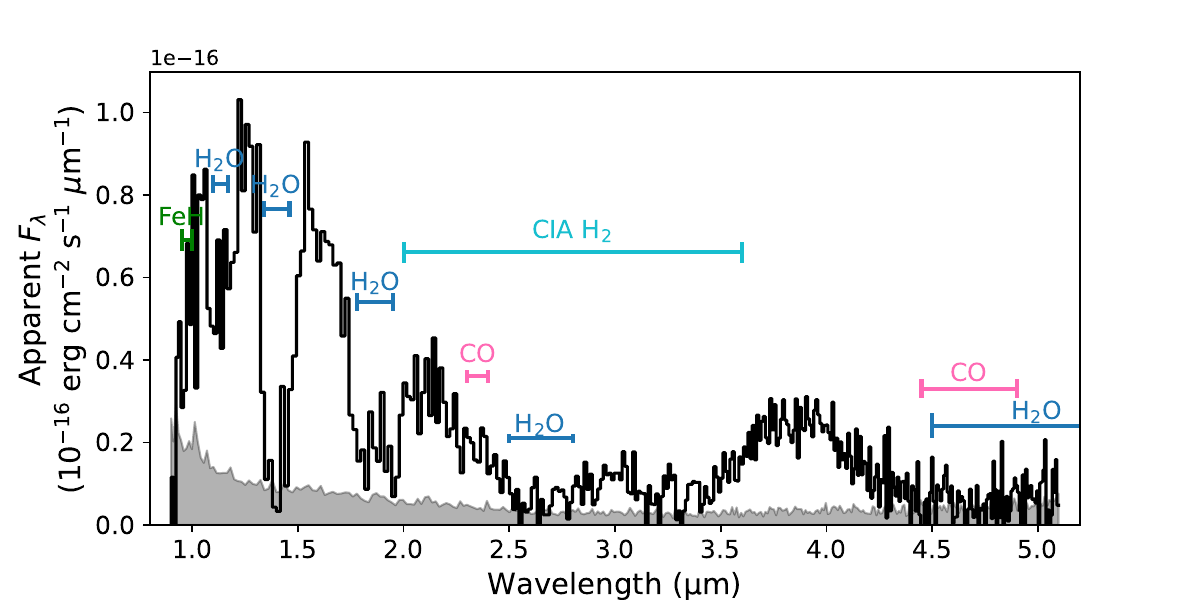}\hfill
    \includegraphics[width=.5\textwidth]{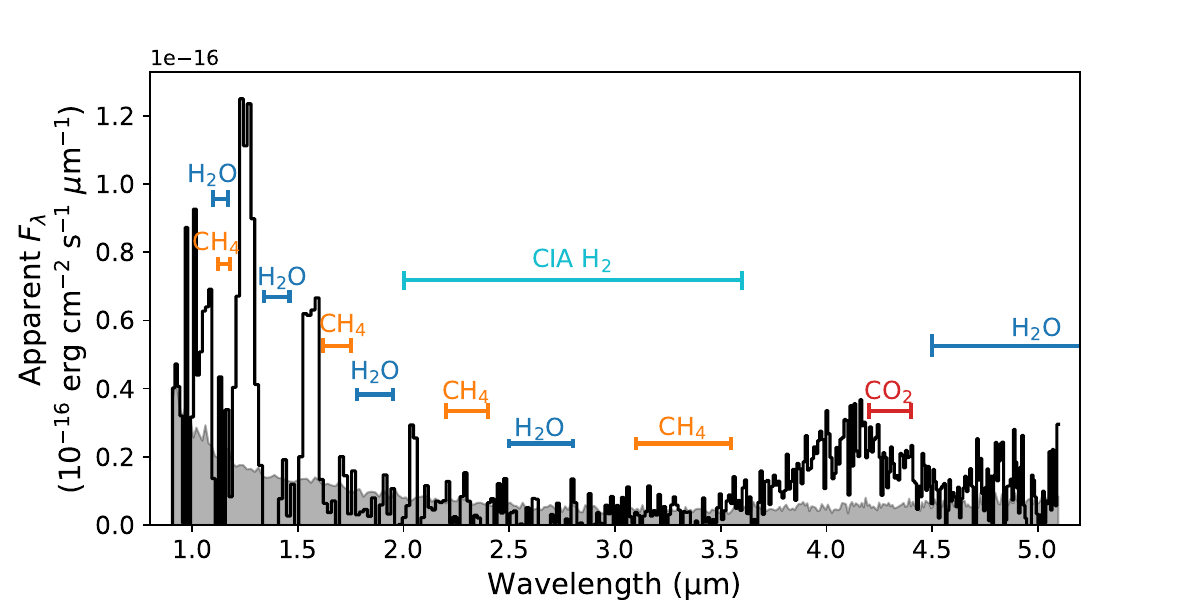}\hfill
    \\[\smallskipamount]
    RUBIES-BD-7 \\
    \includegraphics[width=.5\textwidth]{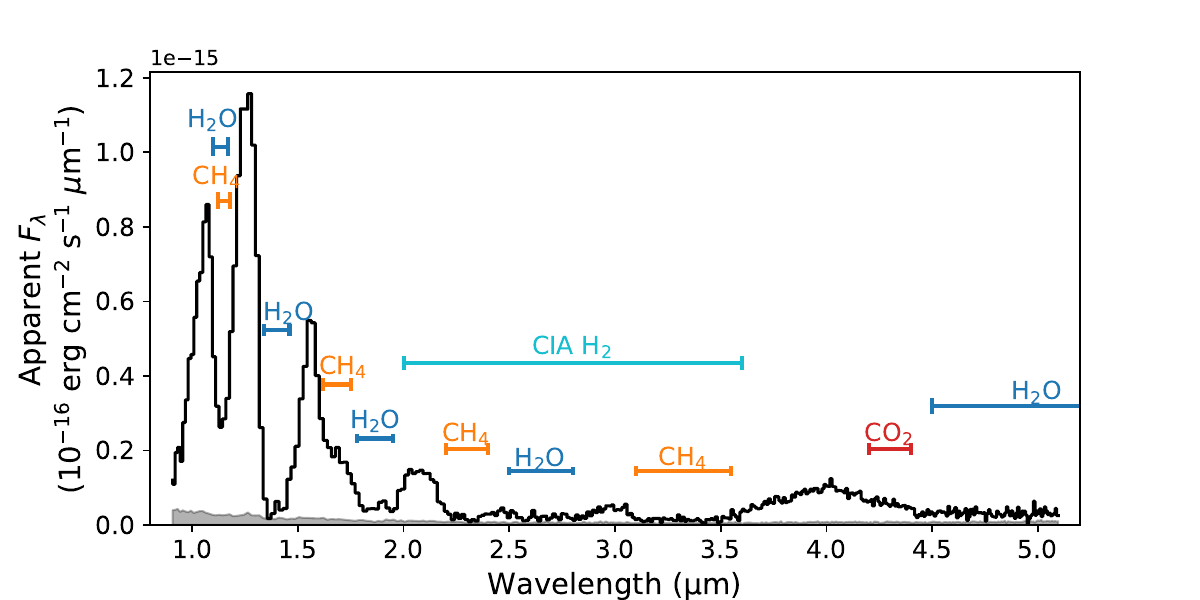}\hfill
    \caption{NIRSpec/Prism spectra of the seven RUBIES brown dwarfs identified in this study (black lines).  
    %(RUBIES-BD-1 through RUBIES-BD-7 from left to right, top to bottom) observed by the RUBIES program 
    Spectra are shown in apparent $F_\lambda$ flux densities after scaling to their F444W magnitudes, and uncertainties are indicated in grey. Key spectral features present in the 1--5~{\micron} spectra of L and T dwarfs are labeled.}
    \label{fig:sources}
\end{figure}

Figure \ref{fig:sources} displays the NIRSpec/Prism spectra for these sources, flux-calibrated to their apparent F444W magnitudes and presented in $F_\lambda$ flux density units.\footnote{\added{Flux-calibrated spectra for these sources are provided as data behind the figure in the online version of this article.}}
RUBIES-BD-1, -6, and -7 exhibit sharp flux peaks at 1.0, 1.25, 1.6, and 2.1 {\micron}, and a broad peak at 4~{\micron}, \added{all} shaped by strong {\wat} and {\meth} bands that are typical of T dwarfs \citep{Burgasser_2006,2012ApJ...760..151S,2024ApJ...973..107B}. RUBIES-BD-2 displays broader flux peaks at 1.2, 1.65, and 2.15 {\micron} between strong {\wat} absorption bands; and a blue, relatively featureless spectral slope toward longer wavelengths, typical of L dwarfs \citep{2005ApJ...623.1115C,2012ApJ...760..151S}. 
RUBIES-BD-3, -4, and -5 have features intermediate between these morphological types
(only marginally discernible in the spectrum of our faintest target
RUBIES-BD-4), suggesting that they are late-L or early-T dwarfs.
RUBIES-BD-3 also shows a peculiar split in the 1.2~{\micron} band
and a bluer overall spectral morphology which may be associated with subsolar metallicity, as discussed below.

\subsection{Classification and Distance Estimation} \label{sec:classify}

Spectral classifications for these sources were determined by comparing their 0.9--2.4~{\micron} spectra to L and T dwarf spectral standards defined in \cite{2010ApJS..190..100K} and \cite{Burgasser_2006}, respectively. 
Near-infrared spectral data for these standards were drawn from the SpeX Prism Library Analysis Toolkit \citep{2017ASInC..14....7B}; see associated references in
Table~\ref{tab:standards} of the Appendix.
We also selected low-metallicity L and T dwarf spectral standards encompassing mild subdwarf (d/sd), subdwarf (sd), and extreme subdwarf (esd) classes as defined in 
\citet{2017MNRAS.464.3040Z,2018MNRAS.479.1383Z,2019AJ....158..182G}; and \citet{2025ApJ...982...79B}, drawing on published near-infrared spectral data acquired with various instruments for comparison \added{(see Appendix)}. 
All spectra were smoothed and interpolated onto the same wavelength grid corresponding to the NIRSpec/Prism data. 
For our source spectra, we also applied small fractional pixel shifts of up to 1.5~pixels before interpolation, as determined by visual comparison to the best-fit standards, to account for variations in the projection of spectral data on the NIRSpec detector from the MSA slits.

For RUBIES-BD-1, -2, -3, -6, and -7, the best-fitting standard was identified by minimizing the $\chi^2$ statistic:
\begin{equation}\label{eqn:chisq}
    \chi^2 = \sum_{i=1}^{n}\left(\frac{O[\lambda_i]-{\alpha}S[\lambda_i]}{\sigma[\lambda_i]}\right)^2
\end{equation}
where $O[\lambda]$ is the observed flux density of the source, 
$\sigma[\lambda]$ are the associated uncertainties, 
$S[\lambda]$ is the standard spectrum flux density,
and $\alpha$ is an optimal scaling factor that minimizes $\chi^2$:
\begin{equation}\label{eqn:alpha}
    \alpha = \left(\sum_{i=1}^{n}\frac{O[\lambda_i]S[\lambda_i]}{\sigma^2[\lambda_i]}\right)/\left(\sum_{i=1}^{n}\frac{S^2[\lambda_i]}{\sigma^2[\lambda_i]}\right)
\end{equation}
(cf.~\citealt{2005ApJ...623.1115C}).
For RUBIES-BD-4 and -5, the spectral S/N is sufficiently low that the $\chi^2$ statistic cannot differentiate between adjacent standards, so we conducted a ``by-eye'' comparison to the spectral standards to identify an approximate best fit while allowing for generous uncertainties. 

\begin{figure}
    \includegraphics[width=.32\textwidth]{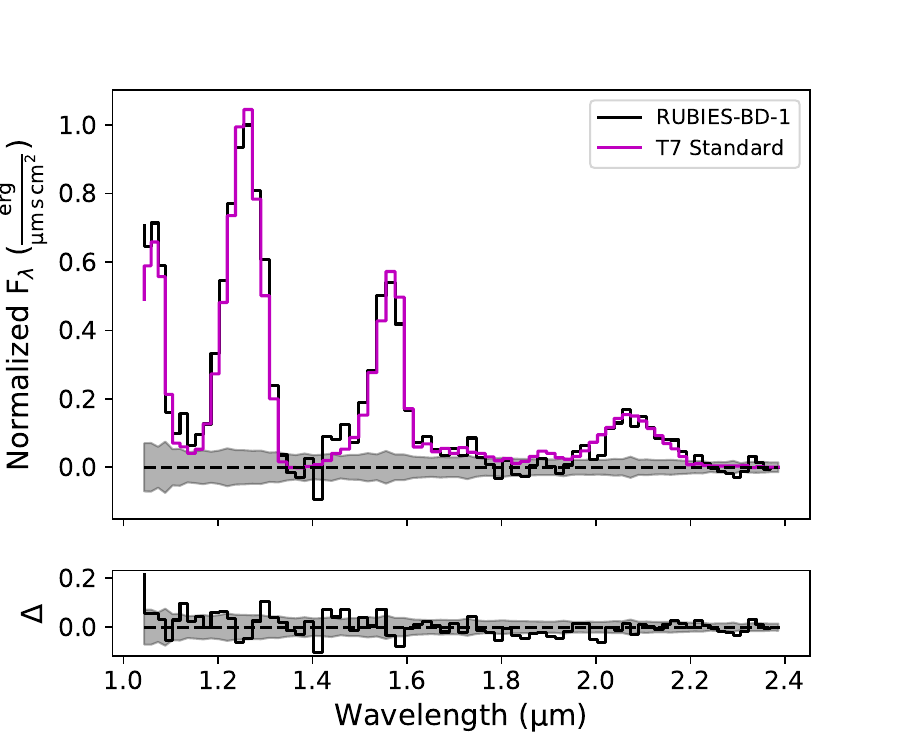}\hfill
    \includegraphics[width=.32\textwidth]{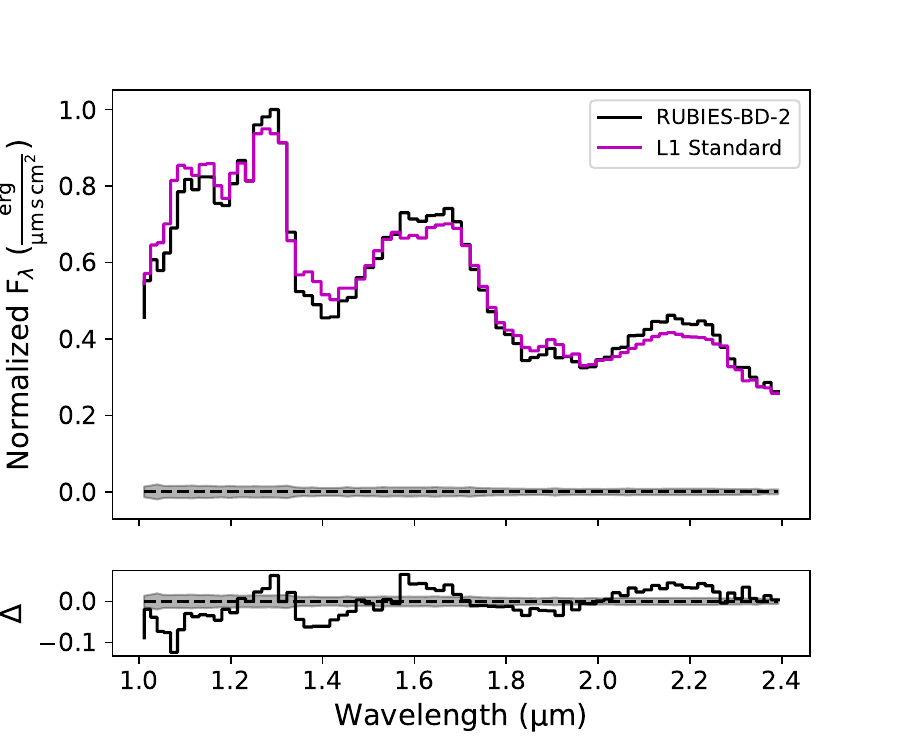}\hfill
    \includegraphics[width=.32\textwidth]{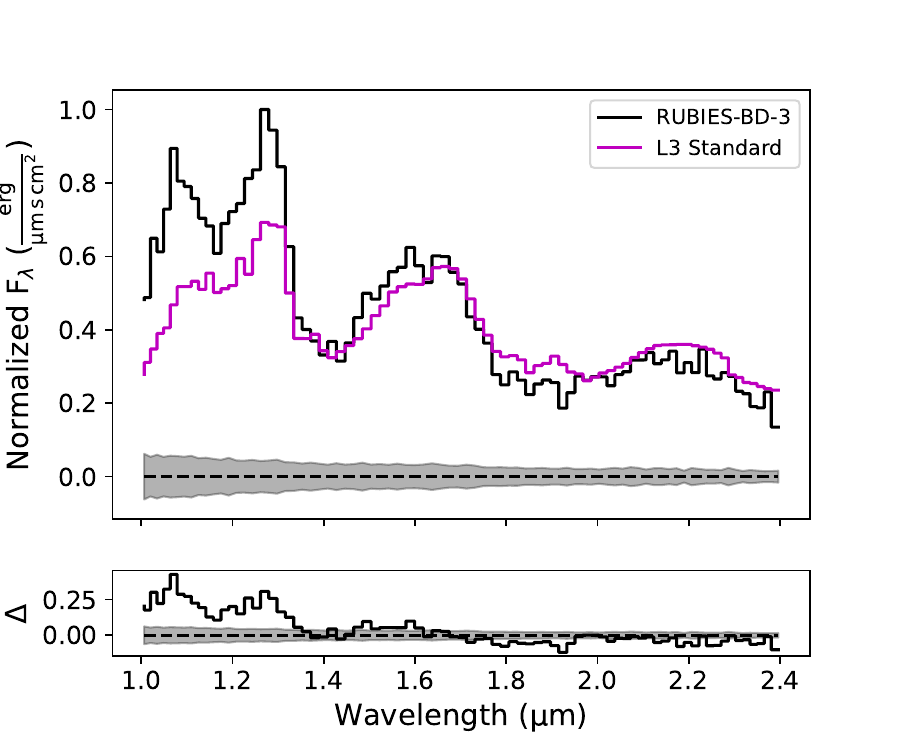}\hfill
    \\[\smallskipamount]
    \includegraphics[width=.32\textwidth]{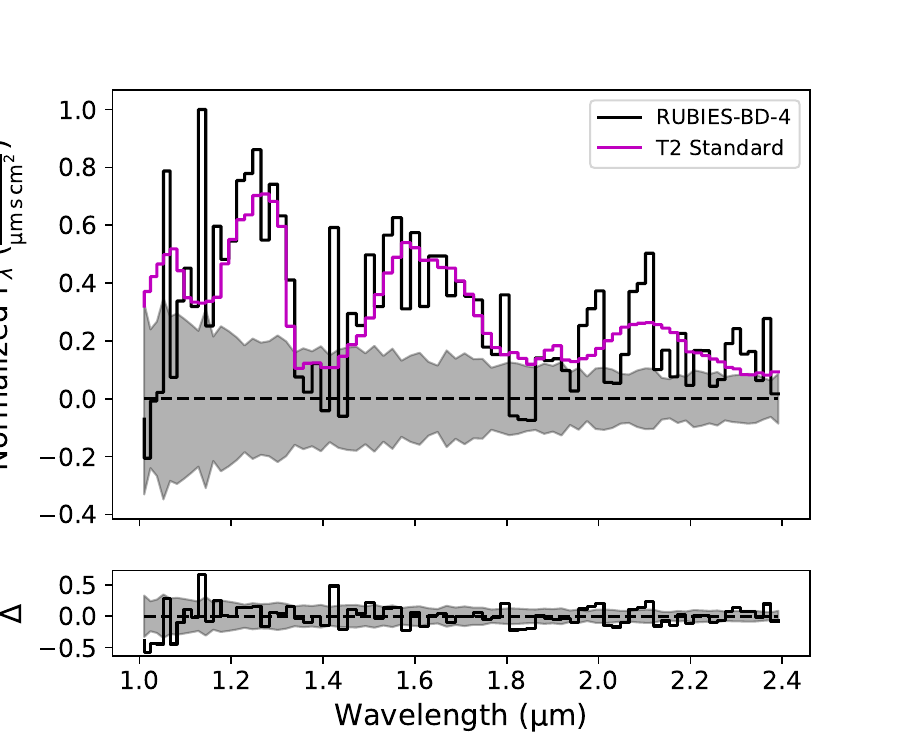}\hfill
    \includegraphics[width=.32\textwidth]{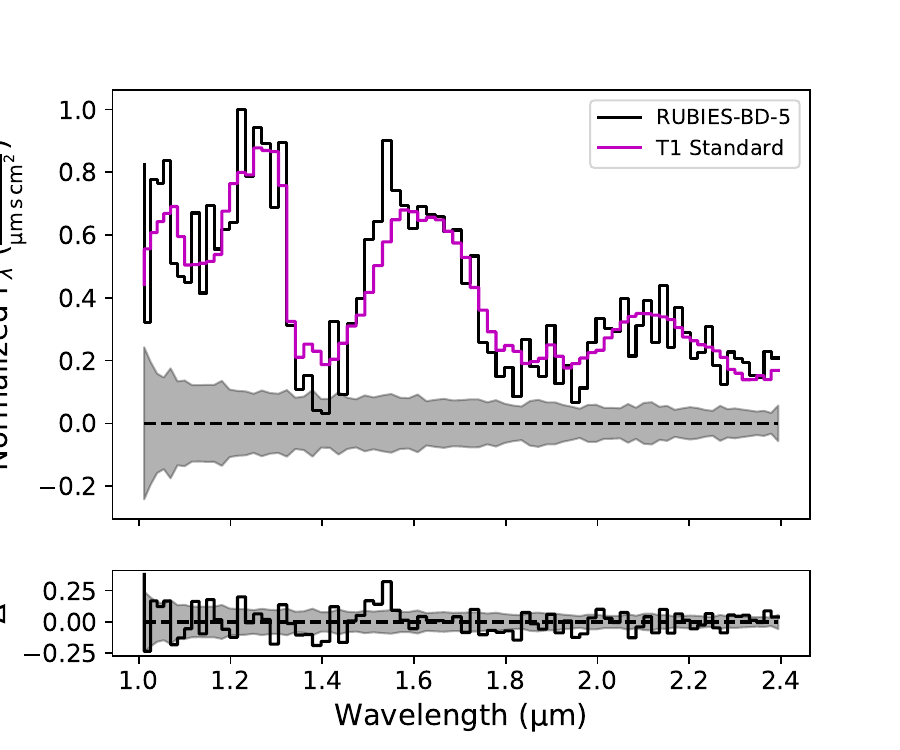}\hfill
    \includegraphics[width=.32\textwidth]{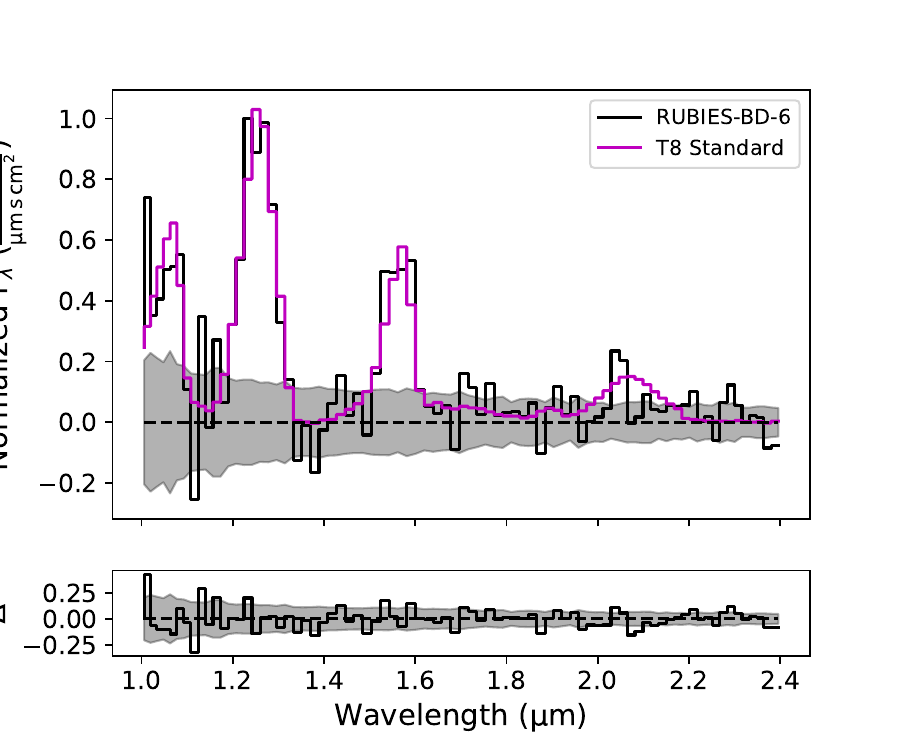}\hfill
    % \\[\smallskipamount]
    \begin{center}
        \includegraphics[width=.32\textwidth]{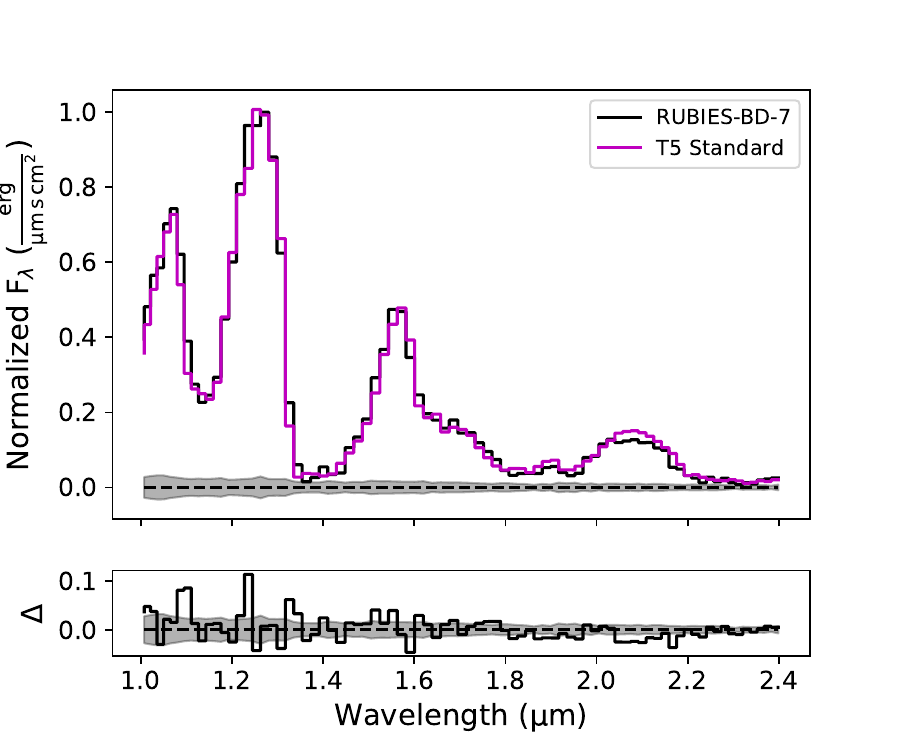}
    \end{center}
    \caption{RUBIES brown dwarf spectra in the 0.9--2.5~{\micron} range (black lines with uncertainties as grey shading) compared to best fit dwarf spectral standards (magenta lines). 
    RUBIES spectra are normalized to the 1.2--1.3~{\micron} peak, and standards are scaled to minimize {\chisquare} (Eqn.~\ref{eqn:alpha}).
    The bottom panels of each plot show the difference between source and standard spectra ($\Delta$) in black, compared to the uncertainty shaded in grey.}\label{fig:spec_class}
\end{figure}

\begin{deluxetable*}{llccccccc}
    \tabletypesize{\footnotesize}
    \tablecaption{Classification and Distances of RUBIES Brown Dwarfs \label{tab:classify}}
    \tablehead{
    \colhead{Name} & \colhead{Unit} & \colhead{BD-1} & \colhead{BD-2} & \colhead{BD-3} & \colhead{BD-4} & \colhead{BD-5} & \colhead{BD-6} & \colhead{BD-7}}
    \startdata 
    \hline
         \multicolumn{9}{c}{Classifications} \\
        \hline
         % RV Shift [km/s] & -1200 & -2350 & -3800 & +5500 & -2150 & 0 & -3500 \\
         $\Delta\lambda$\tablenotemark{a} & ({\micron}) & $-$0.010 & $-$0.020 & $-$0.032 & +0.046 & $-$0.018 & 0 & $-$0.029 \\
         SpT &  & T7$\pm$0.5 & L1$\pm$0.5 & sd:L6$\pm$1 & T2$\pm$3 & T1$\pm$2 & T8$\pm$1 & T5$\pm$0.5\\
         \hline
         \multicolumn{9}{c}{Spectrophotometric Distance Estimates} \\
        \hline
         % $d_{F115W}$ & pc & 370$\pm$60 & 1540$\pm$110 & 930$\pm$120 & 3000$\pm$500 & 2420$\pm$180 & 1000$\pm$400 & 740$\pm$50\\
         % $d_{F150W}$ & pc & 350$\pm$120 & 1560$\pm$120 & 1210$\pm$120 & 3900$\pm$1100 & 2800$\pm$200 & 800$\pm$1400 & 900$\pm$100\\
         % $d_{F200W}$ & pc & 340$\pm$140 & 1460$\pm$80 & 1530$\pm$120 & 5000$\pm$2000 & 3000$\pm$700 & 600$\pm$1300 & 890$\pm$120\\
         % $d_{F277W}$ & pc & 260$\pm$70 & 1080$\pm$90 & 1410$\pm$150 & 5000$\pm$2000 & 2700$\pm$600 & 700$\pm$700 & 850$\pm$110\\
         % $d_{F356W}$ & pc & 440$\pm$90 & 1370$\pm$80 & 1860$\pm$120 & 4700$\pm$1200 & 2700$\pm$600 & 900$\pm$600 & 1000$\pm$100\\
         % $d_{F444W}$ & pc & 340$\pm$40 & 1350$\pm$80 & 1580$\pm$140 & 3700$\pm$900 & 2100$\pm$300 & 800$\pm$300 & 700$\pm$40\\
         % $\langle{d}\rangle$\tablenotemark{b}  & pc & 340$\pm$20 & 1370$\pm$70 & 1420$\pm$140 & 3500$\pm$300 & 2540$\pm$110 & 850$\pm$50 & 770$\pm$40 \\
$d_{F115W}$ & (pc) & $1010_{-270}^{+330}$ & $1580_{-210}^{+250}$ & $1230_{-200}^{+240}$ & $2980_{-220}^{+270}$ & $2790_{-370}^{+430}$ & $1270_{-250}^{+290}$ & $1070_{-110}^{+130}$  \\
$d_{F150W}$ & (pc) & $1000_{-490}^{+890}$ & $1520_{-190}^{+240}$ & $1580_{-270}^{+300}$ & $3270_{-550}^{+690}$ & $2860_{-280}^{+320}$  & $>560$ & $1010_{-180}^{+220}$  \\
$d_{F200W}$ & (pc) & $950_{-490}^{+980}$ & $1410_{-140}^{+160}$ & $2050_{-290}^{+340}$ & $3130_{-1040}^{+1500}$ & $3020_{-550}^{+660}$  & $>450$ & $920_{-190}^{+230}$  \\
$d_{F277W}$ & (pc) & $1070_{-380}^{+700}$ & $1060_{-150}^{+180}$ & $1870_{-170}^{+190}$ & $3400_{-1190}^{+1890}$ & $3080_{-630}^{+780}$  & $>650$ & $960_{-220}^{+310}$  \\
$d_{F356W}$ & (pc) & $1280_{-380}^{+600}$ & $1320_{-130}^{+140}$ & $2290_{-170}^{+180}$ & $3200_{-1020}^{+1400}$ & $2820_{-470}^{+620}$ & $1070_{-250}^{+290}$ & $1050_{-190}^{+220}$  \\
$d_{F444W}$ & (pc) & $960_{-180}^{+220}$ & $1250_{-130}^{+140}$ & $1850_{-160}^{+190}$ & $2700_{-530}^{+640}$ & $2010_{-250}^{+270}$  & $870\pm120$ & $690\pm70$  \\
$\langle{d}\rangle$  & (pc) & 1000$\pm$170 & 1320$\pm$100 & 1870$\pm$180\tablenotemark{a} & 2990$\pm$230 & 2500$\pm$250 & 940$\pm$140 & 830$\pm$90 \\
    \enddata
    \tablecomments{$\Delta\lambda$ corresponds to the wavelength shift applied to the spectrum to align features with the best-fit standard. This shift is applied as a fractional pixel shift before spectra are interpolated onto a common wavelength grid, and accounts for variations in spectral projection on the NIRSpec detector by the MSA pixel mask.}
    \tablenotetext{a}{Significant variance among the individual filter distance estimates.}
%    \tablenotetext{c}{Vertical scaleheight based on average distance and target galactic latitude.}
\end{deluxetable*}

\begin{figure}[ht]
    \centering
    % First column
    \begin{minipage}[t]{0.32\textwidth}
        \vspace{0pt}
        \includegraphics[width=\textwidth]{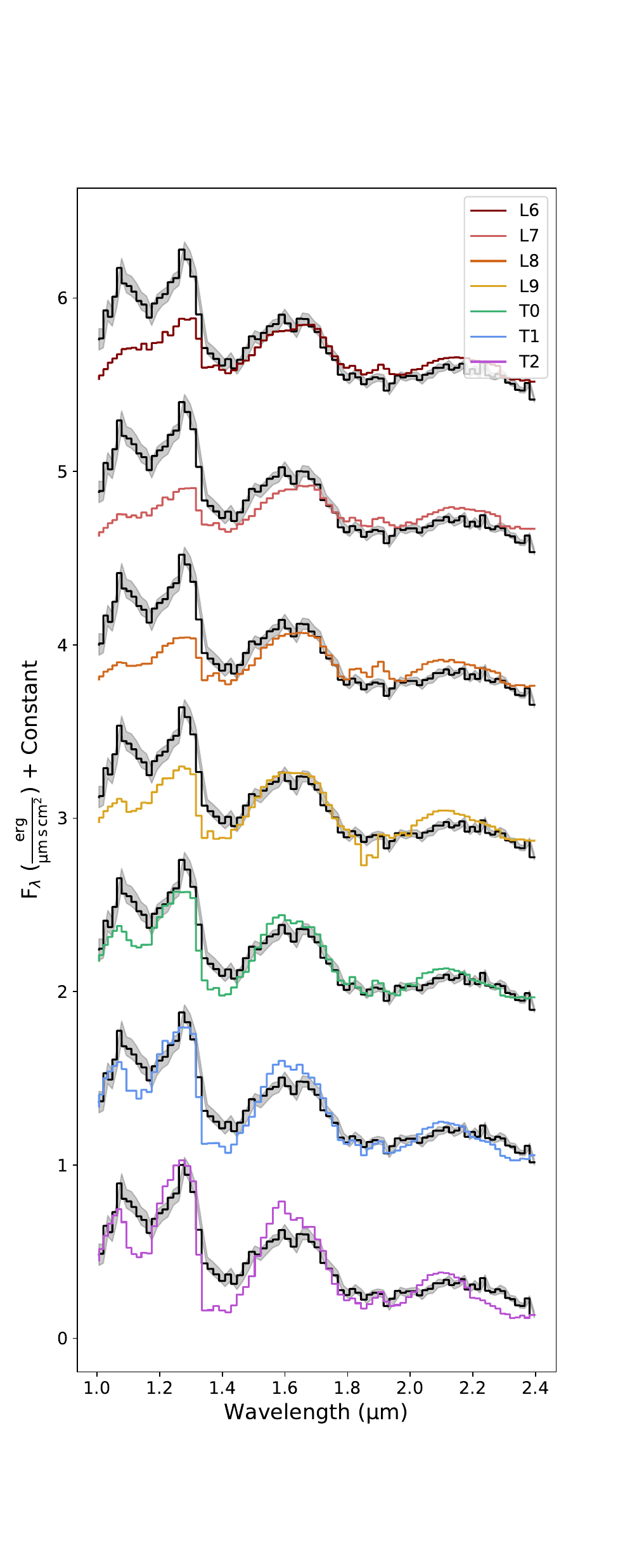}
    \end{minipage}
    \hfill
    % Second column
    \begin{minipage}[t]{0.32\textwidth}
        \vspace{0pt}
        \includegraphics[width=\textwidth]{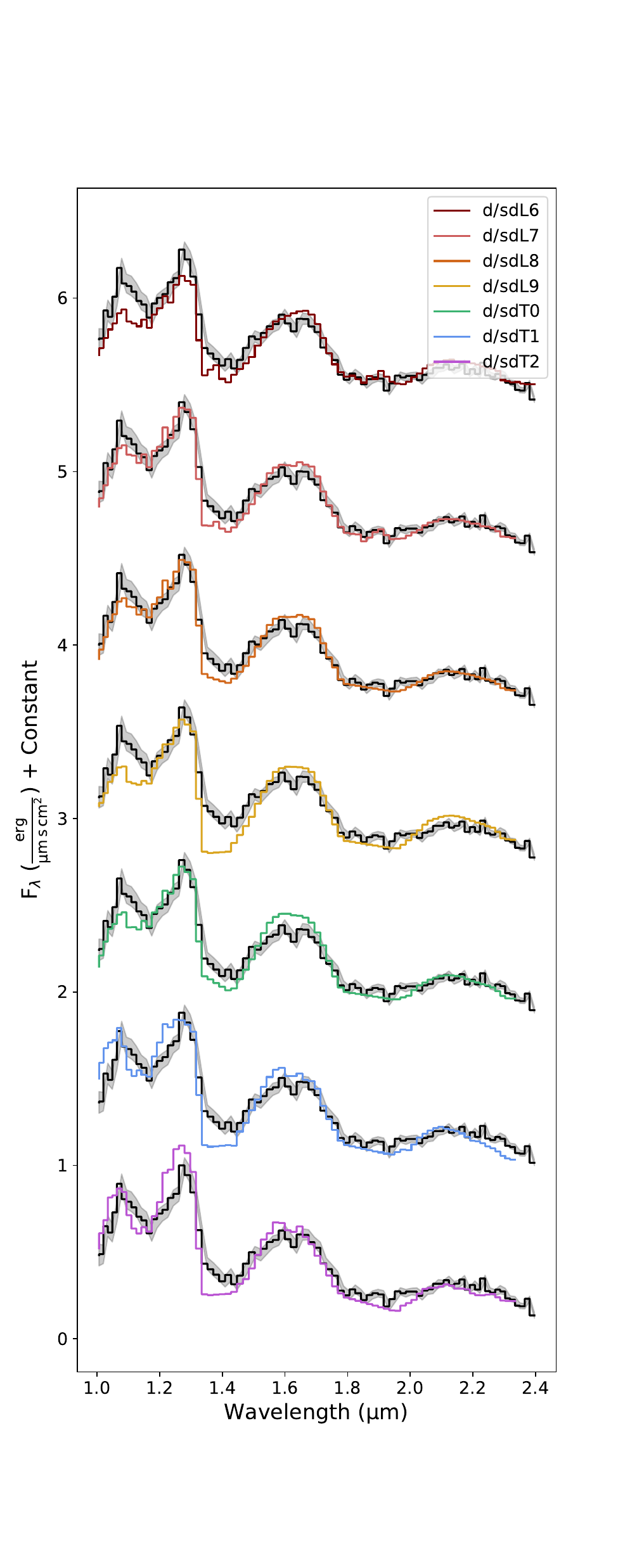}
    \end{minipage}
    \hfill
    % Third column (two stacked plots)
    \begin{minipage}[t]{0.31\textwidth}
        \vspace{0pt}
        \begin{minipage}[t]{\textwidth}
            \includegraphics[width=\textwidth]{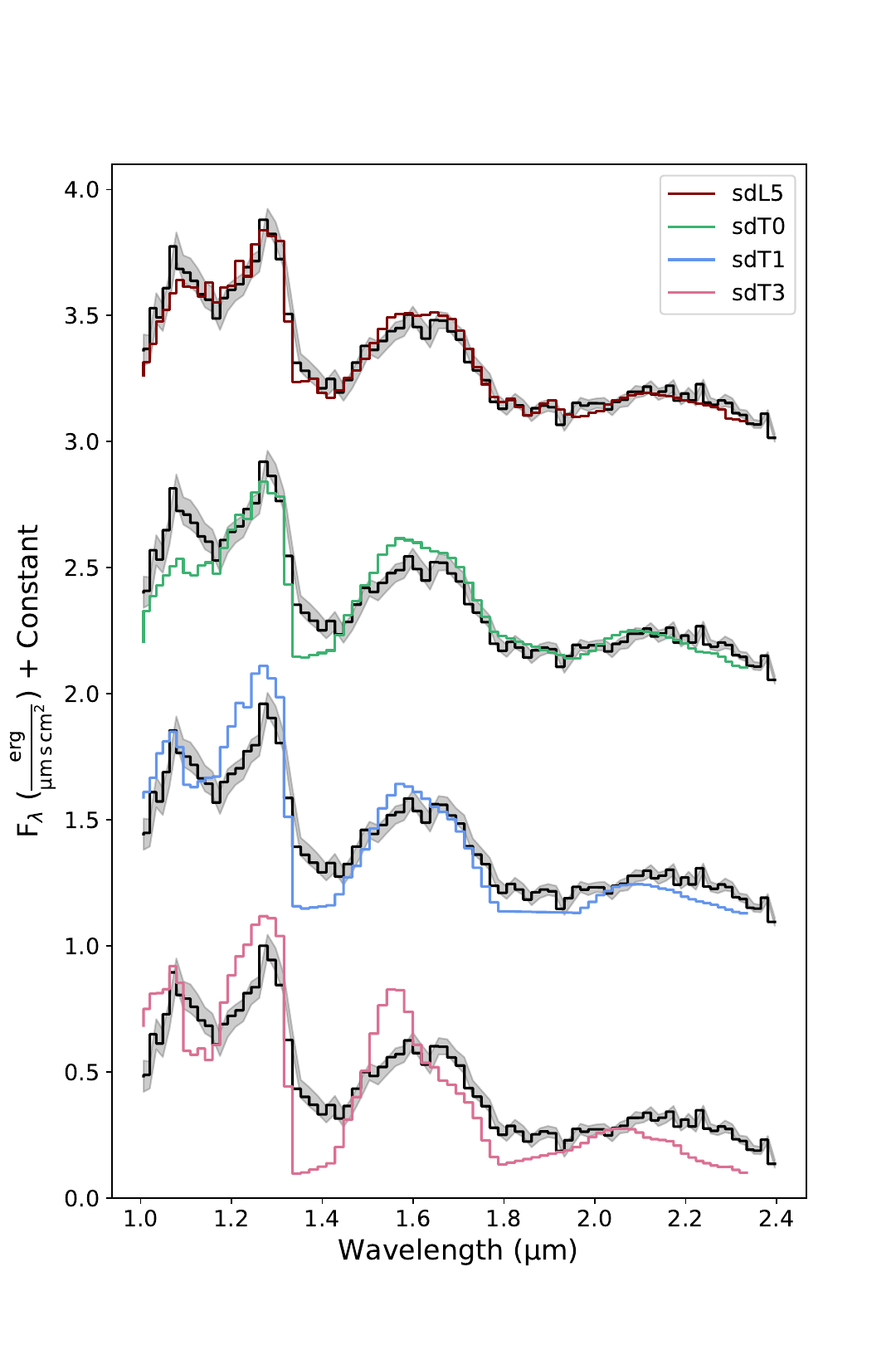}
            \includegraphics[width=\textwidth]{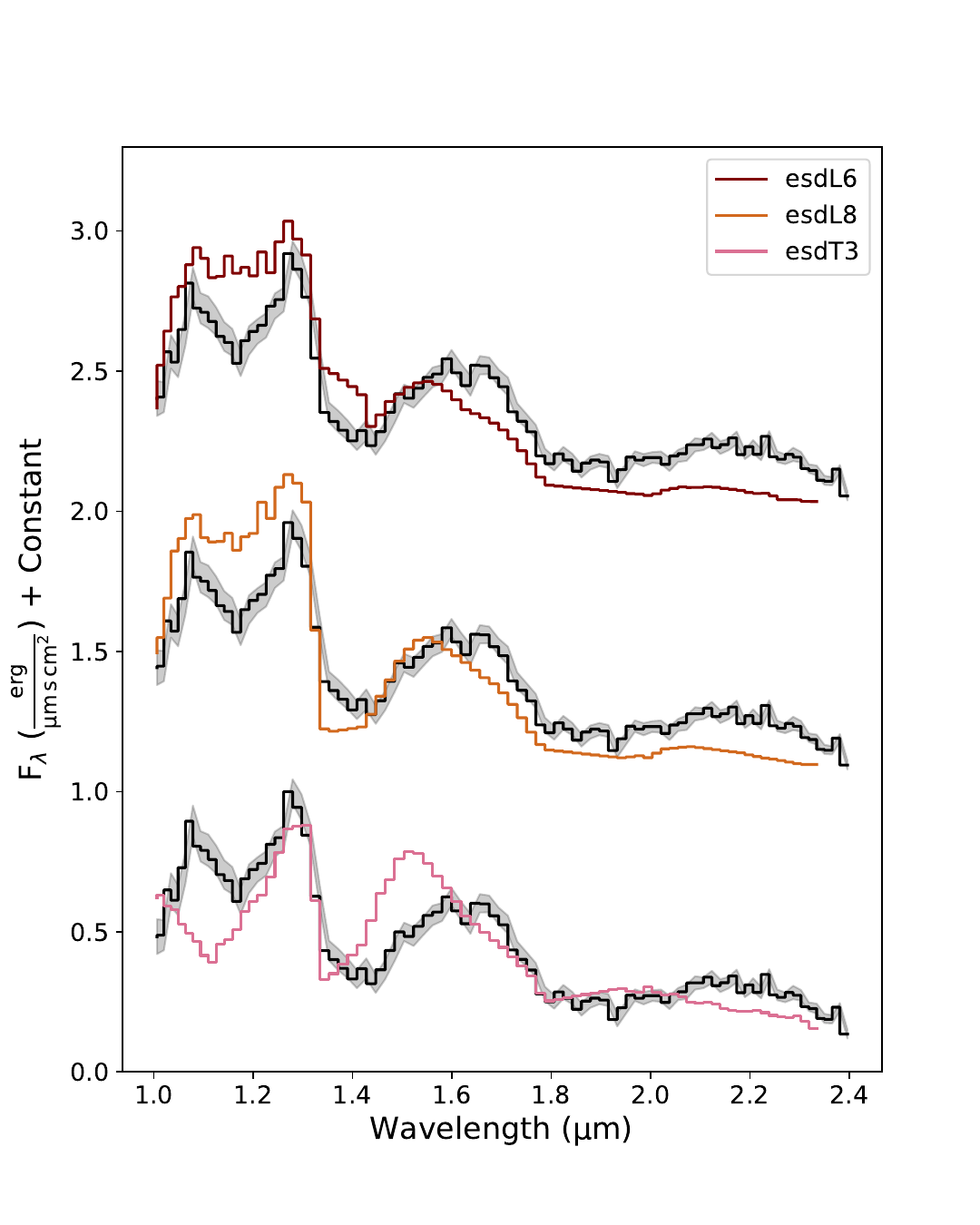}
        \end{minipage}
    \end{minipage}
    \caption{Comparison of the 0.9--2.5~{\micron} NIRSpec/Prism spectrum of RUBIES-BD-3 (black line) to
    L6 to T2 dwarf standards (left panel); 
    d/sdL6 to d/sdT2 mild subdwarf standards (center panel);
    sdL5, sdT0, sdT1, and sdT3 subdwarf standards (top right panel); and
    esdL6, esdL8, and esdT3 extreme subdwarf standards (bottom right panel).
    Standards are color-coded by spectral type. 
    All spectra are normalized and offset by constants to facilitate comparison.}
    \label{fig:rubbd3_classification}
\end{figure}

Sources and their associated best-fit standards are shown in Figure \ref{fig:spec_class} and summarized in Table~\ref{tab:classify}.
RUBIES-BD-1, -2, -6, and -7 are well-matched to T7, L1, T8, and T5 dwarf standards, respectively,
and we estimate uncertainties of 0.5 subtypes for these sources.
For RUBIES-BD-4 and -5, our best ``by-eye'' fits are to the T2 and T1 standards, respectively, with deviations generally consistent with the lower S/N of the spectra. For these sources, we estimate larger classification uncertainties of $\pm$3 and $\pm$2 subtypes, respectively.
For RUBIES-BD-3, the minimum $\chi^2$ among the dwarf standards is achieved for the L3 dwarf, but there are significant \added{discrepancies} in the 1.0--1.4~{\micron} $J$-band region.
We therefore compared this spectrum to the broader subdwarf grid (Figure~\ref{fig:rubbd3_classification}), finding the d/sdL7 mild subdwarf and sdL5 subdwarf standards to be the best fits (minimum {\chisquare}).  We adopt a classification of sd:L6$\pm$1 for this source.

With these classifications, we estimated distances to each source \added{from} NIRCam photometry, using the spectral type/absolute magnitude conversions from \citet{2024ApJ...962..177B}
based on the Sonora Bobcat models \citep{2021ApJ...920...85M}. 
For RUBIES-BD-1, -2, -4, -5, -6, and -7 we assumed a metallicity [M/H] = 0.0; 
for RUBIES-BD-3 we assumed [M/H] = $-$0.5.
Uncertainties include contributions from the measured photometry and spectral classifications, and were combined using Monte Carlo techniques.
Distance estimates are listed in Table~\ref{tab:classify}, and are in agreement across all six filters for all of the sources with the exception of RUBIES-BD-3, which varies from $1230_{-200}^{+240}$~pc for F115W photometry to $2290_{-170}^{+180}$~pc for F356W photometry. This 3.6$\sigma$ discrepancy is likely associated with the metal-poor nature of the source. 
We computed uncertainty-weighted average distance estimates by using the higher of the lower and upper error bounds as the weighting factor ($w_i = 1/{\sigma_i^2}$), and computed the combined uncertainty as
\begin{equation}\label{eqn:uncweighted}
    \langle\sigma^2\rangle = \left(\frac{\chi^2}{n-1}+1\right)/\left(\sum_{i=1}^n\frac{1}{\sigma_i^2}\right)
\end{equation}
where $n$ = 3--6 is the number of filters with distance estimates. This expression accounts for both the scatter among distance estimates and individual distance uncertainties.
Distances range from 830$\pm$90~pc for RUBIES-BD-7 to nearly 3~kpc for RUBIES-BD-4, with five of the sources having estimated distances exceeding 1~kpc. 
At the Galactic latitude $|b| \approx 60^\mathrm{o}$ of the EGS and UDS fields, these distances correspond to vertical offsets from the disk \added{plane $|Z|$ = $d\sin{|b|}$ = 720 to 2600~pc},
extending into the volume dominated by thick disk objects \citep{2008ApJ...673..864J,2024ApJ...962..177B}.

\section{Spectral Model Fits} \label{sec:methods}

\subsection{Model Grids} \label{sec:modelgrids}

To characterize the physical properties of these sources, we compared the NIRSpec spectra to three different model sets, allowing us to assess the quality of the fits across different model assumptions. 
Table \ref{table:models} summarizes the models and associated parameter ranges explored.
The Sonora Elf Owl models \citep{2024ApJ...963...73M} are cloud-free radiative-convective equilibrium models that include the effects of disequilibrium chemistry due to vertical mixing, and allow for variations in temperature ({\teff}), surface gravity ({\logg}), bulk metallicity ([M/H]), the C/O elemental abundance ratio,
and the vertical diffusion coefficient ($\log\kappa_{zz}$). 
We use the version of these models that have artificially suppressed {\phhh} abundances (see \citealt{2024ApJ...973...60B} \added{and \citealt{2025RNAAS...9..108W}}).
The Sonora Diamondback models \citep{Morley_2024} are based on the same framework but include refractory clouds in both the chemistry and atmospheric opacity using the approach of \citet{2001ApJ...556..872A}. These models are computed in chemical equilibrium and include an {\fsed} parameter to account for the efficiency of condensate sedimentation. 
Here, we adopt a value of {\fsed} = 10 to denote models without clouds.
The SAND (Spectral ANalog of Dwarfs) model set \citep{2024ApJ...971...65G, 2024RNAAS...8..134A} is based on the PHOENIX code \citep{1997ApJ...483..390H}, and was designed to explore the non-solar abundances of brown dwarfs in globular clusters and the Milky Way’s thick disk and halo. This grid samples variations in alpha element enrichment ([$\alpha$/Fe]), accounts for self-consistent condensation and gravitational settling of clouds following \cite{2012RSPTA.370.2765A}, and assumes chemical equilibrium. 

\begin{deluxetable}{lccccc}
    \tablecaption{Model Parameters \label{table:models}}
    \tablehead{
    \colhead{Parameter} & \colhead{Unit} & \colhead{Elf Owl [1]} & \colhead{Diamondback [2]} & \colhead{SAND [3,4]} & \colhead{$\sigma_{var}$\tablenotemark{a}}
    }
    \startdata 
    {\teff} & (K) & 500 to 2400 & 900 to 2400 & 700 to 3000 & 35\\
    {\logg} & (cm~s$^{-2}$) & 4.5 to 5.5 & 4.5 to 5.5 & 4.5 to 6.0 & 0.2 \\
    $[M/H]$ & (dex) & $-$0.5 to +0.5 & $-$0.5 to +0.5 & $-$0.35 to +0.3 & 0.2 \\
    {\kzz} & (cm$^2$~s$^{-1}$) & 2.0 to 9.0 & \nodata & \nodata & 0.25 \\
    C/O\tablenotemark{b} & \nodata &  0.5 to 1.5 & \nodata & \nodata & 0.05 \\
    {\fsed}\tablenotemark{c} & \nodata & \nodata & 1 to 10 & \nodata & 0.25 \\
    $[\alpha/Fe]$&  (dex) & \nodata & \nodata & $-$0.05 to +0.40 & 0.1 \\
    \enddata
\tablenotetext{a}{Width of normal distribution for parameter updated in MCMC fitting.}
\tablenotetext{b}{Relative the Sun, for which we adopt a value of C/O = 0.458 \citep{2003ApJ...591.1220L}.}
\tablenotetext{c}{A value of {\fsed} = 10 is adopted for Diamondback models without clouds; see \citet{Morley_2024}.}
\tablerefs{
[1] \citet{2024ApJ...963...73M};
[2] \citet{Morley_2024};
[3] \citet{2024RNAAS...8..134A};
[4] \citet{2024ApJ...971...65G}.
}
\end{deluxetable}

All data and models were interpolated onto a common wavelength scale that matches the variable resolution and wavelength sampling of the NIRSpec/Prism mode, and were evaluated in $F_\lambda$ flux density units scaled to each source's apparent F444W magnitudes. Because the models are computed in surface flux densities, the optimal scaling factory $\alpha$ between data and models (Eqn.~\ref{eqn:alpha}) is equal to the ratio $\left(R/d\right)^2$, where $R$ is the radius of the source and $d$ its distance. 
We assumed a fixed radius of 0.8~{\rjup}, \added{as}
expected for an evolved, relatively massive brown dwarf.\footnote{\added{The assumed radius value was estimated from evolutionary models to account for the ages and observed properties of our sources. We simulated 10,000 low-mass stars and brown dwarfs with 0.01~M$_\odot$ $\leq$ M $\leq$ to 0.1~M$_\odot$, assuming a uniform age distribution of 7--12~Gyr appropriate for the thick disk and halo \citep{2022Natur.603..599X}, a mass distribution that scales $\frac{dN}{d{\rm M}} \propto {\rm M}^{-0.6}$ \citep{2021ApJS..253....7K}, and solar metallicity evolutionary models from \citet{2001RvMP...73..719B} and \citet{2003A&A...402..701B} that are used to evolve observables to the present day. We further selected only those simulated sources with current effective temperatures 800~K $\leq$ {\teff} $\leq$ 2200~K that encompasses our sample. The median radii of these samples are
0.078$^{+0.005}_{-0.001}$~R$_\odot$ and
0.080$^{+0.004}_{-0.001}$~R$_\odot$ 
for the two evolutionary models, respectively, indicating a 5\% systematic uncertainty on our distance estimates, although there may be a larger systematic bias if model radii are incorrect.}} \citep{2021ApJ...920...85M} \added{This} scale factor provides a second estimate of the source distance 
\added{of} $d$ = 1.85$\times$10$^{-9}\alpha^{-0.5}$~pc.
%, which can
%be compared to the distances inferred from spectrophotometry.

\subsection{Markov Chain Monte Carlo Fitting} \label{sec:mcmcmethod}

We used the {\tt ucdmcmc} package \added{v1.2} \citep{ucdmcmc} to fit the three model sets to our brown dwarf spectra. 
Spectra and models were compared using the same
chi-square statistic as Eqn.~\ref{eqn:chisq}.
A first fit was conducted to identify the best individual model within a model set based on the minimum $\chi^2$ value.
These model parameters were then used to initiate a
Metropolis-Hastings Markov Chain Monte Carlo (MCMC) algorithm \citep{1953JChPh..21.1087M,HASTINGS01041970} to determine the overall best-fit parameters and their uncertainties from the posterior distributions.
Model logarithmic flux densities were linearly interpolated between grid points over logarithmic parameter values (e.g., $T_{eff}$ interpolates as log $T_{eff}$). 
The MCMC algorithm proposes a new set of parameters by varying the previous set using normal distributions with the fixed widths $\sigma_{var}$ listed in Table~\ref{table:models}.
The difference between the $\chi^2$ values of the proposed ($i+1$) and previous ($i$) parameters were compared 
to the overall minimum $\chi^2$, which in turn was compared to a random number drawn from a uniform distribution between 0 and a threshold value $T$:
\begin{equation}\label{eqn:chicomp}
    \frac{\chi^2_{i+1} - \chi^2_{i}}{\chi^2_{min}} \leq U(0,T).
\end{equation}
If this condition was satisfied, the proposed parameters were adopted; otherwise, the current parameters were retained.
Through experimentation, we found an optimal value of $T$ = 0.1. The MCMC algorithm proceeds over 5,000 steps in a single chain, and the first 25\% of the chain was removed for burn-in. 
Convergence was assessed based on the consistency of the chains and the existence of single optimal values for well-fit parameters in the posterior distributions.
Parameters and their uncertainties were computed as weighted quantiles of the marginalized distribution across the chain, using the weighting function:
\begin{equation}\label{eqn:mcmcweight}
    w_i = \frac{DOF}{DOF+\chi^2_{i} - \chi^2_{min}}
\end{equation}
which accounts for the degrees of freedom $DOF$ = (740 spectral data points) - (5--6 model parameters) for each model fit. 
We adopted the 50\% quantile (median) as the fit value and the 16\% and 84\% quantiles as estimates of the lower and upper uncertainty bounds. 
For each set of models, we conducted separate fits for both the 0.9--2.4~{\micron} near-infrared (NIR) range and the full 0.9--5.1~{\micron} range of the NIRSpec/Prism data.

\section{Results} \label{sec:results}

Best fit models for each source and model set are displayed in 
Figures~\ref{fig:nir} and~\ref{fig:nirspec} and fit parameters 
are listed in Tables~\ref{table:nir} and~\ref{table:nirspec} 
for the 0.9--2.4 {\micron} (NIR) and 0.9--5.1~{\micron} (NIRSpec) regions, respectively.  
A summary of fit parameters across models is provided in Table~\ref{tab:summary}, while
a full list of parameter values and distribution visualizations are provided in a separate Zenodo link.\footnote{\added{\url{https://zenodo.org/records/17585135}.}}
Here, we describe the outcomes of the fits and the physical interpretation of the fit parameters for individual sources. 

\subsection{RUBIES-BD-1} \label{sec:bd1}

The best NIR and NIRSpec fits for this T7 dwarf are from the Elf Owl and Diamondback models, with a slight preference for the former with modest vertical mixing ({\kzz} = 3.1$\pm$1.0 averaged between NIR and NIRSpec fits). The Diamondback models require very efficient sedimentation or no clouds ({\fsed} = 9.5$\pm$0.4). 
Averaging the results of the NIRSpec fits for these models yields
{\teff} = 930$^{+100}_{-80}$~K, {\logg} = 4.9$\pm$0.4, and [M/H] = +0.1$^{+0.3}_{-0.4}$, \added{with} Elf Owl models \added{indicating} a solar C/O ratio.
The inferred effective temperature is slightly higher but consistent with previously established trends in spectral type (e.g., {\teff} = 830$\pm$130~K for a T7$\pm$0.5 dwarf based on \citealt{2015ApJ...810..158F}). 
This modest overshoot may be due to the best-fitting models failing to match the depth of the 1.6~{\micron} CH$_4$ band or the strong CO$_2$ band at 4.1~{\micron} (see discussion below). Nevertheless, the agreement in classification and model fit parameters indicate that this is a normal late T-type brown dwarf. 

RUBIES-BD-1 was \added{reported previously} as CEERS-EGS-BD-4 in \citet{2024ApJ...964...66H}, identified from multi-band JWST/NIRCam photometry. The photometric measurements were fit to three sets of atmosphere models: Sonora Bobcat and Sonora Cholla \citep{2021ApJ...923..269K}, precursors to Diamondback and Elf Owl models; and ATMO \citep{2020A&A...637A..38P}. The fits yield {\teff} = 1000--1050~K and {\logg} = 5.0, but no evaluation of metallicity. The authors infer a distance of 1350--1831~pc for estimated radii of 0.92--1.2~{\rjup}, somewhat further than our estimate of 1030$\pm$140~pc using a smaller radius estimate.
% this is worrisome!
RUBIES JWST/NIRSpec spectral data of this source were reported in
\citet{2025ApJ...980..230T}, where the source is identified as o006\_s00089.
These authors compared the NIRSpec spectrum to Elf Owl and ATMO 2020++ \citep{2023AJ....166...57M} model sets, the latter an updated version of the ATMO models. For the \added{Elf Owl fits}, the authors inferred a similar {\teff} = 955$^{+16}_{-17}$~K but a considerably smaller {\logg} = 4.34$^{+0.18}_{-0.20}$, \added{the latter} outside the range considered in these fits, albeit formally consistent with our uncertainties. \added{The authors} also infer a much larger distance for this source, 2.0$\pm$0.2~kpc, by assuming a radius of 1.2~{\rjup} as suggested by the low surface gravity, but estimate a shorter distance of 1.4--1.7~kpc if the radius is constrained to 0.8--0.9~{\rjup}. The authors find a significantly supersolar metallicity of [M/H] = +0.41$\pm$0.05 from their model fits, above but again formally consistent with our fits. \citet{2025ApJ...980..230T} note that their parameter uncertainties, which are considerably smaller than those reported here, may be underestimated.

% Specifically, o005_s41280 is classified as
% approximately M9–L1 type, o006_s00089 as T5–T7 type, and
% o006_s35616 as L3–L4 type.

\subsection{RUBIES-BD-2} \label{sec:bd2}

The NIR spectrum for this L1 dwarf is best fit by the Diamondback models, while the full NIRSpec spectrum has equal $\chi_r^2$ values between the Diamondback and SAND models. As both models include a prescription for cloud condensation and scattering,  it is likely the presence of clouds that rules out the condensate-free Elf Owl models as a reliable fit. Indeed, Diamondback model fits to both the NIR and NIRSpec spectra require relatively thick clouds ({\fsed} = 1.8$\pm$1.0). 
However, despite providing the best fits, Figure~\ref{fig:nirspec} shows that the Diamondback and SAND models have difficulty reproducing the 0.9--2.4~{\micron} spectral slope of RUBIES-BD-2, the former underestimating the 1.6 and 2.1~{\micron} flux peaks and the latter overestimating the 1.2 and 1.6~{\micron} peaks. 
We also find that the fits between the NIR and NIRSpec ranges yield marginally distinct temperatures, with {\teff} =  1760$^{+140}_{-80}$~K for the former (Diamondback only) and {\teff} = 2040$^{+60}_{-80}$~K for the latter (average of Diamondback and SAND), a 1.7$\sigma$ difference.  
The higher temperature is more in line with local L1$\pm$0.5 dwarfs ({\teff} = 2100$\pm$130~K from \citealt{2015ApJ...810..158F}).
We also find a 3$\sigma$ difference in the inferred {\logg} value between the SAND models ({\logg} = 4.22$^{+0.26}_{-0.18}$) and Diamondback models ({\logg} = 5.30$^{+0.20}_{-0.25}$) based on fits to the NIRSpec data; and a 4$\sigma$ discrepancy between our model-scaled distance estimate of 860$^{+50}_{-60}$~pc and our spectrophotometric distance estimate of 1320$\pm$100~pc. 
These deviations suggest that the Elf Owl, Diamondback, and SAND models all have difficulty reproducing the spectrum of RUBIES-BD-2, indicating potentially missing elements to the models. 

RUBIES JWST/NIRSpec spectral data of this source were \added{reported previously} in
\citet{2025ApJ...980..230T}, where the source is identified as o006\_s35616.
The authors compared these data to Elf Owl and BT-Settl CIFIST models \citep{2012RSPTA.370.2765A,2011SoPh..268..255C},
finding {\teff} = 1810--1910~K, lower but consistent with our NIRSpec fits. However, the authors again infer a very low {\logg} = 3.25$\pm$0.01 and a supersolar [M/H] = +1.00$\pm$0.01 from their Elf Owl fits, inconsistent with our findings.
\citet{2025ApJ...980..230T} note significant residuals in their model fits, recognizing that neither the Elf Owl nor BT-Settl models include condensate opacity.
They introduce a reddening parameter based on the interstellar extinction law of \citet{2024ApJ...964L...3W}, and find that $A_V$ = 3~mag substantially improves their Elf Owl model fits while yielding a substantially higher {\teff}  = 2286$^{+9}_{-8}$~K and subsolar [M/H] = $-$0.71$\pm$0.02.
It is likely that a combination of atmospheric condensate opacity and interstellar reddening are required to accurately reproduce these data.

\subsection{RUBIES-BD-3} \label{sec:bd3}

As discussed in Section~\ref{tab:classify}, RUBIES-BD-3 is better matched to subdwarf standards, suggesting that it should be well-matched to metal-poor atmosphere models.
However, both NIR and NIRSpec fits yield consistently better fits to {supersolar} metallicity models. In the NIR range, Elf Owl, Diamondback, and SAND models all yield comparable best-fit $\chi^2_r$ values; in the NIRSpec range, the Elf Owl models are somewhat less reliable. Primary fit parameters of {\teff}, {\logg}, and [M/H] are in agreement across all three models and fit ranges, with a relatively high {\teff} = 2170$\pm$130~K from the NIRSpec fits, notably warmer than the earlier-type L dwarf RUBIES-BD-2. 
Secondary parameters are similar to NIRSpec fits for RUBIES-BD-2, with all three models indicating relatively low surface gravities ({\logg} = 4.7$\pm$0.4), and for the Diamondback models thick clouds ({\fsed} = 2.1$^{+1.7}_{-0.9}$).
The Elf Owl model fits also indicate a high C/O = 0.73$^{+0.08}_{-0.10}$ and modest vertical mixing ({\kzz} = 3.6$^{+2.2}_{-1.3}$). 
However, like RUBIES-BD-2, all three models have clear flaws in reproducing the observed spectrum, particularly in the NIR range where Elf Owl models show excess flux shortward of 1.3~{\micron}, and  SAND models fail to match the shape of the 1.6~{\micron} peak. None of the models accurately reproduce the 1.2~{\micron} ``notch'' absorption feature that is characteristic of L- and T-type subdwarfs (Figure~\ref{fig:rubbd3_classification}). 
%There is also a similar discrepancy in the estimated distances based on the model scaling (2240$\pm$250~pc) and the spectophotometry (1420$\pm$140~pc), in this case with the model-based distance being 2.9$\sigma$ larger.
Again, the models deployed for this seemingly metal-poor L dwarf may be missing a critical element of atmospheric chemistry, composition, or dynamics that may be skewing the fits toward higher metallicities.

\citet{2025ApJ...980..230T} reported RUBIES JWST/NIRSpec spectral data for this source, identified as o005\_s41280, and compared the data to Elf Owl and BT-Settl models.
The authors find a similar {\teff} = 2160--2260~K from these fits without taking into account reddening, with an approximately 200~K increase when a modest reddening of $A_V$ = 1.0-1.5~mag is taken into account. In both cases, the authors find a very low surface gravity for this source, {\logg} = 3.3--3.5, well outside our fitting parameter range. As with RUBIES-BD-2, such a low surface gravity is highly unlikely for a field brown dwarf as it would require an age $<$1~Myr \citep{2023A&A...671A.119C}.

\subsection{RUBIES-BD-4} \label{sec:bd4}

The relatively low S/N of the RUBIES-BD-4 spectrum results in equivalent matches among all three models in both NIR and NIRSpec ranges, and relatively large uncertainties in inferred parameters.
The poorly-constrained {\teff} = 1800$^{+300}_{-700}$~K is higher but not inconsistent with trends among solar metallicity dwarfs for its T2$\pm$3 classification ({\teff} = 1180$\pm$200~K; \citealt{2015ApJ...810..158F}).
Notably, SAND model fits to the NIRSpec data yield a considerably lower {\teff} = 1050$^{+260}_{-230}$~K as compared to the other models, even though the best-fit model has a {\teff} = 1506~K (Figure~\ref{fig:nirspec}).
Secondary parameters indicate efficient vertical mixing 
({\kzz} = 7.8$\pm$1.0 from Elf Owl models),
thin clouds ({\fsed} = 8.4$\pm$0.7 from Diamondback models),
and near-solar elemental abundances ([M/H] = $-$0.1$^{+0.6}_{-0.5}$ from all models, C/O = 0.54$^{+0.21}_{-0.19}$ from Elf Owl models, and [$\alpha$/M] = +0.17$^{+0.17}_{-0.13}$ from SAND models). The large range in temperatures across the models results in a comparably large uncertainty in the estimated distance, from $\sim$2000~pc for SAND to $\sim$6400~pc for Diamondback, which straddle the \added{highly} uncertain 5400$^{+2200}_{-3400}$~pc distance estimate inferred from its spectral type and NIRCam photometry (Table~\ref{tab:classify}).

\subsection{RUBIES-BD-5} \label{sec:bd5}

RUBIES-BD-5 is best fit in both the NIR and NIRspec ranges by the SAND models, yielding an atmospheric temperature {\teff} = 1270$^{+80}_{-60}$~K that is in line with its T1$\pm$2 classification
({\teff} = 1220$\pm$150~K; \citealt{2015ApJ...810..158F}). Figure~\ref{fig:nirspec} shows that the SAND models do a better job reproducing the 1--2.2~{\micron} and 4.0~{\micron} peaks compared to the Elf Owl and Diamondback models, both of which produce overly blue spectra. 
The NIRSpec SAND fits also yield a significantly subsolar metallicity ([M/H] = $-$0.7$^{+0.4}_{-0.3}$) and modest alpha enrichment ([$\alpha$/M] = +0.19$^{+0.14}_{-0.17}$), both indicative of a thick disk or halo origin. 

\subsection{RUBIES-BD-6} \label{sec:bd6}

RUBIES-BD-6, like RUBIES-BD-4, has a relatively low S/N spectrum and is equally-well fit by all three models in both NIR and NIRSpec ranges, with a slight preference for the Elf Owl models in the latter.
The NIRSpec fits yield an average {\teff} = 830$\pm$130~K, high but consistent with its T8$\pm$1 classification
({\teff} = 680$\pm$190~K; \citealt{2015ApJ...810..158F}).
Given the large uncertainties, there is no evidence of a metallicity or elemental compositions significantly distinct from solar, while the Diamondback models require efficient sedimentation ({\fsed} = 8.0$^{+1.4}_{-0.7}$) as is typical for late T dwarfs.

\subsection{RUBIES-BD-7} \label{sec:bd7}

RUBIES-BD-7 is best fit in both the NIR and NIRSpec ranges by the Elf Owl and Diamondback models, the latter requiring very efficient sedimentation ({\fsed} = 9.6$\pm$0.4, consistent with no clouds). The inferred {\teff} = 1180$\pm$80~K is consistent with other T5$\pm$0.5 dwarfs
({\teff} = 1030$\pm$120~K; \citealt{2015ApJ...810..158F}), while the metallicity ([M/H] = $-$0.2$\pm$0.3) and Elf Owl C/O = 0.48$^{+0.16}_{-0.12}$ are consistent with solar composition. The excellent agreement between the model scaling and spectrophotometric distances further indicate that this is likely a normal thin disk brown dwarf.

\begin{longrotatetable}
\begin{deluxetable*}{cccccccccc}
    \tablecaption{MCMC Parameters for 0.9--2.4~{\micron} Fits (NIR Range) \label{table:nir}}
    \tablehead{
    \colhead{Model} & \colhead{\teff} & \colhead{\logg} & \colhead{[$M/H$]} & \colhead{C/O}  & \colhead{\kzz} & \colhead{\fsed} & \colhead{[$\alpha/Fe$]} & \colhead{$d$} & \colhead{Min.~$\chi^2_r$}\\
    & \colhead{(K)} & \colhead{(cm~s$^{-2}$)} & \colhead{(dex)} & & \colhead{(cm$^2$~s$^{-1}$)} & & \colhead{(dex)} 
     & \colhead{(pc)} & 
    }
\startdata
\multicolumn{10}{c}{RUBIES-BD-1 (T7$\pm$0.5)} \\ 
\hline 
E & $890_{-80}^{+70}$ & $4.9\pm0.3$ & $0.2\pm0.2$ & $0.44_{-0.10}^{+0.16}$ & $2.9_{-0.7}^{+1.3}$ &  \nodata &  \nodata & $980_{-170}^{+180}$ & 2.8 \\ 
D & $960_{-50}^{+70}$ & $4.9_{-0.3}^{+0.4}$ & $0.1\pm0.3$ &  \nodata &  \nodata & $9.5_{-0.4}^{+0.5}$ &  \nodata & $1150_{-140}^{+200}$ & 3.0 \\ 
S & $770_{-70}^{+130}$ & $4.8_{-0.3}^{+0.8}$ & $0.12_{-0.03}^{+0.13}$ &  \nodata &  \nodata &  \nodata & $0.00_{-0.03}^{+0.10}$ & $660_{-130}^{+370}$ & 3.5 \\ 
\hline 
\multicolumn{10}{c}{RUBIES-BD-2 (L1$\pm$0.5)} \\ 
\hline 
E & $2120_{-100}^{+110}$ & $4.9_{-0.3}^{+0.4}$ & $0.37_{-0.22}^{+0.13}$ & $0.65_{-0.06}^{+0.04}$ & $2.8_{-0.7}^{+1.5}$ &  \nodata &  \nodata & $920_{-90}^{+100}$ & 77 \\ 
D & $1760_{-80}^{+140}$ & $5.30_{-0.26}^{+0.17}$ & $0.31_{-0.25}^{+0.18}$ &  \nodata &  \nodata & $2.0_{-0.8}^{+1.3}$ &  \nodata & $630_{-60}^{+90}$ & 19 \\ 
S & 1830$\pm$30 & $5.51_{-0.07}^{+0.06}$ & $0.14_{-0.04}^{+0.09}$ &  \nodata &  \nodata &  \nodata & $-0.01_{-0.03}^{+0.04}$ & 650$\pm$20 & 29 \\ 
\hline 
\multicolumn{10}{c}{RUBIES-BD-3 (sd:L6$\pm$1)} \\ 
\hline 
E & $2220_{-90}^{+110}$ & $4.7_{-0.2}^{+0.4}$ & $0.33_{-0.21}^{+0.17}$ & $0.62_{-0.09}^{+0.06}$ & $3.1_{-0.9}^{+1.8}$ &  \nodata &  \nodata & $2690_{-210}^{+270}$ & 3.8 \\ 
D & $1990_{-120}^{+160}$ & $5.0\pm0.3$ & $0.2\pm0.3$ &  \nodata &  \nodata & $2.8\pm1.3$ &  \nodata & $2150_{-230}^{+350}$ & 4.0 \\ 
S & $2250_{-130}^{+140}$ & $4.70_{-0.17}^{+0.26}$ & $0.16_{-0.31}^{+0.14}$ &  \nodata &  \nodata &  \nodata & $-0.01\pm0.04$ & $2760_{-290}^{+330}$ & 4.2 \\ 
\hline 
\multicolumn{10}{c}{RUBIES-BD-4 (T2$\pm$3)} \\ 
\hline 
E & 1870$\pm$360 & $5.0\pm0.4$ & $-0.1_{-0.3}^{+0.4}$ & $0.47_{-0.15}^{+0.14}$ & $6_{-3}^{+2}$ &  \nodata &  \nodata & $6200_{-2300}^{+2600}$ & 1.8 \\ 
D & $1900_{-370}^{+300}$ & $5.0\pm0.4$ & $-0.0\pm0.4$ &  \nodata &  \nodata & $8.4_{-1.0}^{+0.9}$ &  \nodata & $6400_{-2200}^{+2100}$ & 1.7 \\ 
S & $1510_{-340}^{+630}$ & $5.2_{-0.5}^{+0.6}$ & $-0.0_{-0.2}^{+0.3}$ &  \nodata &  \nodata &  \nodata & $0.04_{-0.09}^{+0.14}$ & $4100_{-1800}^{+4100}$ & 1.7 \\ 
\hline 
\multicolumn{10}{c}{RUBIES-BD-5 (T1$\pm$2)} \\ 
\hline 
E & $1840_{-190}^{+220}$ & $5.0\pm0.4$ & $0.2\pm0.3$ & $0.48\pm0.17$ & $7.9\pm0.8$ &  \nodata &  \nodata & $4720_{-960}^{+1230}$ & 2.1 \\ 
D & $1640_{-200}^{+280}$ & $4.9\pm0.3$ & $0.1_{-0.4}^{+0.3}$ &  \nodata &  \nodata & $8.5_{-1.1}^{+1.0}$ &  \nodata & $3790_{-890}^{+1360}$ & 2.1 \\ 
S & $1370_{-80}^{+190}$ & $5.6_{-0.5}^{+0.2}$ & $0.0_{-0.3}^{+0.2}$ &  \nodata &  \nodata &  \nodata & $0.11_{-0.14}^{+0.08}$ & $2540_{-290}^{+830}$ & 1.6 \\ 
\hline 
\multicolumn{10}{c}{RUBIES-BD-6 (T8$\pm$0.5)} \\ 
\hline 
E & $840_{-180}^{+210}$ & $5.1_{-0.4}^{+0.3}$ & $-0.0_{-0.3}^{+0.4}$ & $0.52_{-0.18}^{+0.11}$ & $4.1_{-1.4}^{+1.9}$ &  \nodata &  \nodata & $1460_{-610}^{+900}$ & 1.0 \\ 
D & $990_{-80}^{+180}$ & $5.1_{-0.4}^{+0.3}$ & $-0.1\pm0.3$ &  \nodata &  \nodata & $9.3_{-1.3}^{+0.6}$ &  \nodata & $2110_{-360}^{+970}$ & 1.1 \\ 
S & $890_{-180}^{+160}$ & $5.6_{-0.4}^{+0.3}$ & $0.13_{-0.21}^{+0.14}$ &  \nodata &  \nodata &  \nodata & $0.11_{-0.12}^{+0.07}$ & $1660_{-750}^{+770}$ & 1.0 \\ 
\hline 
\multicolumn{10}{c}{RUBIES-BD-7 (T5$\pm$0.5)} \\ 
\hline 
E & 1140$\pm$80 & $4.9\pm0.4$ & $-0.2_{-0.2}^{+0.3}$ & $0.41_{-0.12}^{+0.15}$ & $4.1_{-1.4}^{+1.2}$ &  \nodata &  \nodata & 710$\pm$100 & 13 \\ 
D & 1219$\pm$100 & $5.1_{-0.4}^{+0.3}$ & $-0.2_{-0.2}^{+0.3}$ &  \nodata &  \nodata & $9.5_{-0.4}^{+0.5}$ &  \nodata & 850$\pm$150 & 15 \\ 
S & $1100_{-150}^{+90}$ & $5.75_{-0.24}^{+0.18}$ & $-0.1_{-0.2}^{+0.3}$ &  \nodata &  \nodata &  \nodata & $0.12_{-0.17}^{+0.08}$ & $670_{-200}^{+120}$ & 18 \\ 
\enddata
\tablecomments{Model labels are: E = Elf Owl \citep{2024ApJ...963...73M},
D = Diamondback \citep{Morley_2024}, and
S = SAND \citep{2024ApJ...971...65G, 2024RNAAS...8..134A}.}
%Models with an asterisk (*) indicate the best overall fit for the spectrum.}
\end{deluxetable*}
\end{longrotatetable}

\begin{longrotatetable}
\begin{deluxetable*}{cccccccccc}
    \tablecaption{MCMC Parameters for 0.9--5.1{\micron} Fits (NIRSpec Range) \label{table:nirspec}}
    \tablehead{
    \colhead{Model} & \colhead{\teff} & \colhead{\logg} & \colhead{[$M/H$]} & \colhead{C/O} & \colhead{\kzz} & \colhead{\fsed} & \colhead{[$\alpha/Fe$]} & \colhead{$d$} & \colhead{Min.~$\chi^2_r$}\\
    & \colhead{(K)} & \colhead{(cm~s$^{-2}$)} & \colhead{(dex)} & & \colhead{(cm$^2$~s$^{-1}$)} & & \colhead{(dex)} 
     & \colhead{(pc)} & 
    }
    \startdata
\multicolumn{10}{c}{RUBIES-BD-1 (T7$\pm$0.5)} \\ 
\hline 
E & 890$\pm$80 & $5.0\pm0.4$ & $0.2_{-0.3}^{+0.4}$ & $0.46_{-0.12}^{+0.15}$ & $3.3_{-1.0}^{+1.6}$ &  \nodata &  \nodata & 980$\pm$170 & 2.6 \\ 
D & $980_{-60}^{+90}$ & $4.9_{-0.6}^{+0.5}$ & $0.0_{-0.4}^{+0.3}$ &  \nodata &  \nodata & $9.4_{-0.7}^{+0.5}$ &  \nodata & $1180_{-150}^{+250}$ & 3.1 \\ 
S & $960_{-80}^{+100}$ & $5.7\pm0.2$ & $0.21_{-0.16}^{+0.09}$ &  \nodata &  \nodata &  \nodata & $0.03_{-0.05}^{+0.13}$ & $1190_{-190}^{+280}$ & 4.0 \\ 
\hline 
\multicolumn{10}{c}{RUBIES-BD-2 (L1$\pm$0.5)} \\ 
\hline 
E & $2150_{-110}^{+140}$ & $4.7_{-0.2}^{+0.4}$ & $0.7_{-0.4}^{+0.3}$ & $0.65_{-0.07}^{+0.04}$ & $3.0_{-0.7}^{+1.8}$ &  \nodata &  \nodata & $940_{-90}^{+120}$ & 49 \\ 
D & $2020_{-100}^{+90}$ & $5.30_{-0.25}^{+0.20}$ & $0.3_{-0.3}^{+0.2}$ &  \nodata &  \nodata & $1.6_{-0.6}^{+1.4}$ &  \nodata & $840_{-80}^{+70}$ & 24 \\ 
S & 2050$\pm$50 & $4.22_{-0.18}^{+0.26}$ & $-0.0_{-0.4}^{+0.2}$ &  \nodata &  \nodata &  \nodata & $-$0.05\tablenotemark{a} & $860_{-30}^{+40}$ & 24 \\ 
\hline 
\multicolumn{10}{c}{RUBIES-BD-3 (sd:L6$\pm$1)} \\ 
\hline 
E & $2220_{-90}^{+120}$ & $4.8_{-0.2}^{+0.4}$ & $0.7_{-0.4}^{+0.3}$ & $0.61_{-0.09}^{+0.06}$ & $3.6_{-1.3}^{+2.2}$ &  \nodata &  \nodata & $2650_{-210}^{+270}$ & 3.3 \\ 
D & $2060_{-110}^{+120}$ & $4.8_{-0.6}^{+0.5}$ & $0.2_{-0.4}^{+0.3}$ &  \nodata &  \nodata & $2.1_{-0.9}^{+1.7}$ &  \nodata & $2290_{-210}^{+260}$ & 2.8 \\ 
S & $2190_{-90}^{+120}$ & $4.4_{-0.3}^{+0.4}$ & $0.1_{-0.5}^{+0.2}$ &  \nodata &  \nodata &  \nodata & $-0.039_{-0.011}^{+0.068}$ & $2600_{-200}^{+260}$ & 2.7 \\ 
\hline 
\multicolumn{10}{c}{RUBIES-BD-4 (T2$\pm$3)} \\ 
\hline 
E & $1890_{-450}^{+300}$ & $5.1_{-0.4}^{+0.3}$ & $0.2_{-0.8}^{+0.7}$ & $0.45_{-0.15}^{+0.17}$ & $8.0_{-0.9}^{+0.8}$ &  \nodata &  \nodata & $6100_{-2400}^{+2000}$ & 2.2 \\ 
D & $1930_{-220}^{+240}$ & $5.0_{-0.6}^{+0.4}$ & $0.1_{-0.4}^{+0.3}$ &  \nodata &  \nodata & $8.5\pm1.1$ &  \nodata & $6400_{-1300}^{+1700}$ & 2.3 \\ 
S & $1050_{-230}^{+260}$ & $5.0_{-0.7}^{+0.6}$ & $-0.4_{-0.6}^{+0.5}$ &  \nodata &  \nodata &  \nodata & $0.17_{-0.13}^{+0.17}$ & $1960_{-590}^{+980}$ & 2.1 \\ 
\hline 
\multicolumn{10}{c}{RUBIES-BD-5 (T1$\pm$2)} \\ 
\hline 
E & $1570_{-240}^{+310}$ & $5.0\pm0.4$ & $0.1\pm0.5$ & $0.42_{-0.15}^{+0.20}$ & $8.1_{-0.8}^{+0.7}$ &  \nodata &  \nodata & $3140_{-850}^{+1430}$ & 3.7 \\ 
D & $1410_{-140}^{+210}$ & $4.2_{-0.2}^{+0.5}$ & $0.2_{-0.4}^{+0.2}$ &  \nodata &  \nodata & $8.0_{-0.7}^{+0.9}$ &  \nodata & $2550_{-440}^{+920}$ & 3.0 \\ 
S & $1270_{-70}^{+80}$ & $5.0_{-0.4}^{+0.5}$ & $-0.7\pm0.4$ &  \nodata &  \nodata &  \nodata & $0.19_{-0.17}^{+0.14}$ & $2050_{-210}^{+280}$ & 1.6 \\ 
\hline 
\multicolumn{10}{c}{RUBIES-BD-6 (T8$\pm$0.5)} \\ 
\hline 
E & $730_{-80}^{+120}$ & $5.0\pm0.3$ & $-0.3_{-0.4}^{+0.6}$ & $0.39_{-0.12}^{+0.19}$ & $3.7_{-1.3}^{+1.8}$ &  \nodata &  \nodata & $1010_{-200}^{+300}$ & 1.1 \\ 
D & $940_{-40}^{+80}$ & $5.0_{-0.5}^{+0.3}$ & $-0.2\pm0.3$ &  \nodata &  \nodata & $8.0_{-0.7}^{+1.4}$ &  \nodata & $1600_{-130}^{+280}$ & 1.4 \\ 
S & $800_{-80}^{+160}$ & $5.7_{-0.4}^{+0.2}$ & $0.11_{-0.53}^{+0.15}$ &  \nodata &  \nodata &  \nodata & $0.04_{-0.06}^{+0.13}$ & $1220_{-220}^{+490}$ & 1.3 \\ 
\hline 
\multicolumn{10}{c}{RUBIES-BD-7 (T5$\pm$0.5)} \\ 
\hline 
E & 1170$\pm$80 & $4.8_{-0.3}^{+0.4}$ & $-0.3\pm0.3$ & $0.40_{-0.10}^{+0.14}$ & $3.3_{-0.9}^{+1.2}$ &  \nodata &  \nodata & $750_{-90}^{+100}$ & 7 \\ 
D & 1186$\pm$90 & $5.0\pm0.4$ & $-0.1\pm0.3$ &  \nodata &  \nodata & $9.6\pm0.4$ &  \nodata & $800_{-110}^{+130}$ & 8 \\ 
S & $1170_{-60}^{+90}$ & $5.79_{-0.19}^{+0.15}$ & $-1.7_{-0.6}^{+1.2}$ &  \nodata &  \nodata &  \nodata & $-$0.05\tablenotemark{a} & $810_{-70}^{+120}$ & 14 \\ 
\enddata
\tablecomments{Model labels are: E = Elf Owl \citep{2024ApJ...963...73M},
D = Diamondback \citep{Morley_2024}, and
S = SAND \citep{2024ApJ...971...65G, 2024RNAAS...8..134A}.}
%Models with an asterisk (*) indicate the best overall fit for the spectrum.}
\tablenotetext{a}{MCMC fits were unable to converge on this parameter.}
\end{deluxetable*}
\end{longrotatetable}

\begin{figure}[h]
\centering
    \fbox{\includegraphics[width=.3\textwidth]{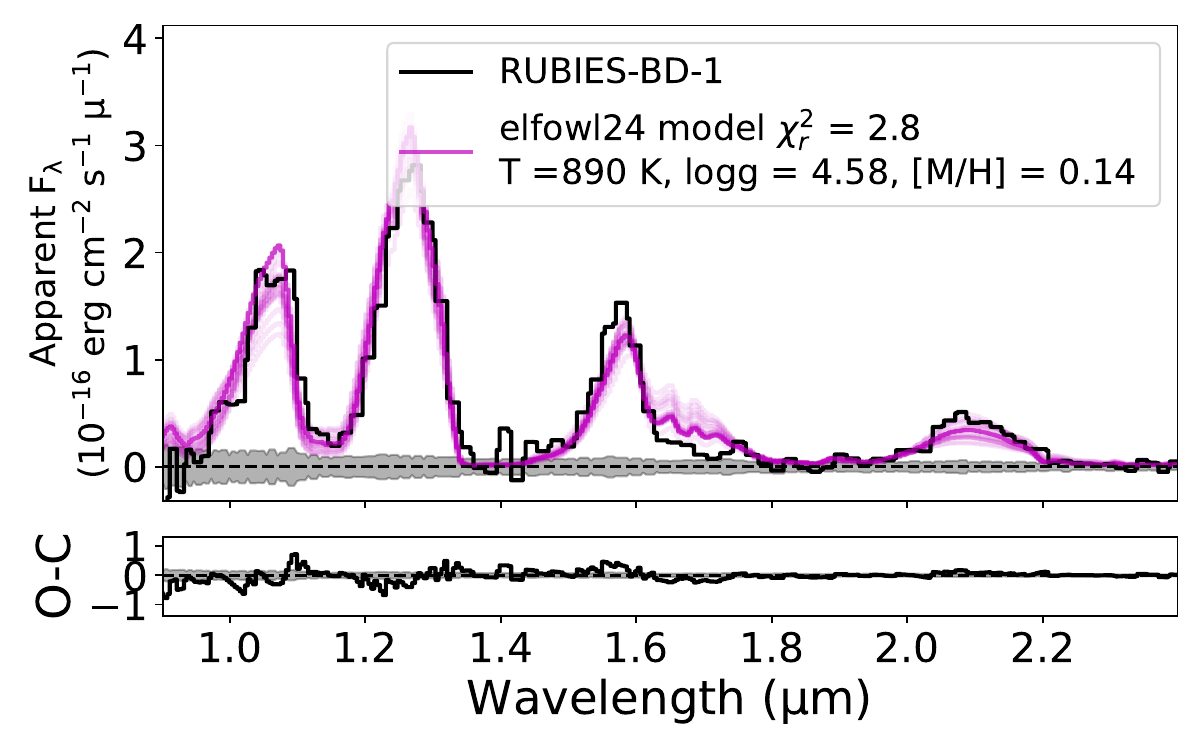}}\hfill
    \includegraphics[width=.3\textwidth]{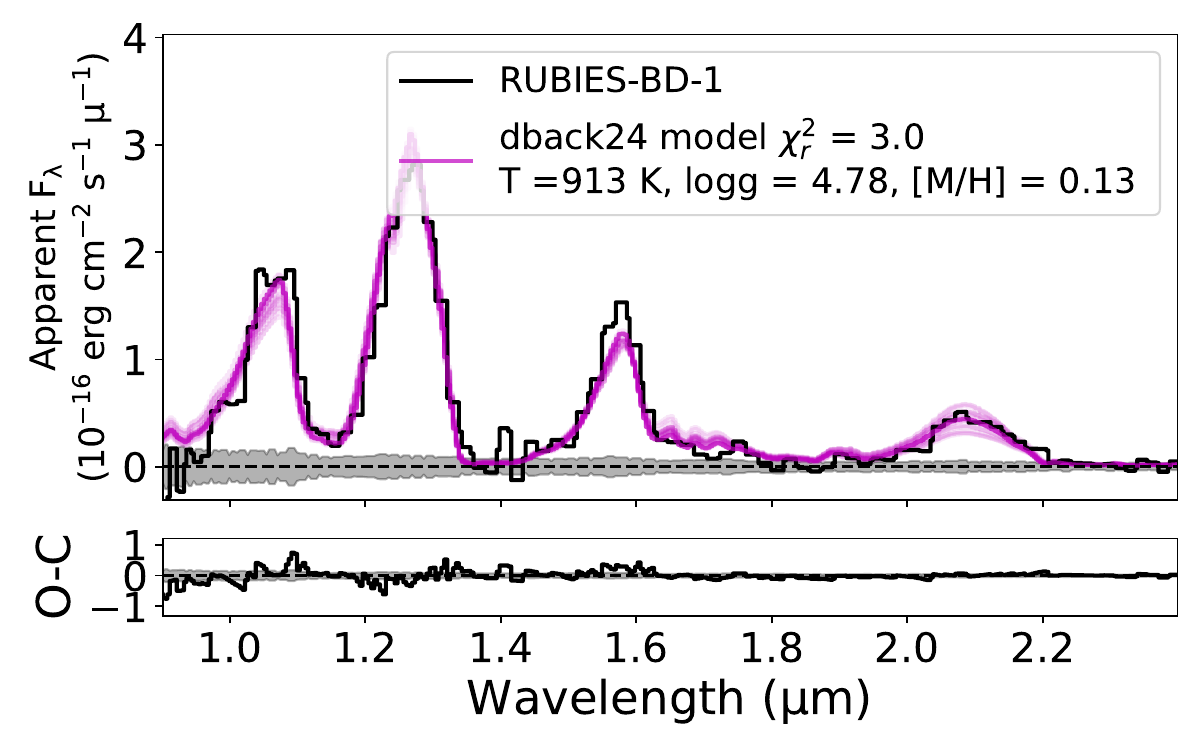}\hfill
    \includegraphics[width=.3\textwidth]{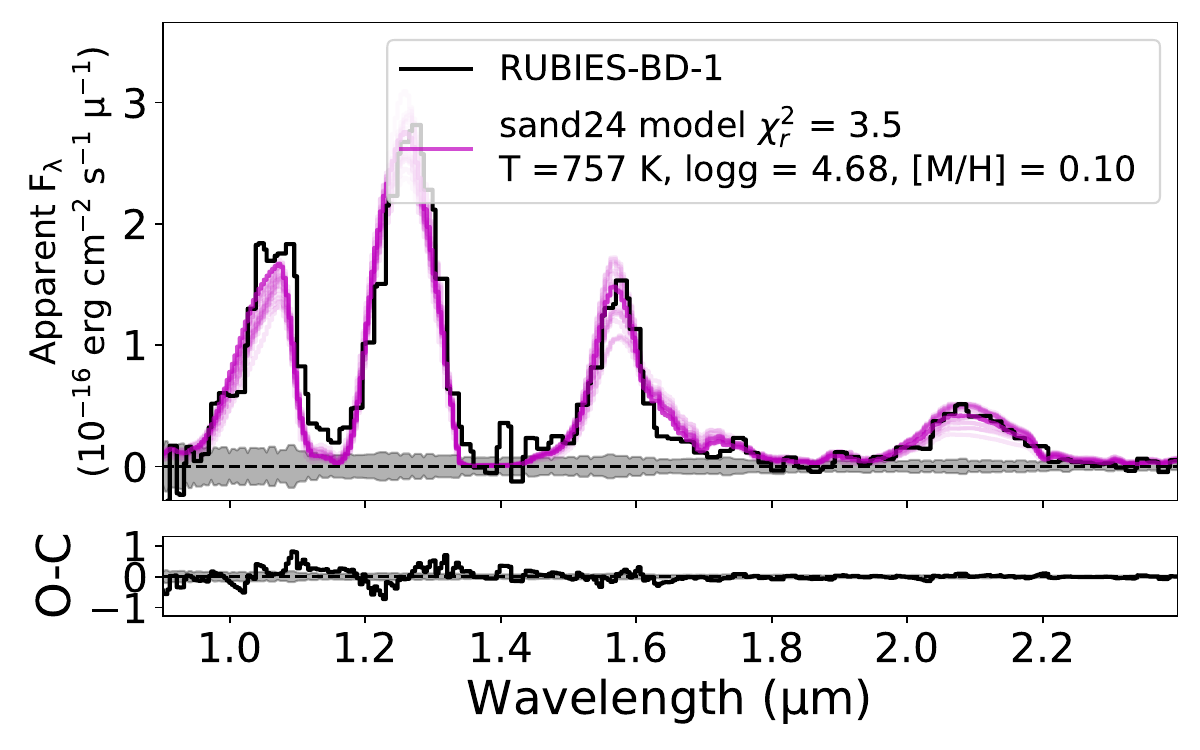}\hfill
    \\[\smallskipamount]
    \includegraphics[width=.3\textwidth]{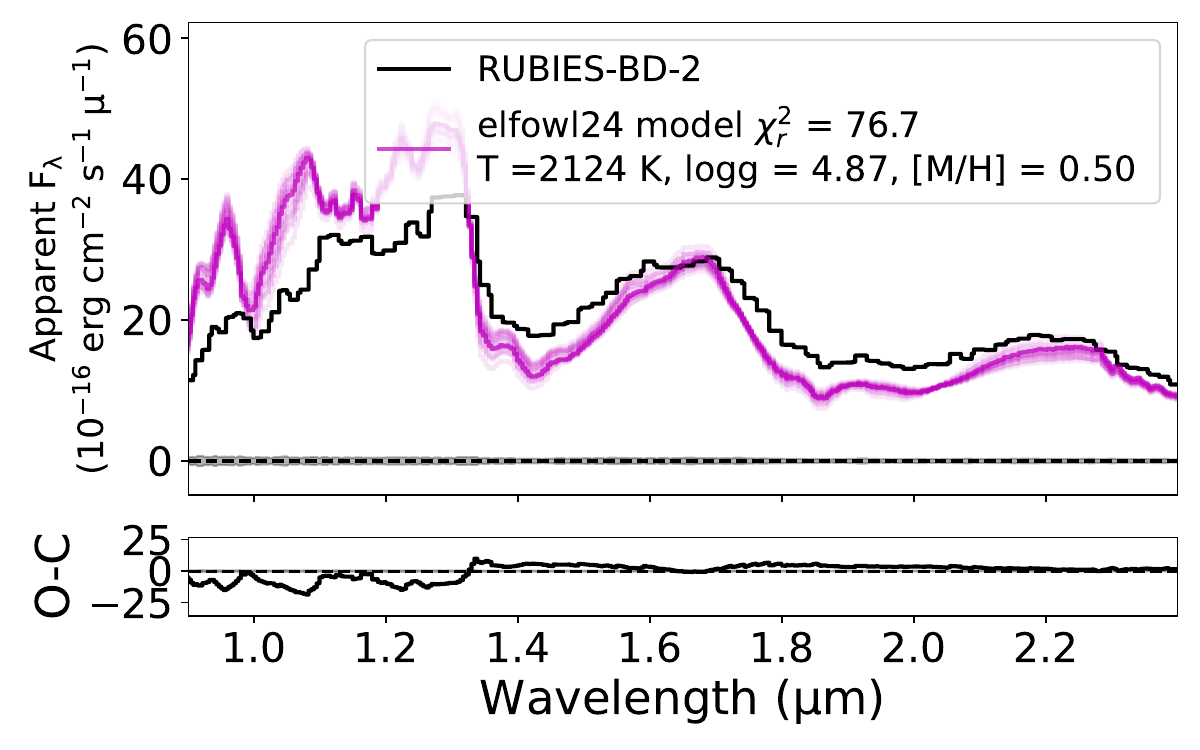}\hfill
    \fbox{\includegraphics[width=.3\textwidth]{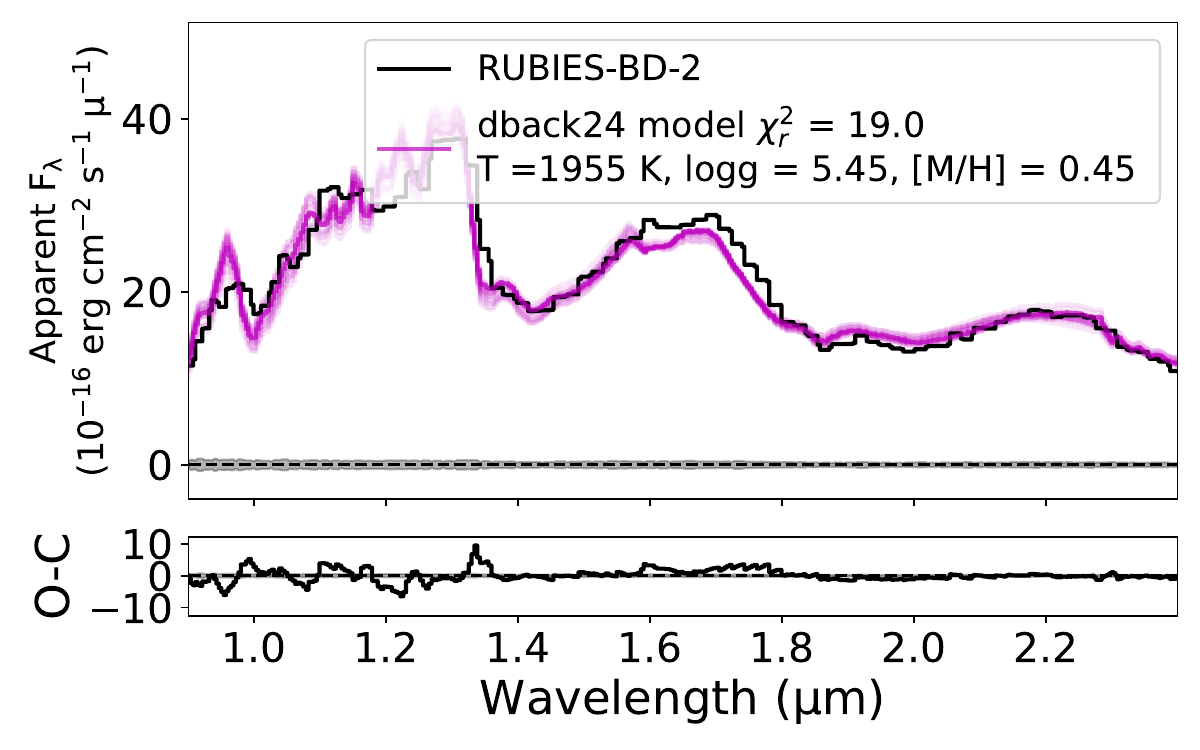}}\hfill
    \includegraphics[width=.3\textwidth]{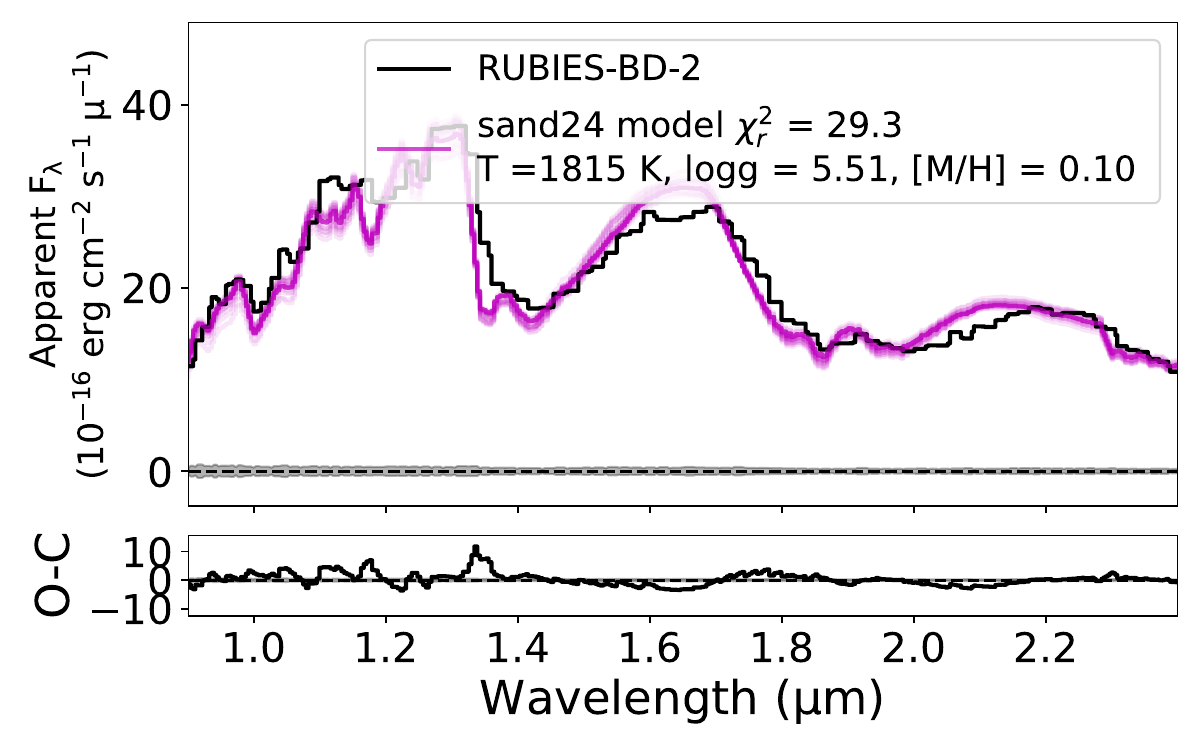}\hfill
    \\[\smallskipamount]
    \fbox{\includegraphics[width=.3\textwidth]{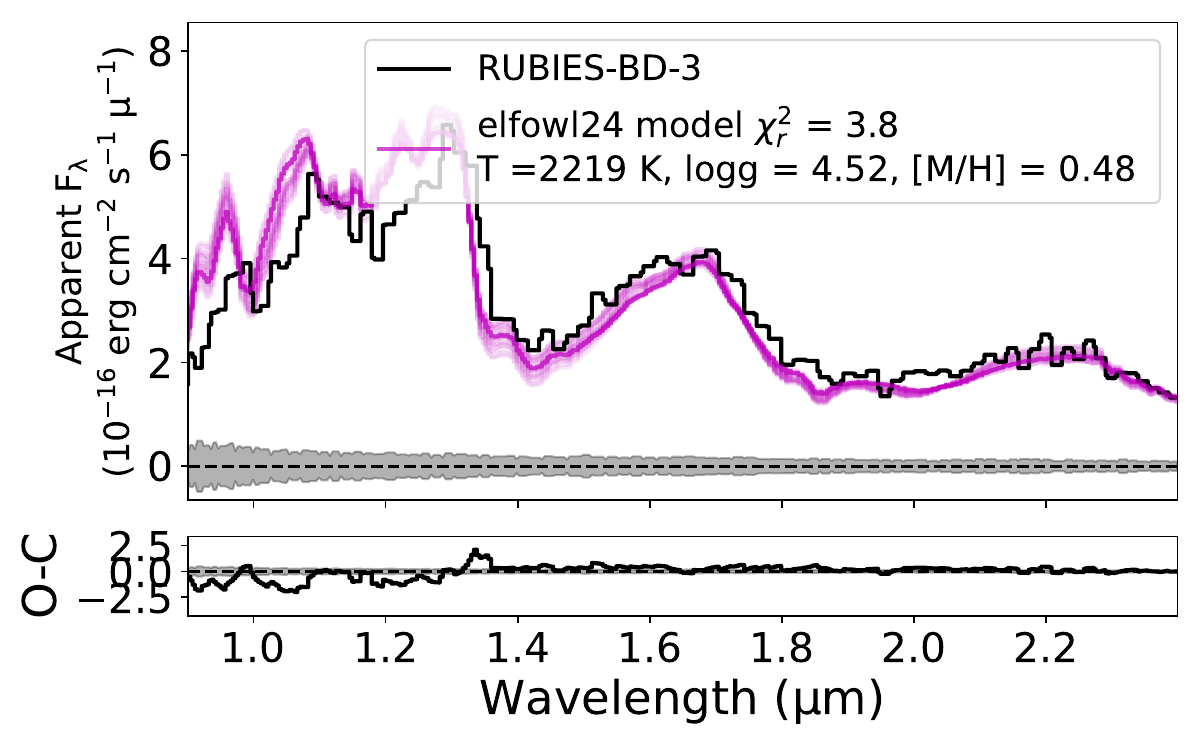}}\hfill
    \includegraphics[width=.3\textwidth]{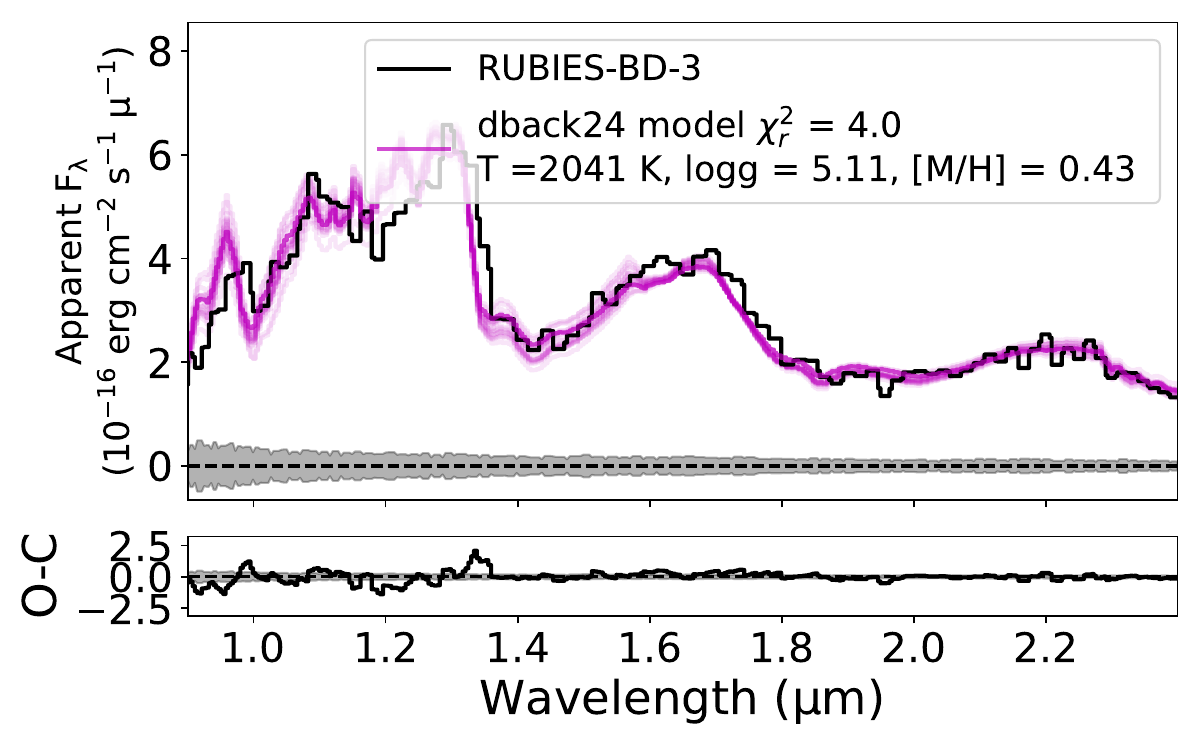}\hfill
    \includegraphics[width=.3\textwidth]{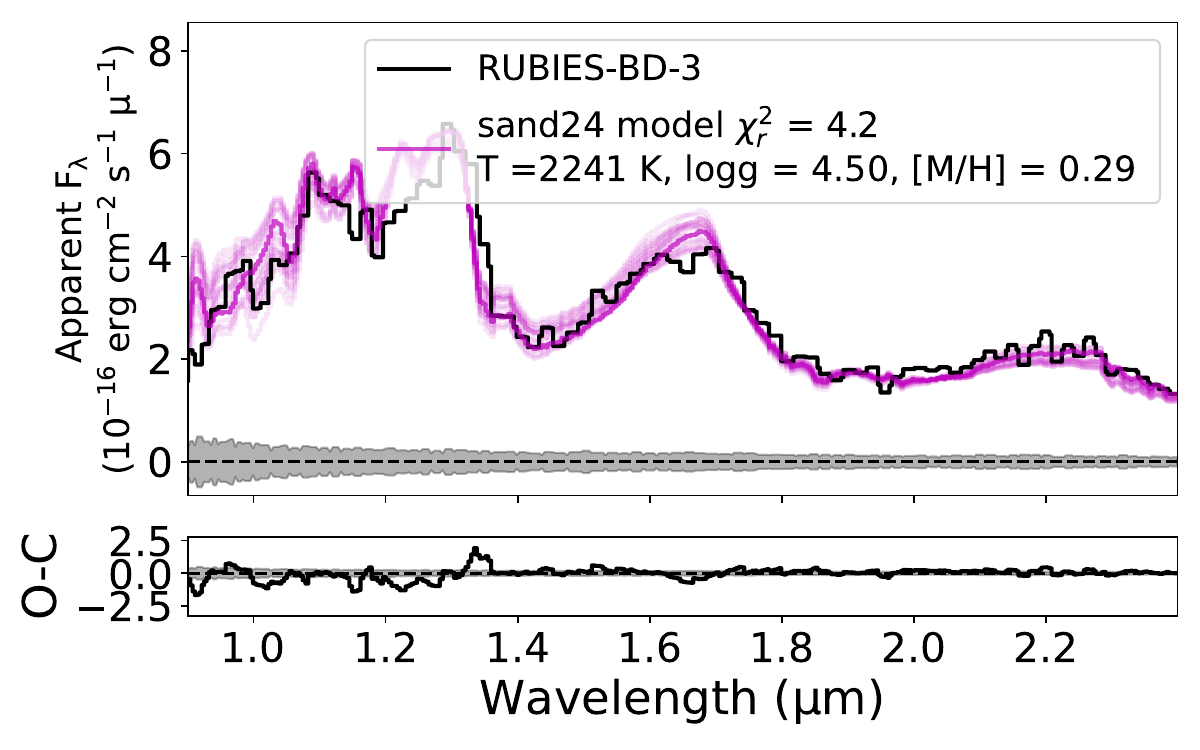}\hfill
    \\[\smallskipamount]
    \fbox{\includegraphics[width=.3\textwidth]{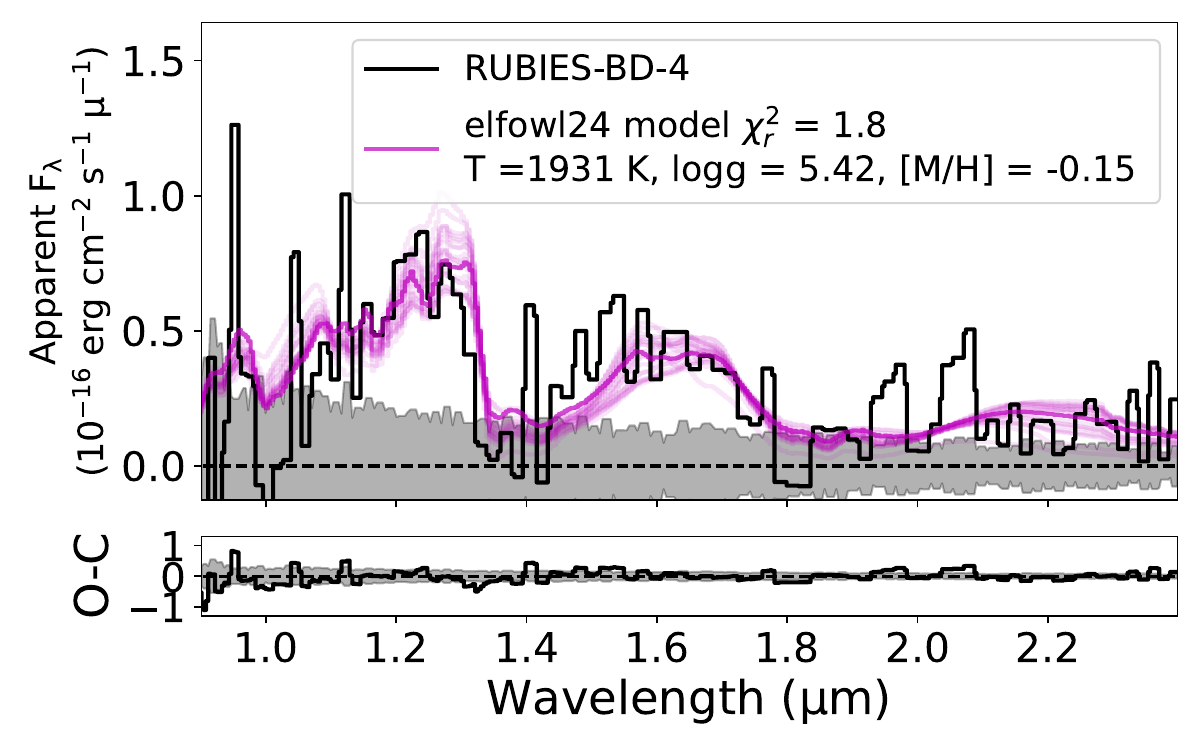}}\hfill
    \fbox{\includegraphics[width=.3\textwidth]{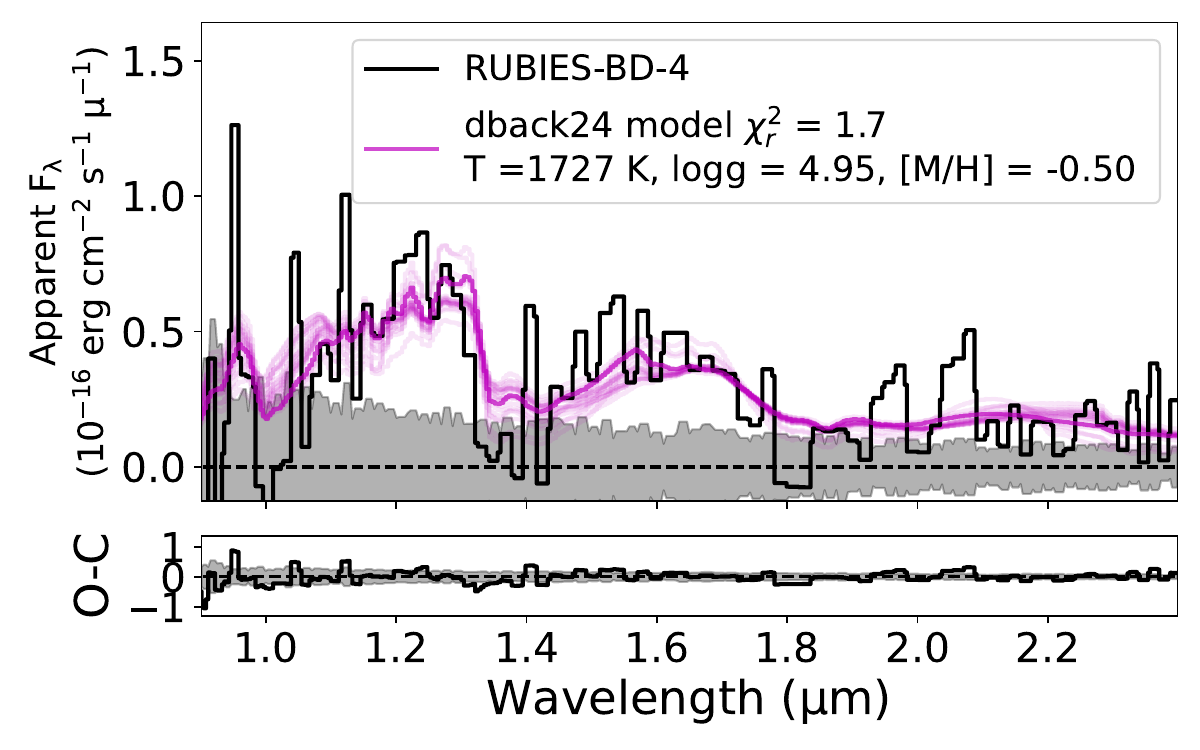}}\hfill
    \fbox{\includegraphics[width=.3\textwidth]{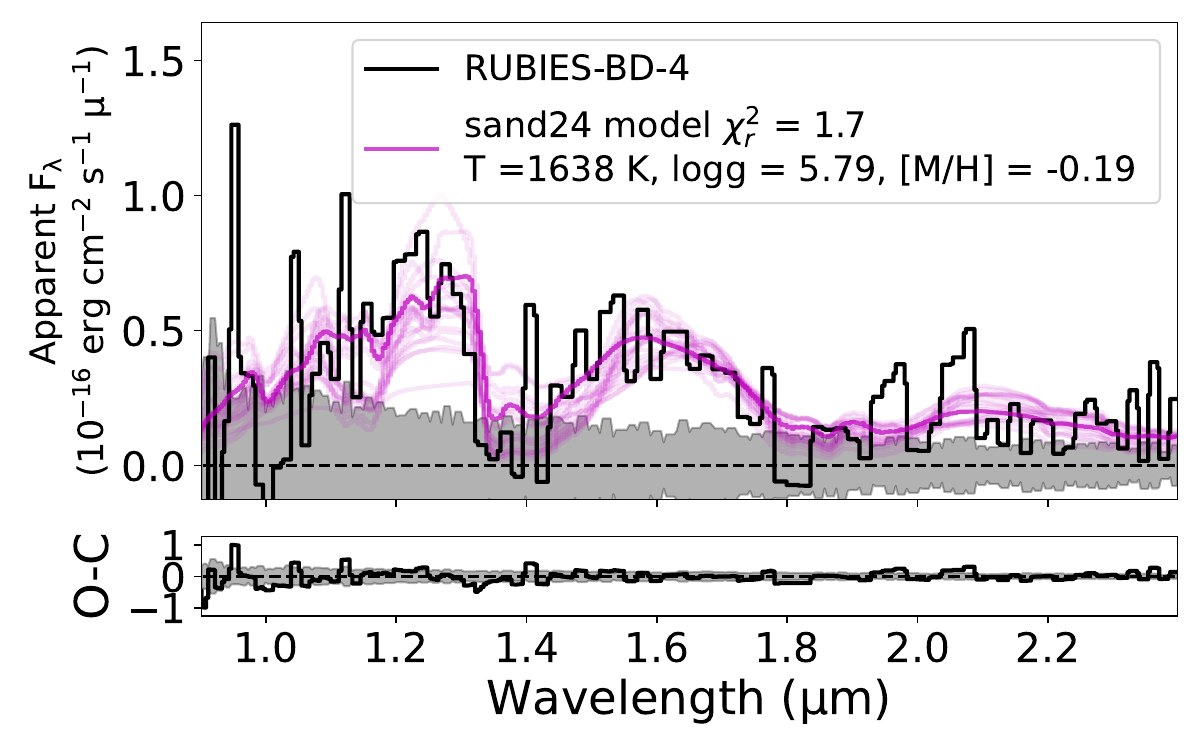}}\hfill
    \\[\smallskipamount]
    \includegraphics[width=.3\textwidth]{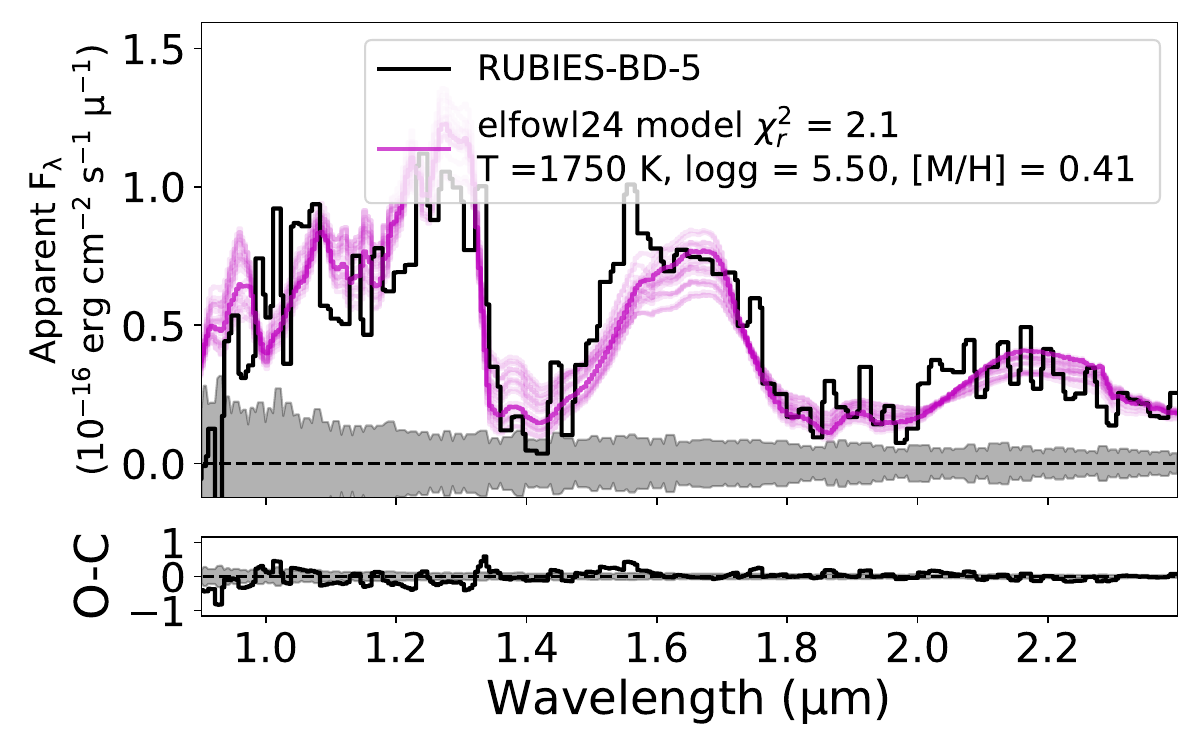}\hfill
    \includegraphics[width=.3\textwidth]{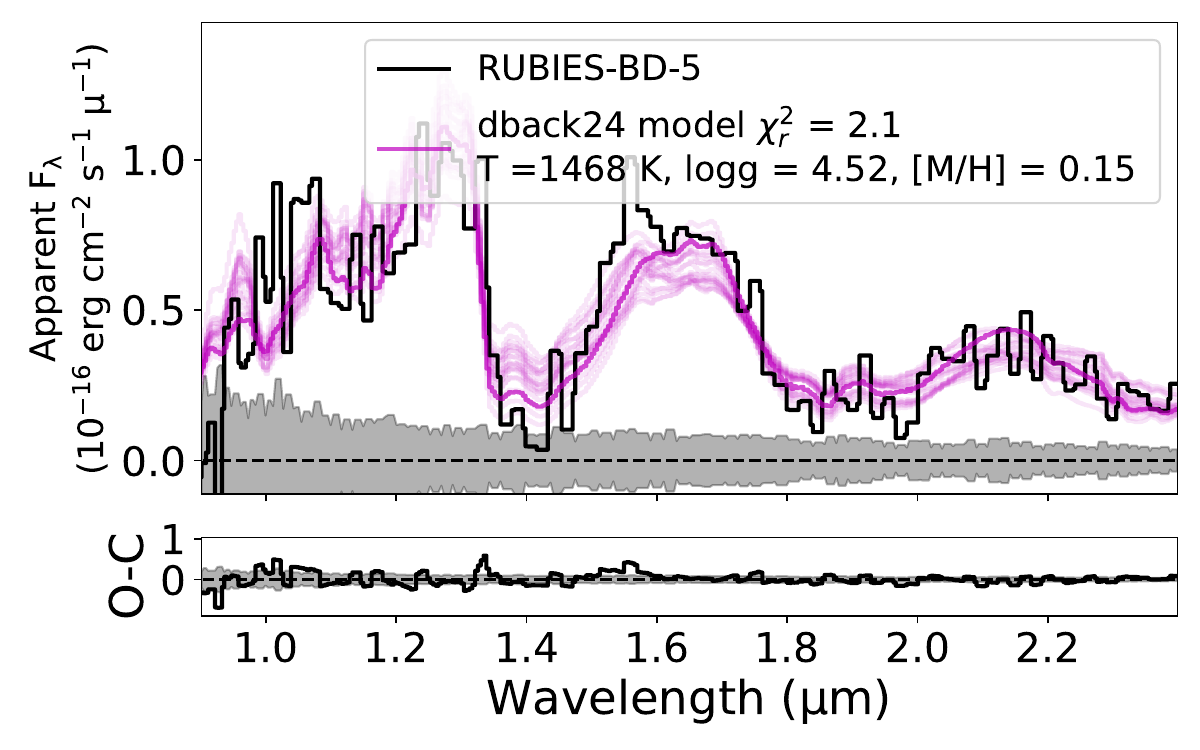}\hfill
    \fbox{\includegraphics[width=.3\textwidth]{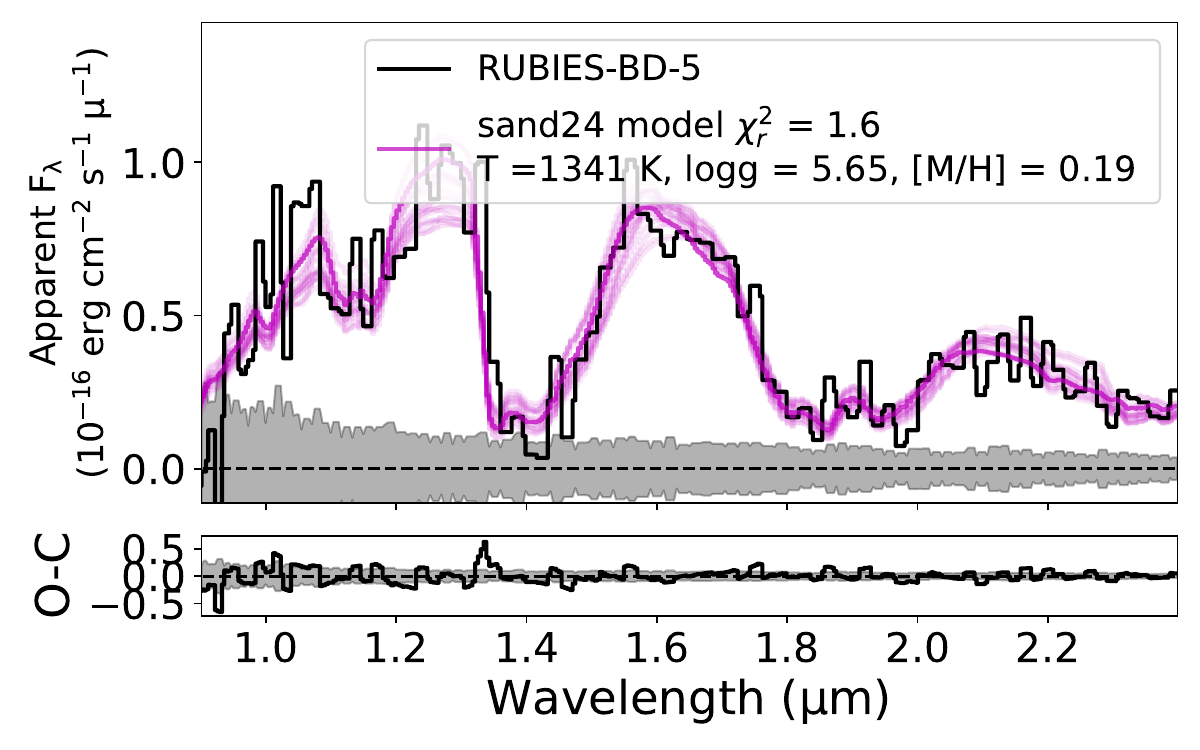}}\hfill
    \caption{\added{Markov Chain Monte Carlo} fits to NIRSpec/Prism data constrained to the NIR (0.9-2.4~{\micron}) range. 
    Each row corresponds to one of the RUBIES brown dwarf spectra scaled to apparent F444W magnitudes (black lines), while the columns separate fits to the Elf Owl (left), Diamondback (middle) and SAND (right) models. Both the best-fit models (solid magenta lines) and draws from the posterior distribution (semi-transparent magenta lines) are shown, and the $\pm$1$\sigma$ spectral uncertainties are indicated by the grey band. 
    Below each spectral comparison, we compare the difference between the source and best-fit model fluxes (O-C as black lines) to the $\pm$1$\sigma$ spectral uncertainties (grey bands).
    Panels that are surrounded by boxes indicate models that provide the best fits, including equivalent cases (see Table~\ref{tab:summary}).}
    \label{fig:nir}
\end{figure}

% Save and override figure counter
\newcounter{savedfigure}
\setcounter{savedfigure}{\value{figure}}
\addtocounter{figure}{-1}
\renewcommand{\thefigure}{\arabic{figure}}

\begin{figure}[h]
    \fbox{\includegraphics[width=.3\textwidth]{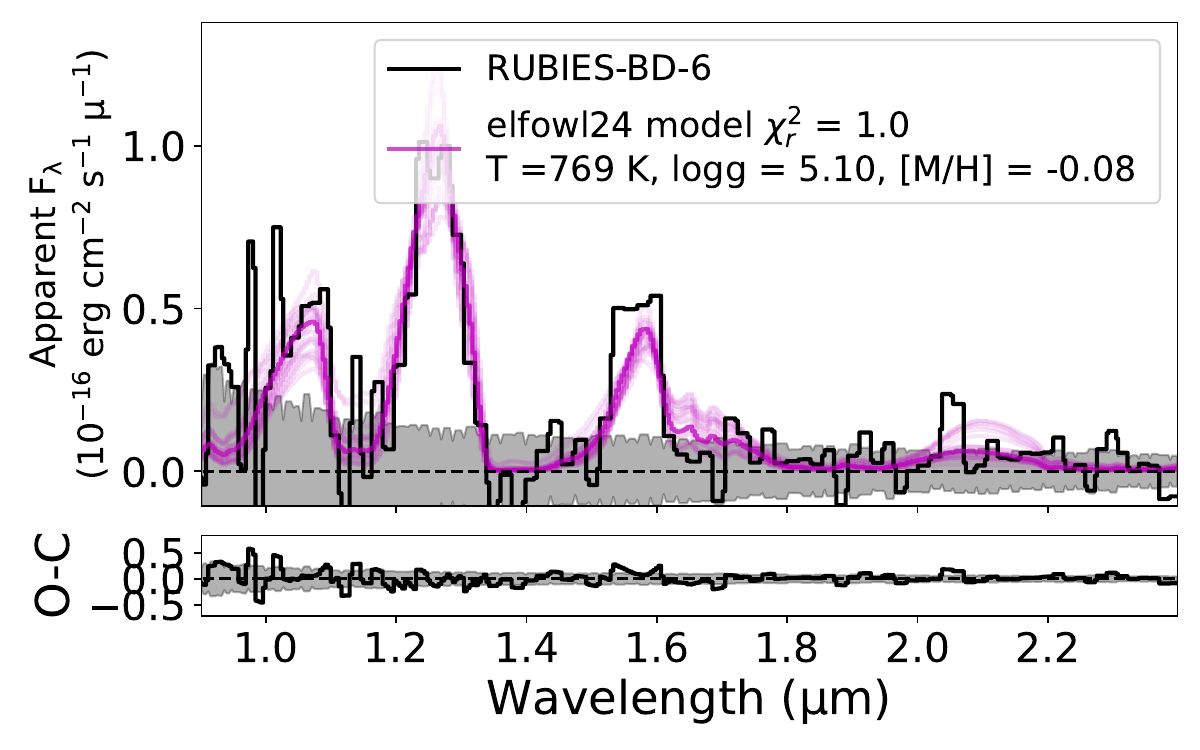}}\hfill
    \fbox{\includegraphics[width=.3\textwidth]{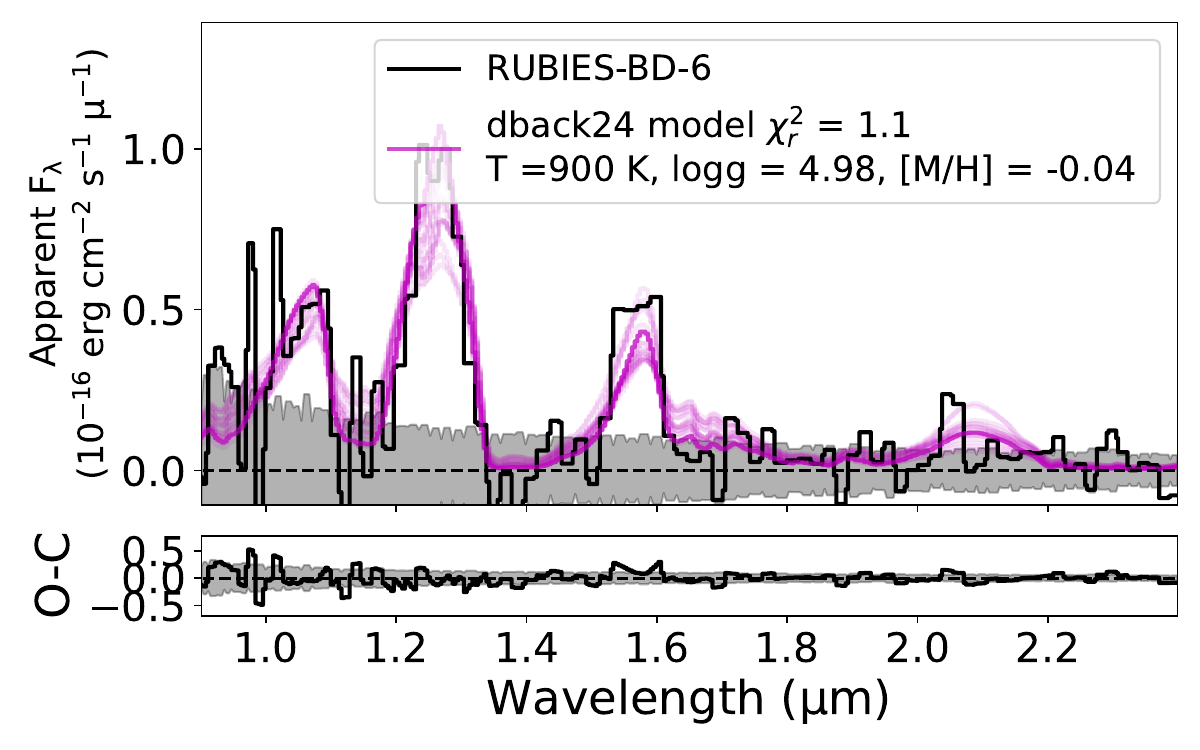}}\hfill
    \fbox{\includegraphics[width=.3\textwidth]{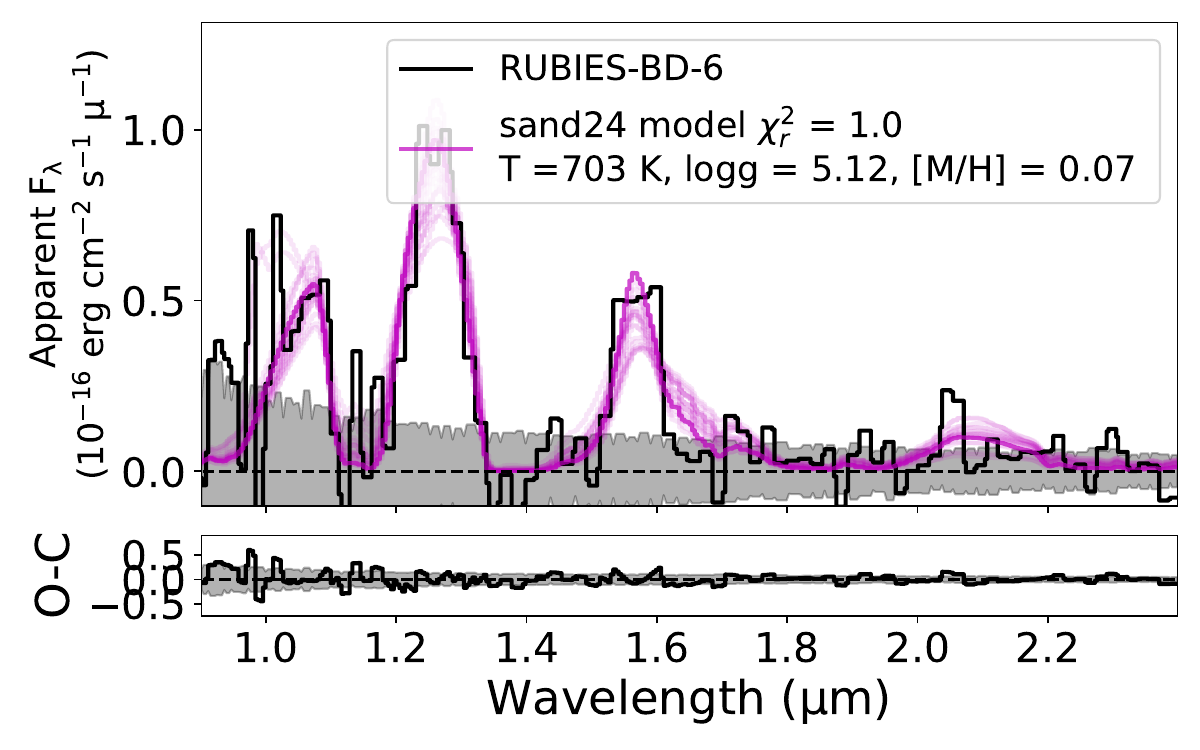}}\hfill
    \\[\smallskipamount]
    \fbox{\includegraphics[width=.3\textwidth]{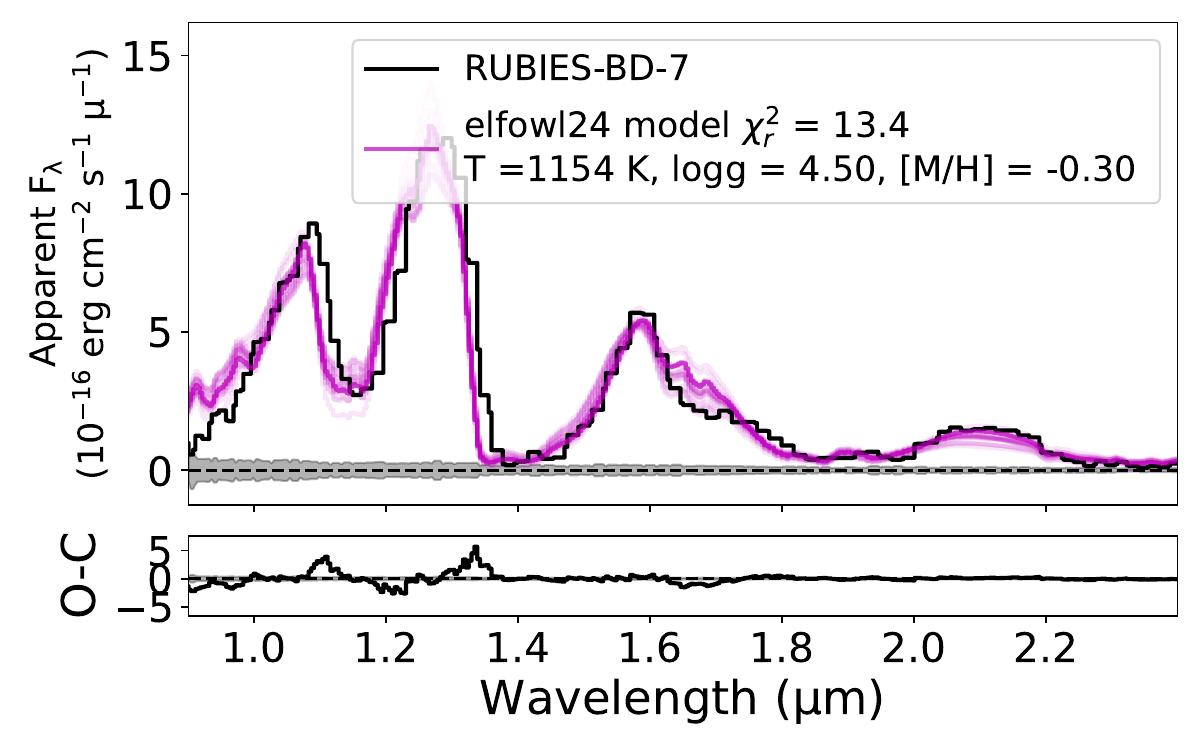}}\hfill
    \includegraphics[width=.3\textwidth]{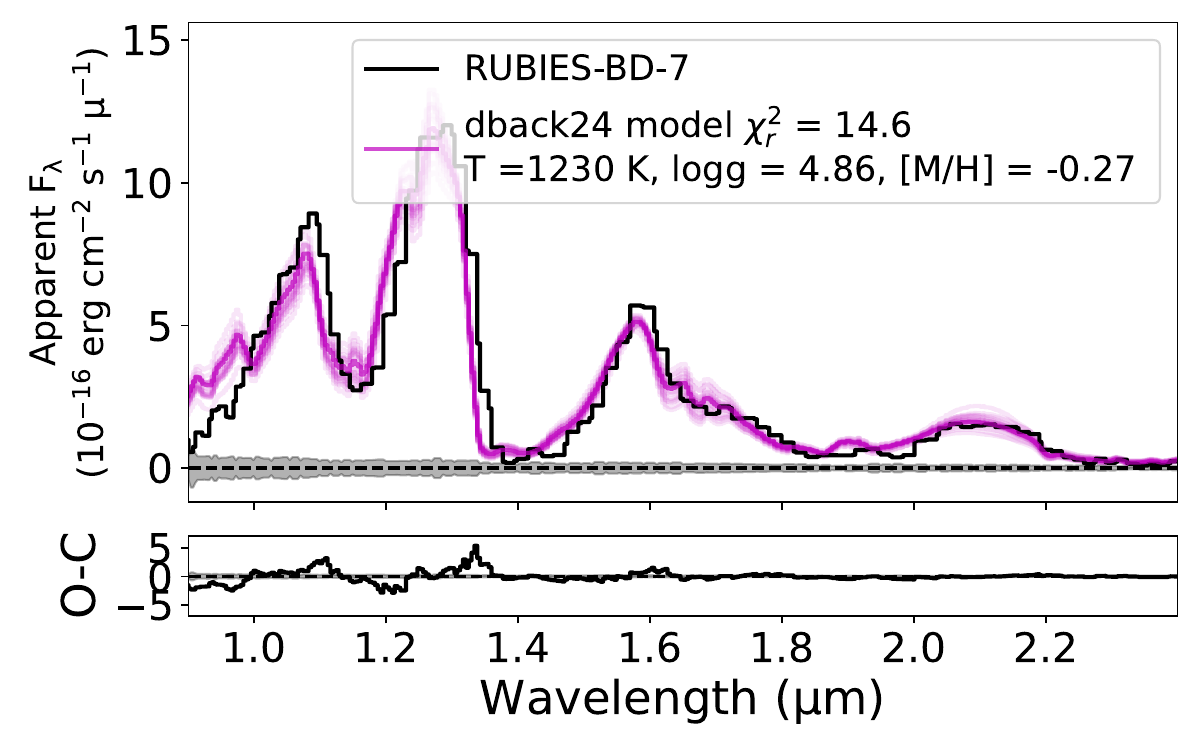}\hfill
    \includegraphics[width=.3\textwidth]{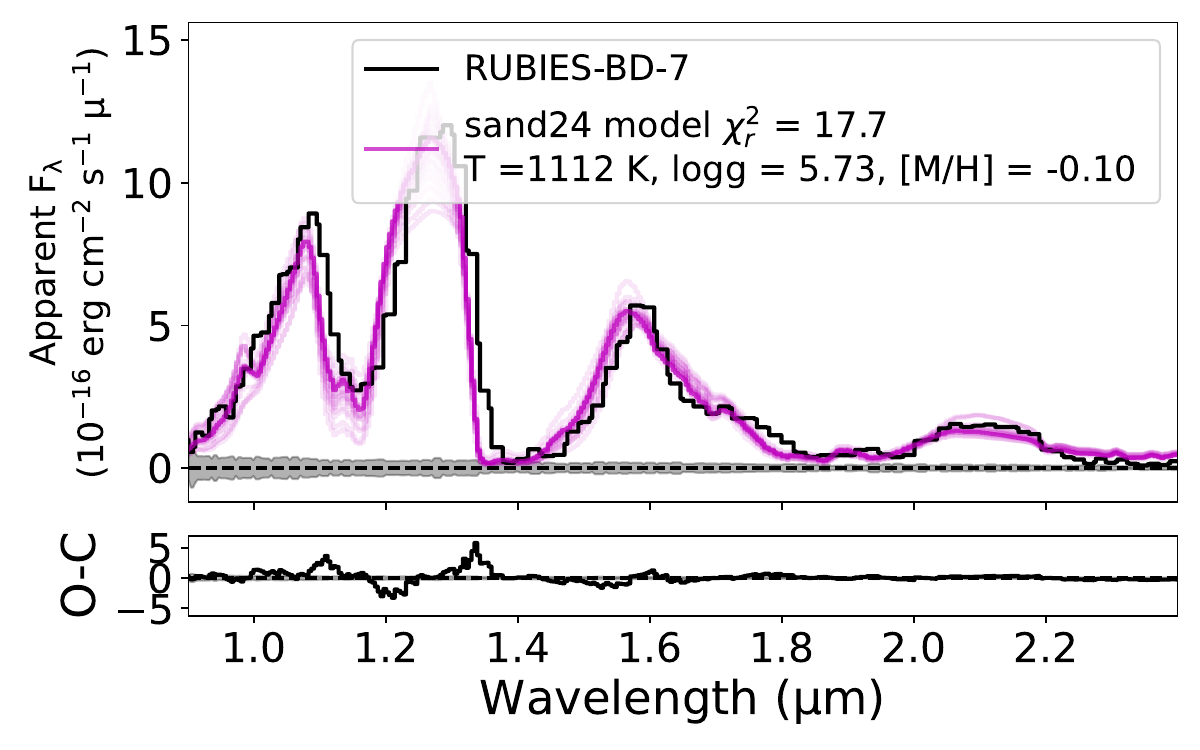}\hfill
    \caption{Continued.}
\end{figure}

% Restore figure numbering
\setcounter{figure}{\value{savedfigure}}
\renewcommand{\thefigure}{\arabic{figure}}

\begin{figure}[h]
    \fbox{\includegraphics[width=.3\textwidth]{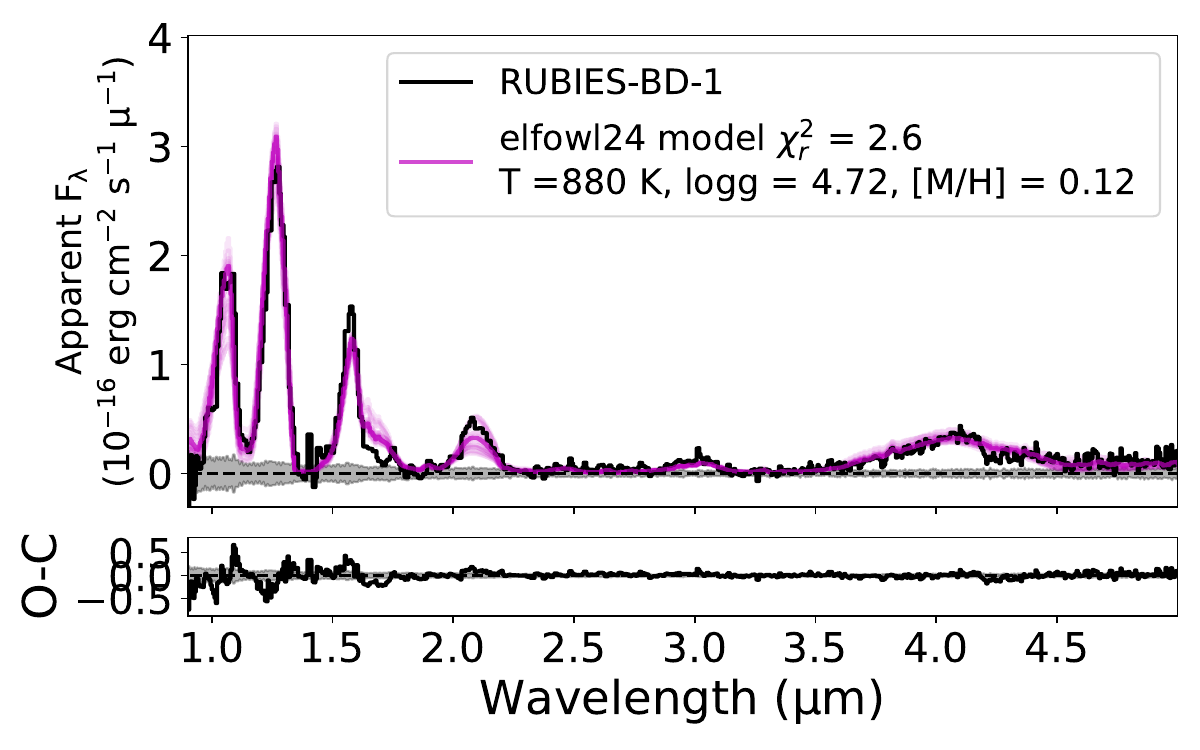}}\hfill
    \includegraphics[width=.3\textwidth]{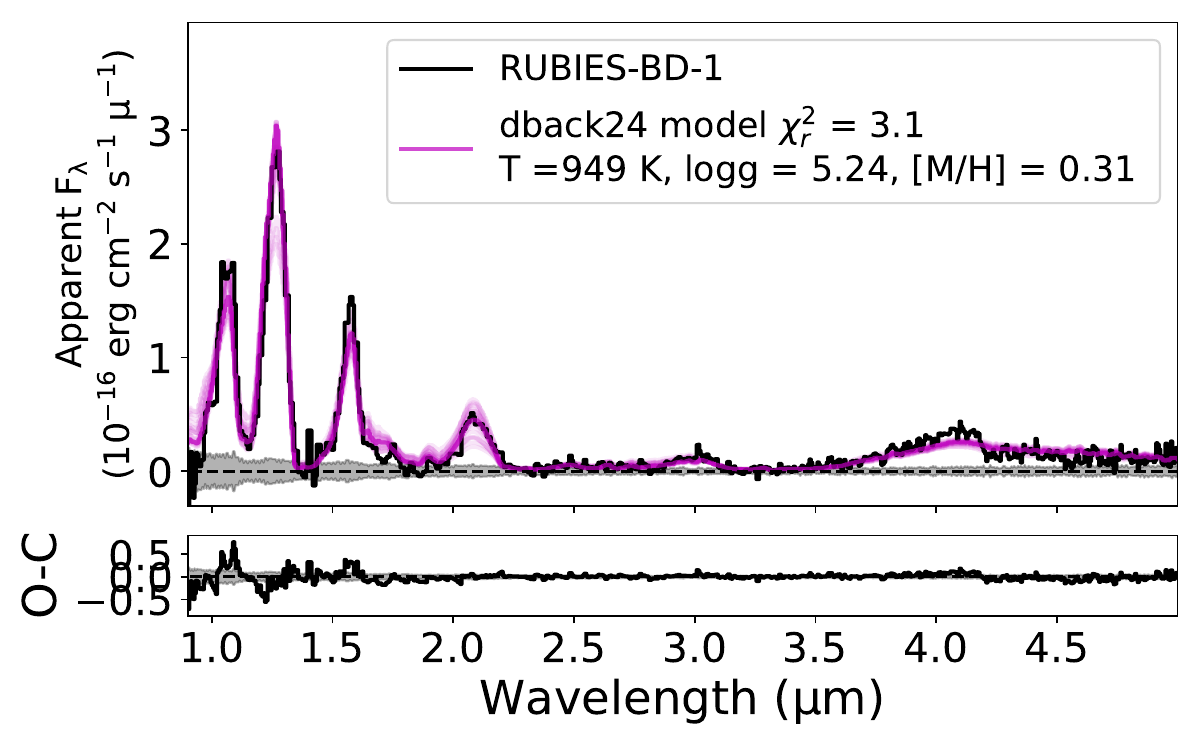}\hfill
    \includegraphics[width=.3\textwidth]{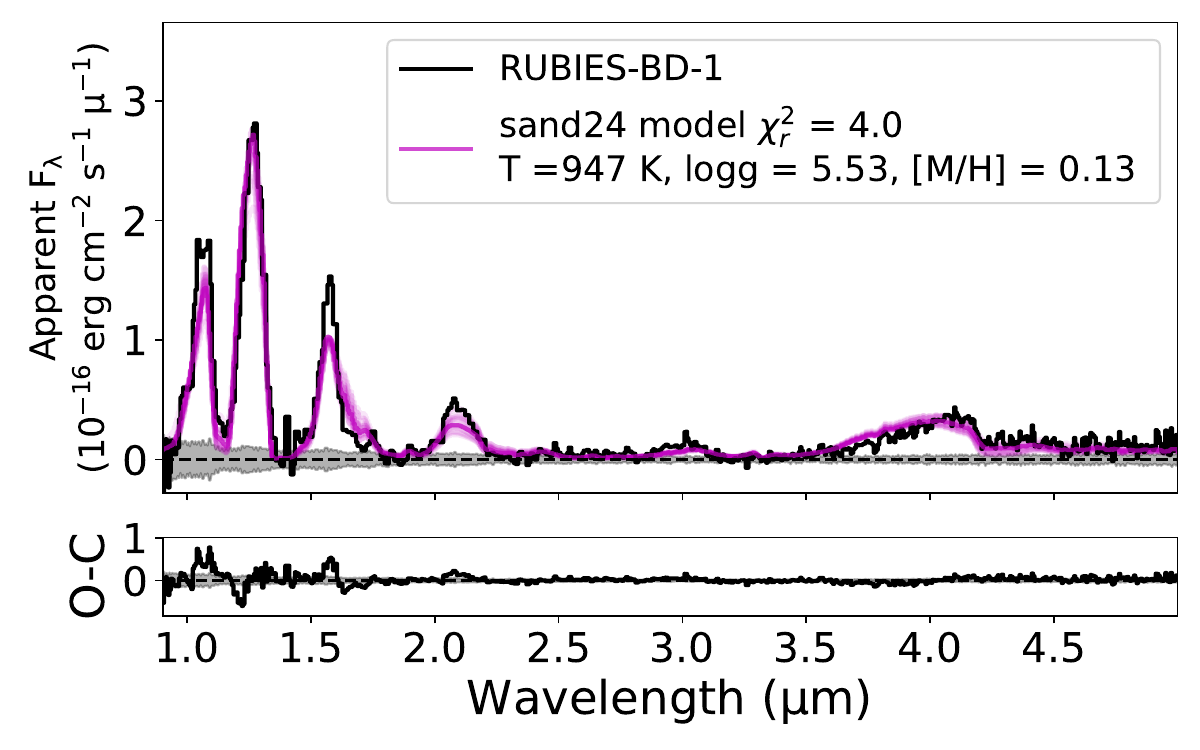}\hfill
    \\[\smallskipamount]
    \includegraphics[width=.3\textwidth]{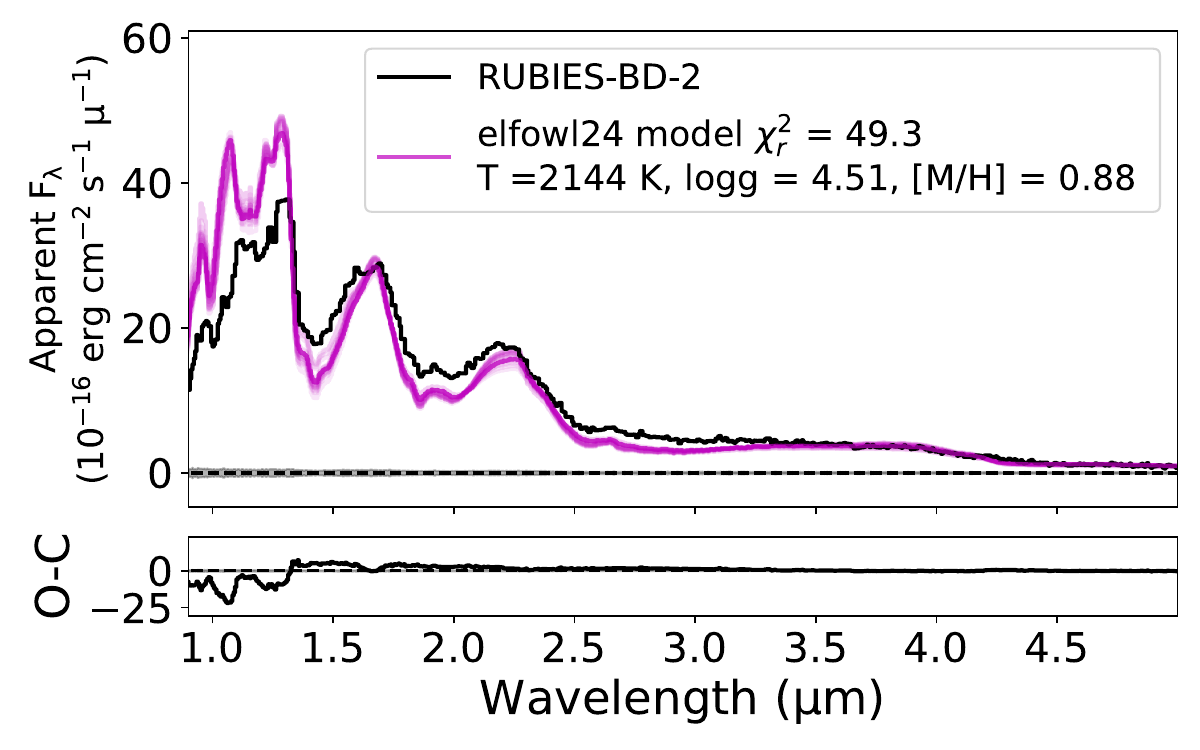}\hfill
    \fbox{\includegraphics[width=.3\textwidth]{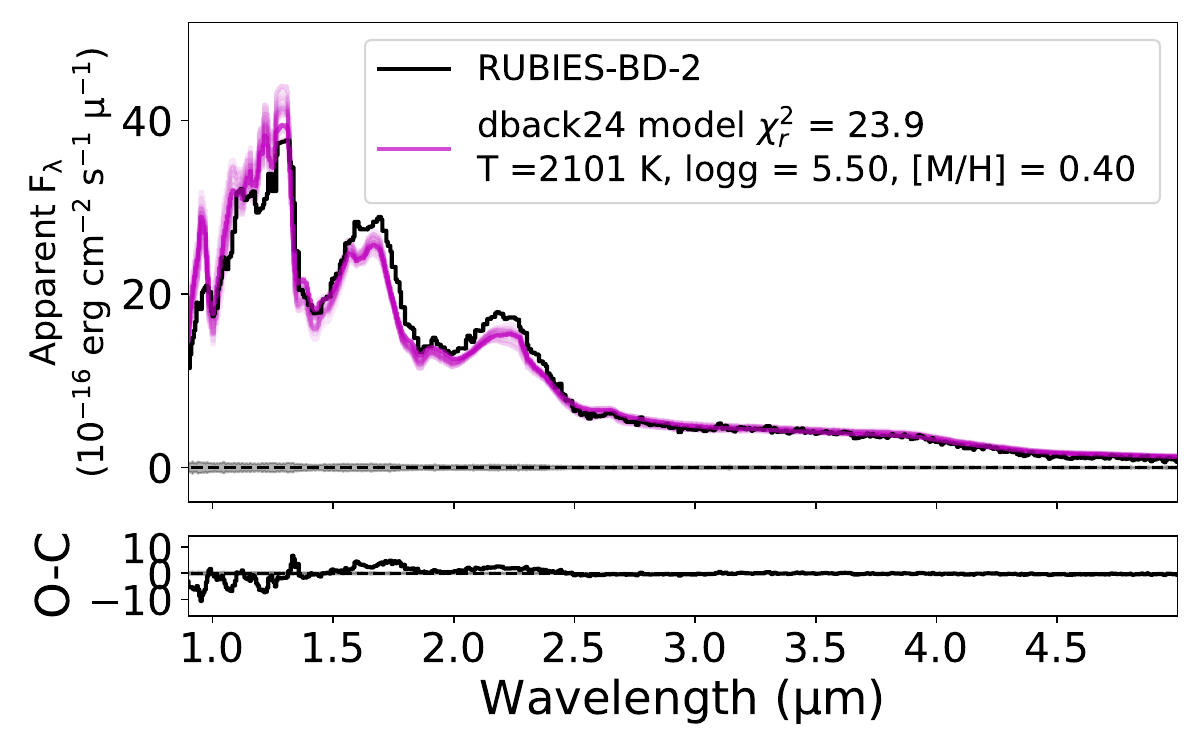}}\hfill
    \fbox{\includegraphics[width=.3\textwidth]{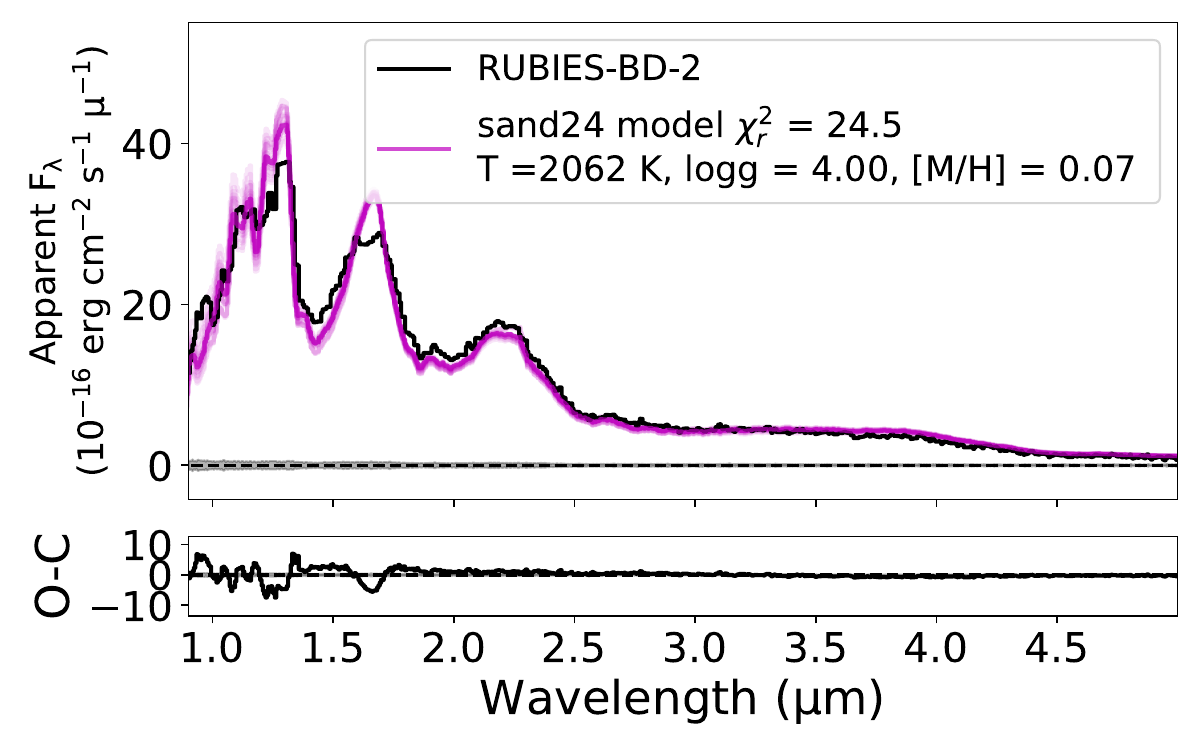}}\hfill
    \\[\smallskipamount]
    {\includegraphics[width=.3\textwidth]{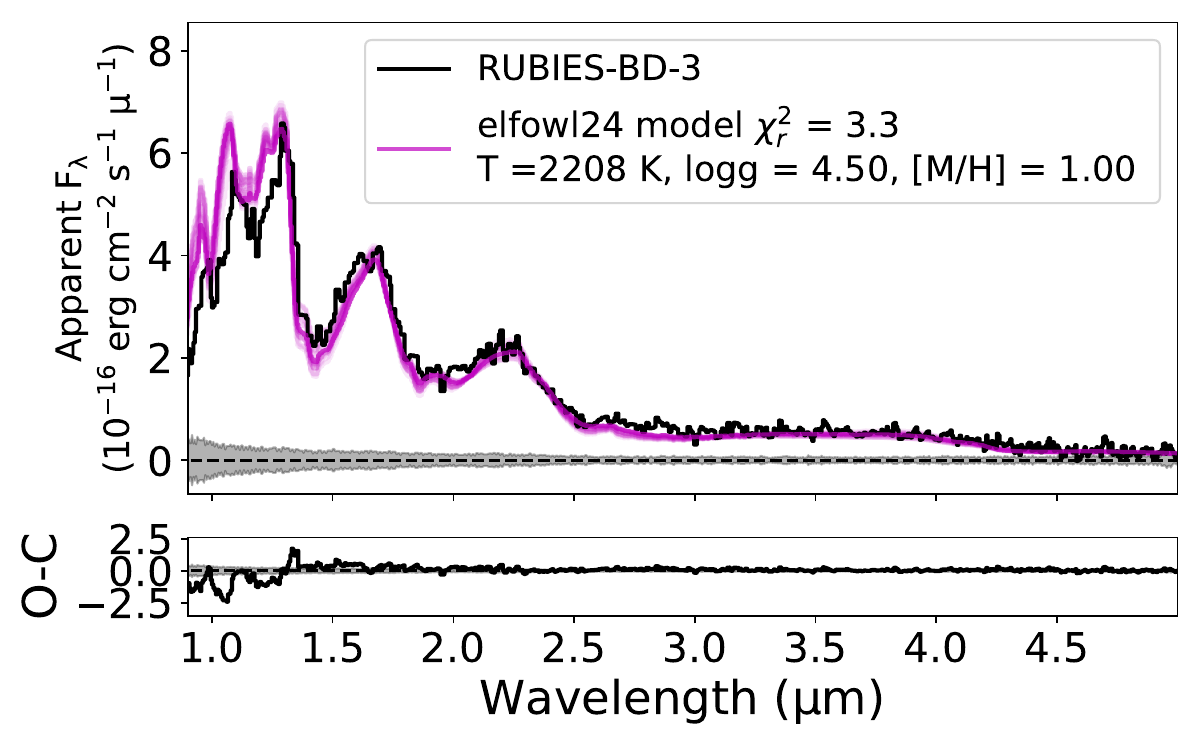}}\hfill
    \fbox{\includegraphics[width=.3\textwidth]{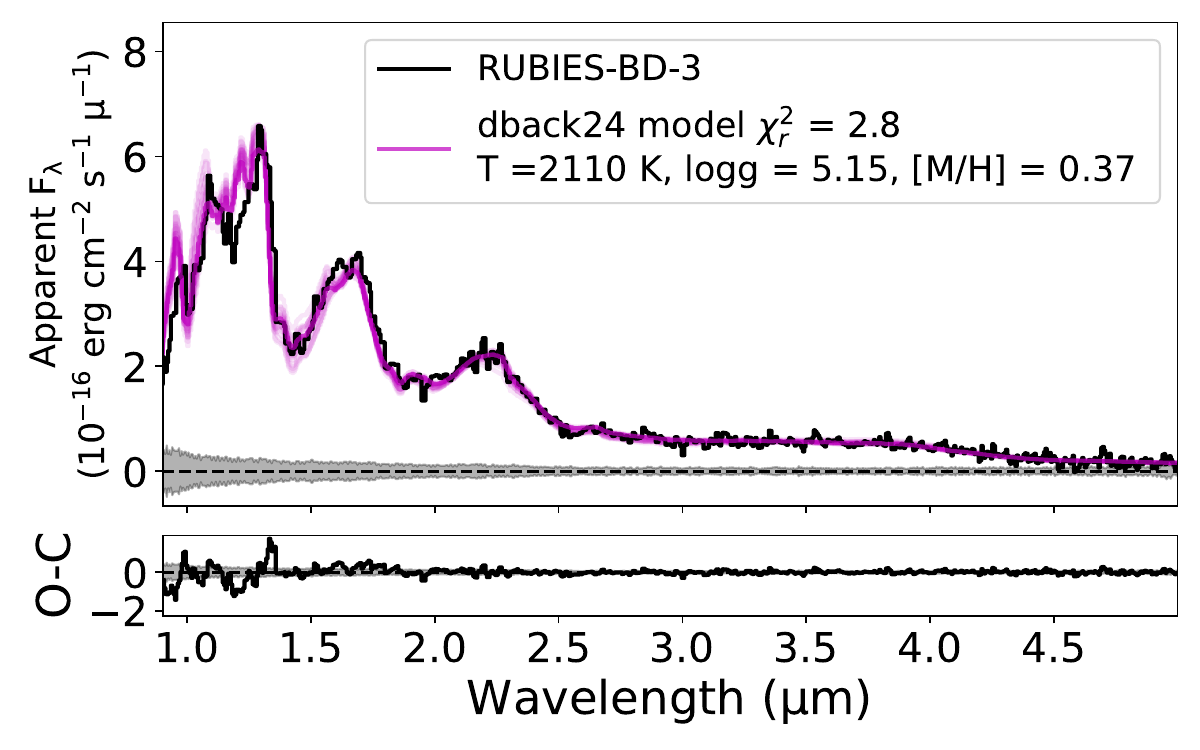}}\hfill
    \fbox{\includegraphics[width=.3\textwidth]{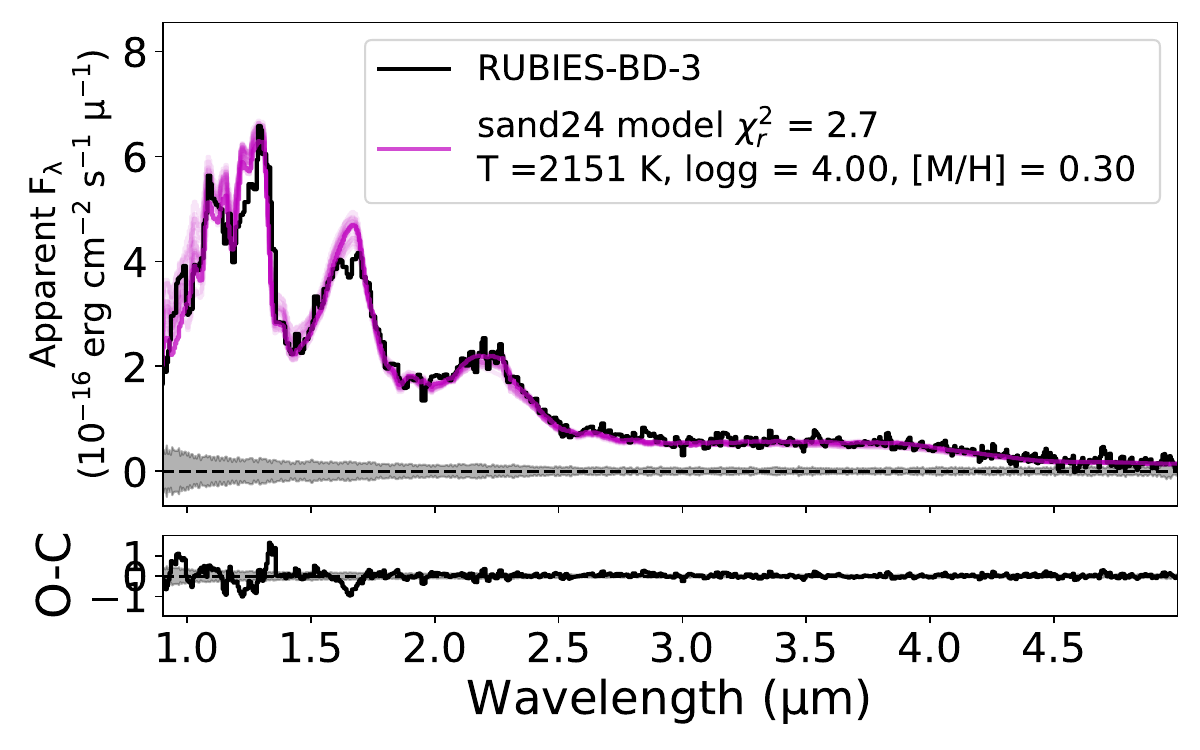}}\hfill
    \\[\smallskipamount]
    \fbox{\includegraphics[width=.3\textwidth]{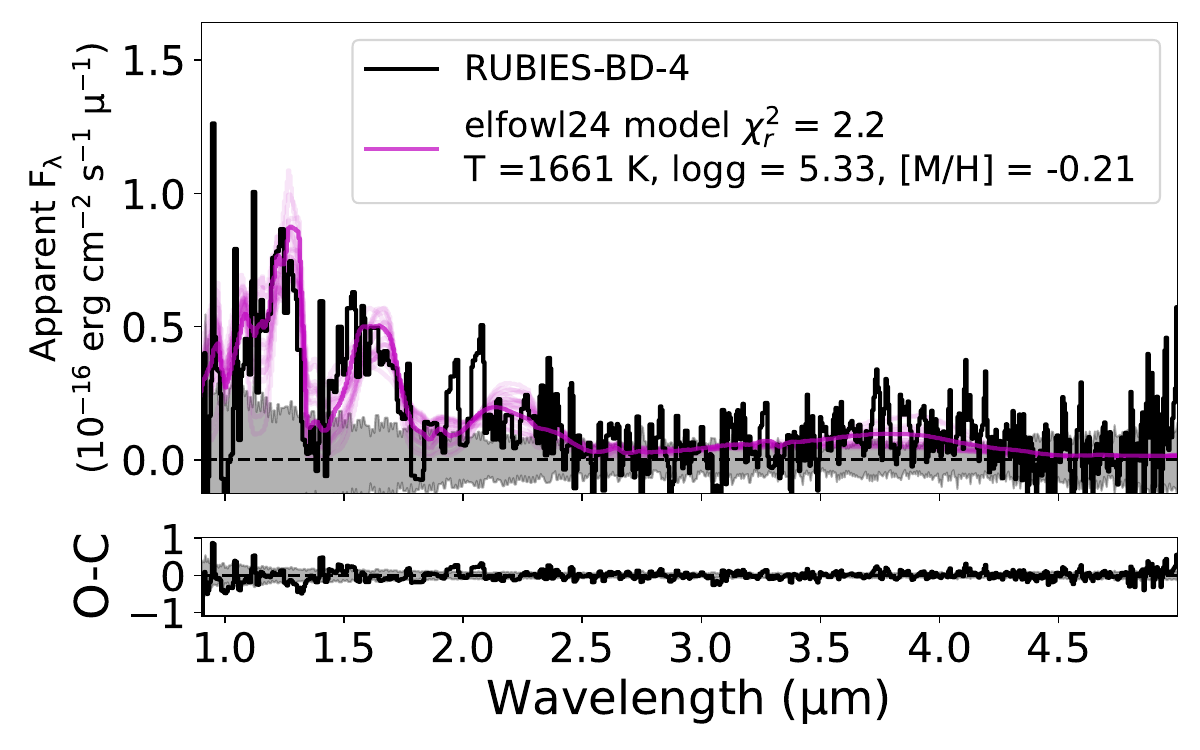}}\hfill
    \fbox{\includegraphics[width=.3\textwidth]{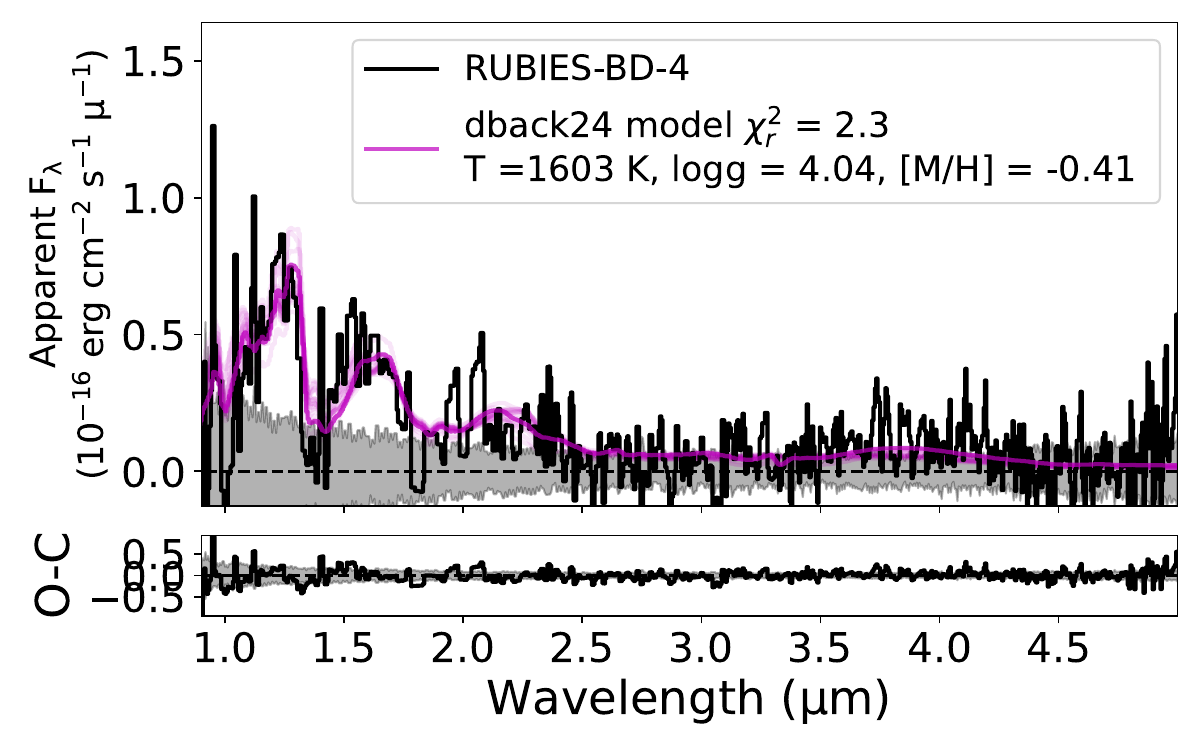}}\hfill
    \fbox{\includegraphics[width=.3\textwidth]{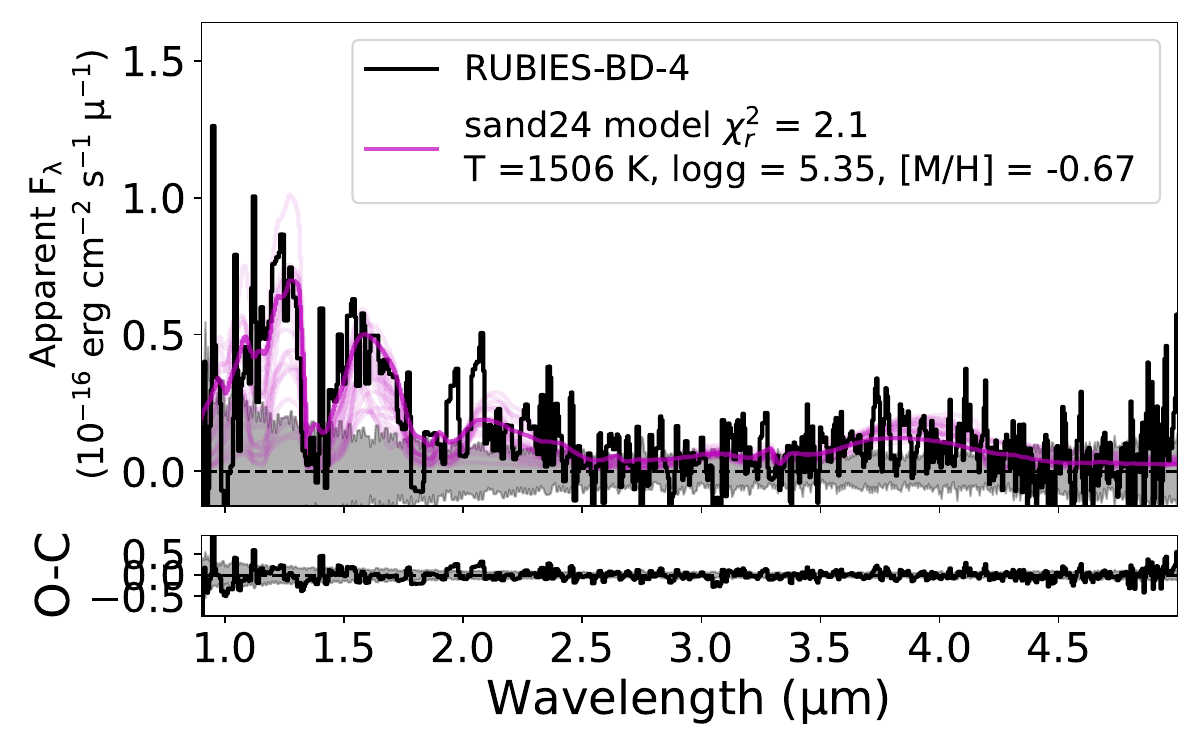}}\hfill
    \\[\smallskipamount]
    \includegraphics[width=.3\textwidth]{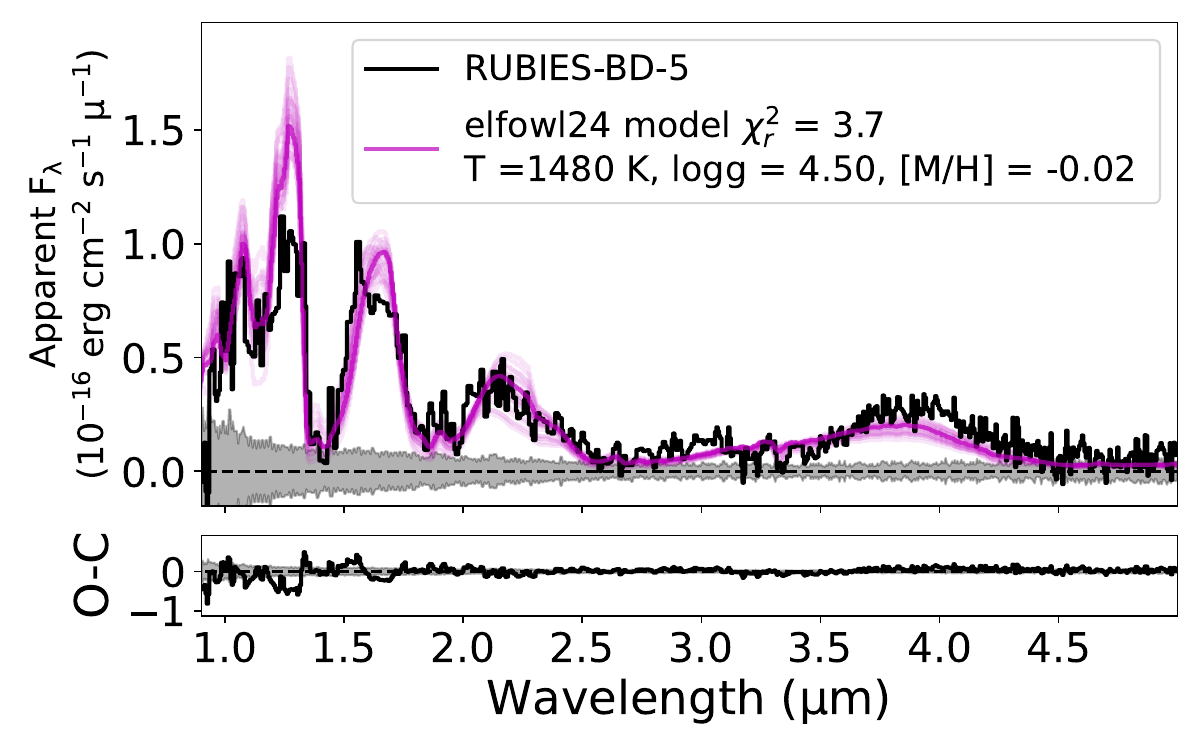}\hfill
    \includegraphics[width=.3\textwidth]{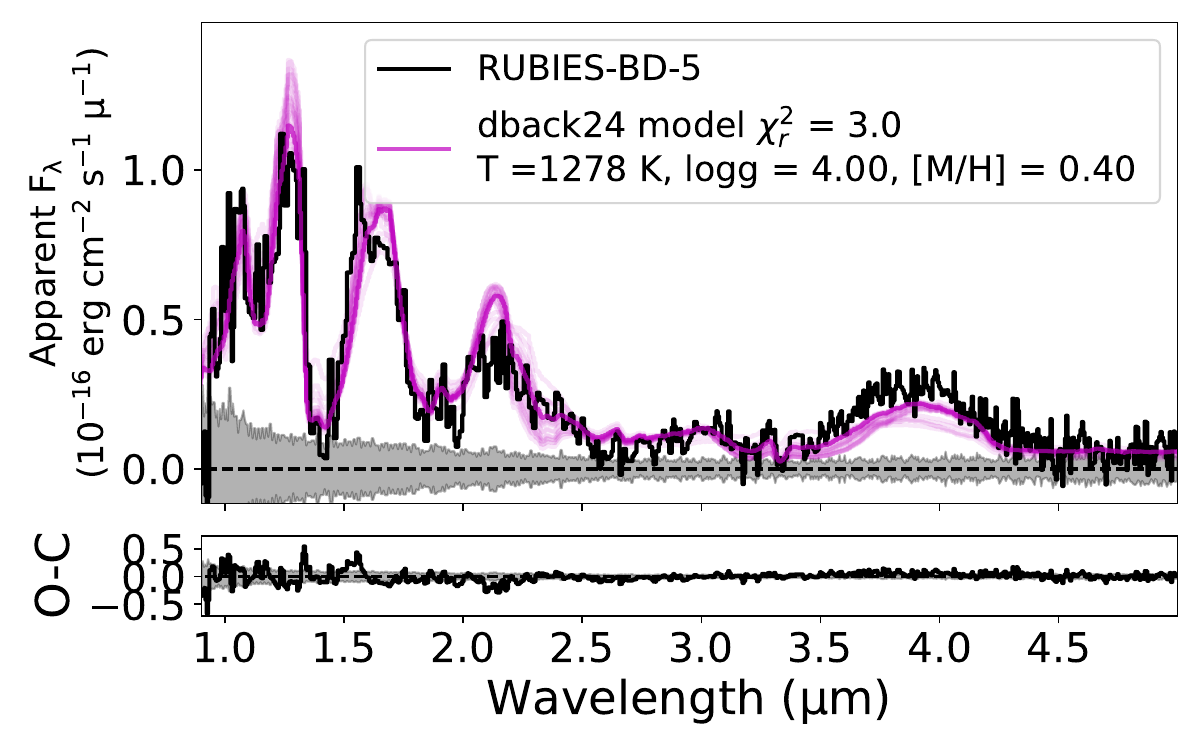}\hfill
    \fbox{\includegraphics[width=.3\textwidth]{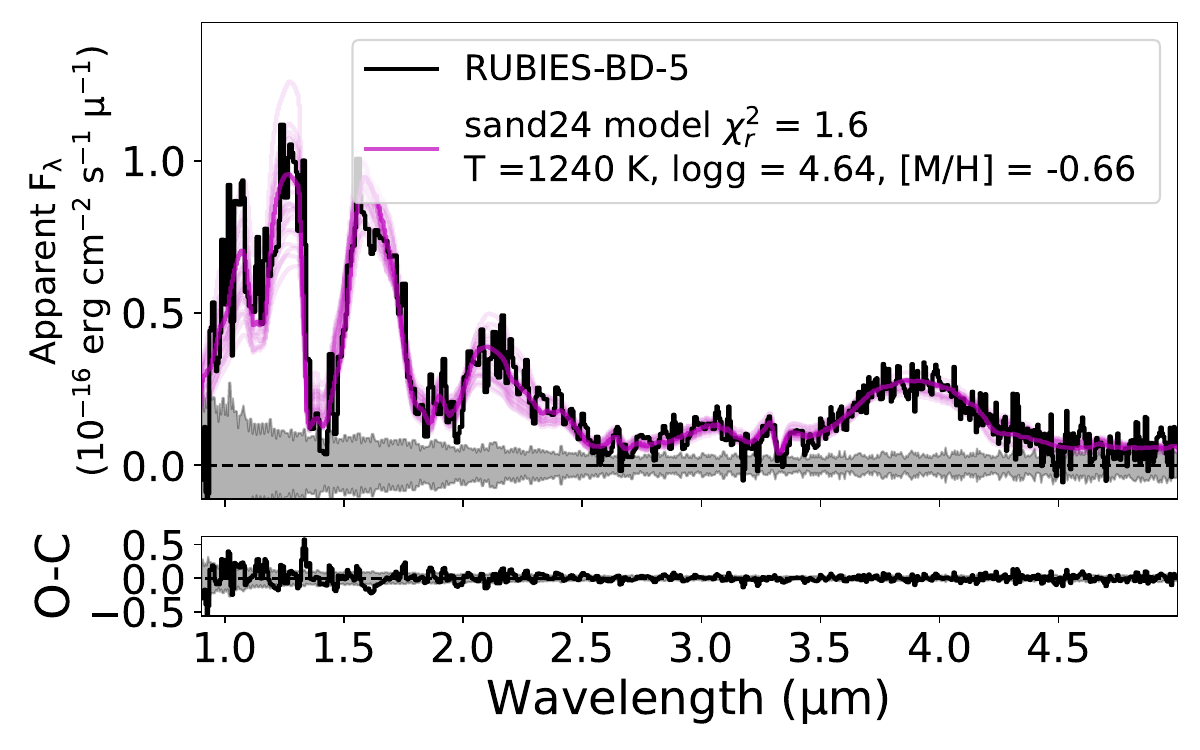}}\hfill
    \caption{Same as Figure~\ref{fig:nir} for the full NIRSpec/Prism 0.9--5.1~{\micron} spectral range.}
%    The first row shows fits for RUBIES-BD-1, the second row for RUBIES-BD-2, and so on. The left column shows SAND model fits, the middle Diamondback, and the right Elf Owl. Sources are plotted in black with grey shading for uncertainty, and models are plotted in magenta. Below each spectral plot is a graph showing the difference between the source and model flux, also with uncertainty shaded in grey.}
    \label{fig:nirspec}
\end{figure}

% Save and override figure counter
% \newcounter{savedfigure}
\setcounter{savedfigure}{\value{figure}}
\addtocounter{figure}{-1}
\renewcommand{\thefigure}{\arabic{figure}}

\begin{figure}[h]
    \fbox{\includegraphics[width=.3\textwidth]{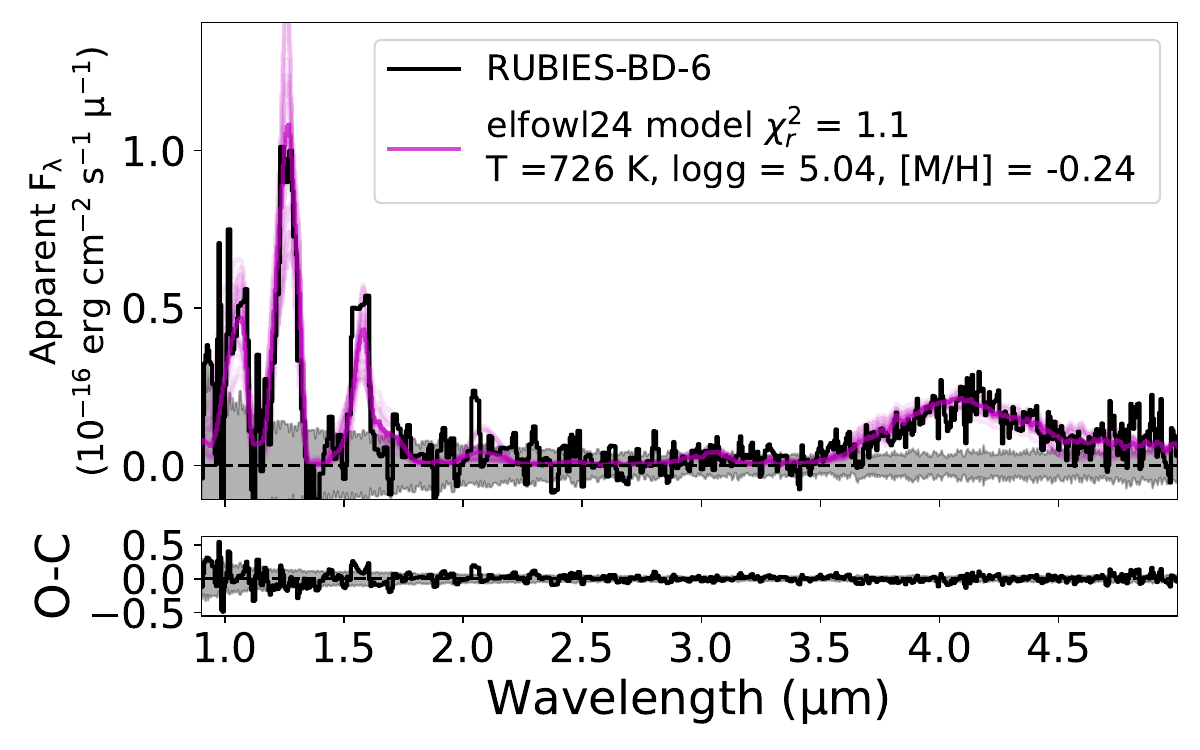}}\hfill
    \includegraphics[width=.3\textwidth]{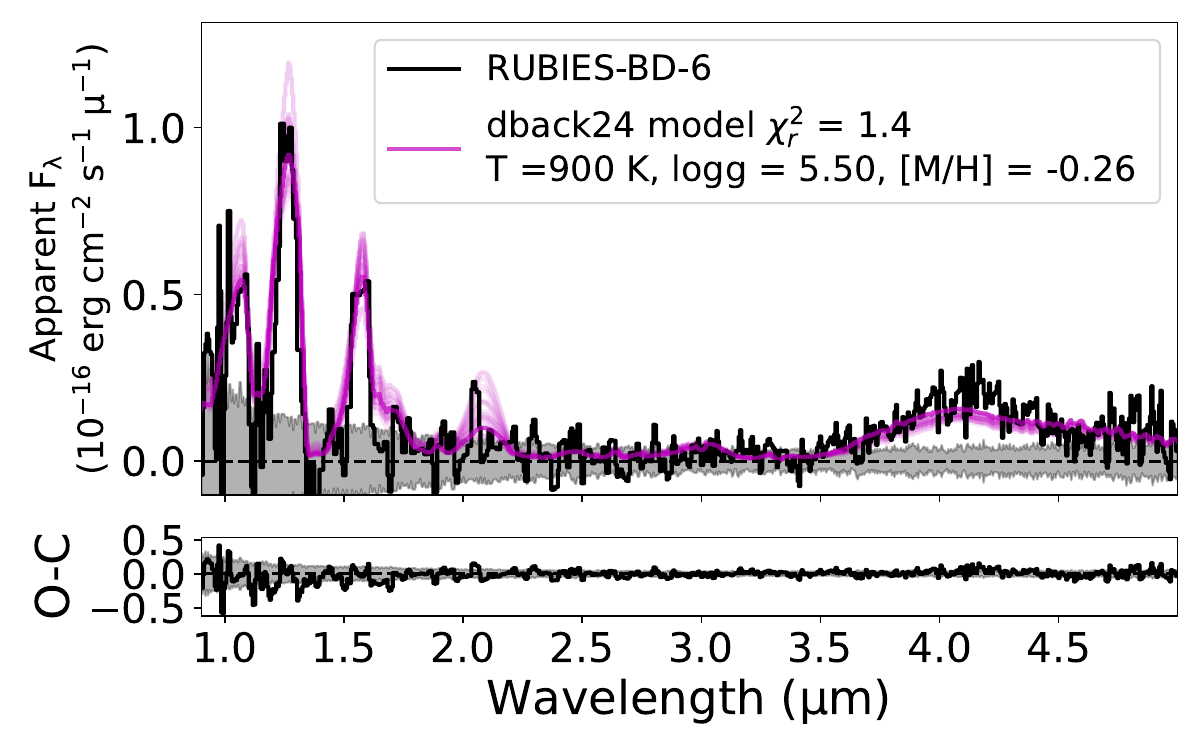}\hfill
    \includegraphics[width=.3\textwidth]{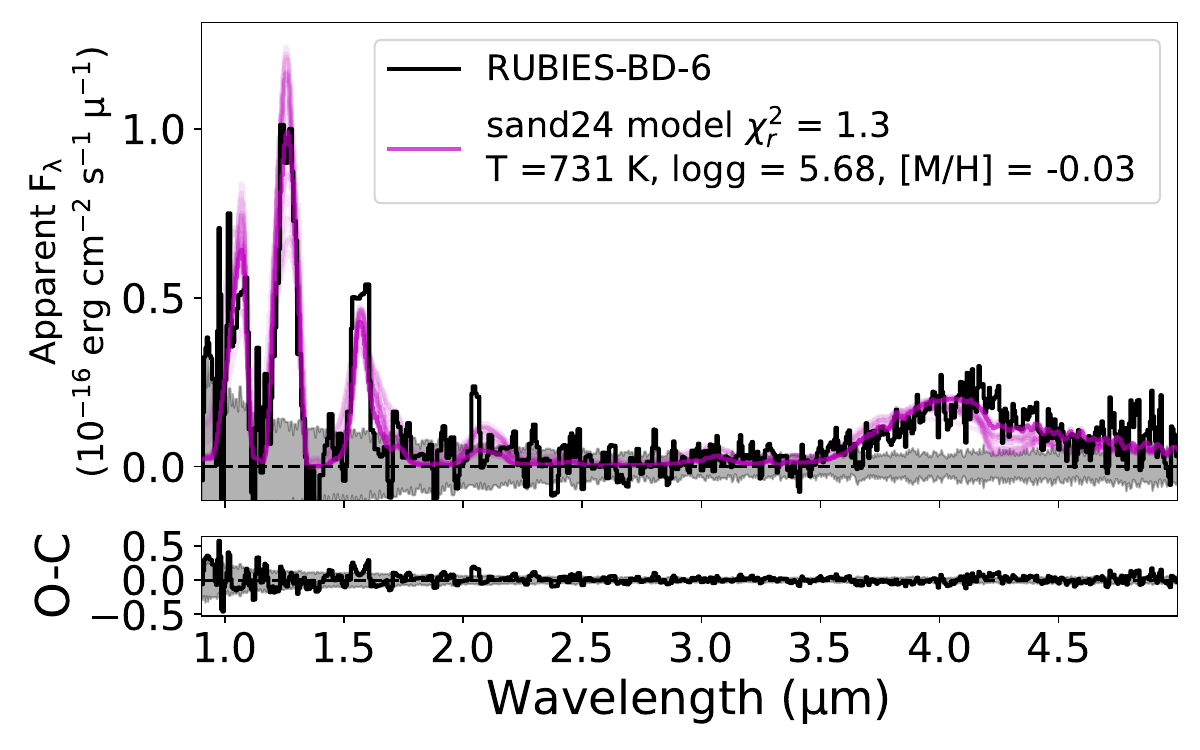}\hfill
    \\[\smallskipamount]
    \fbox{\includegraphics[width=.3\textwidth]{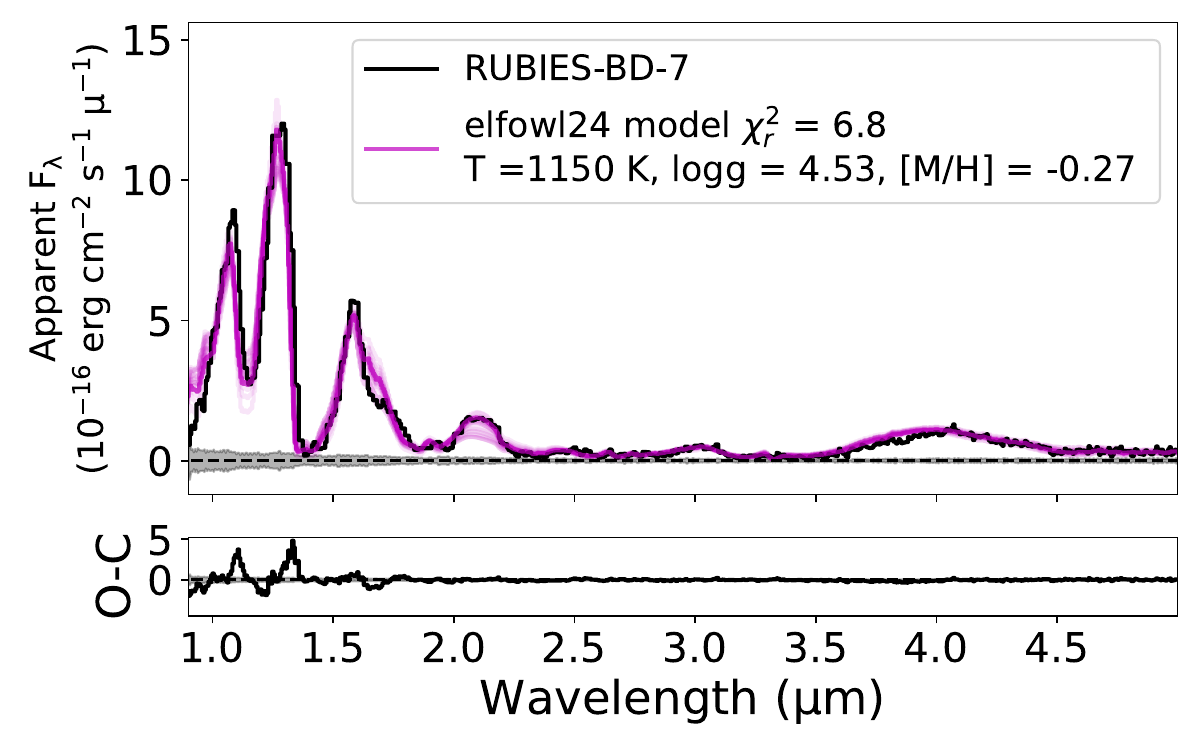}}\hfill
    \includegraphics[width=.3\textwidth]{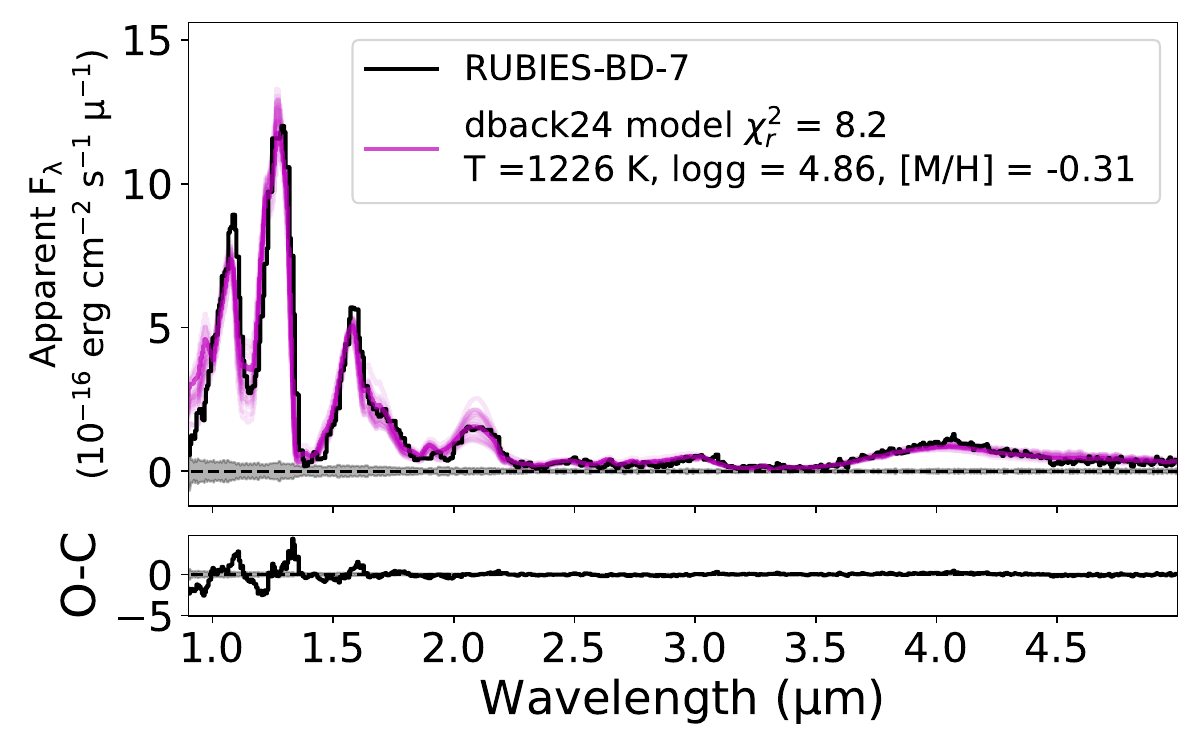}\hfill
    \includegraphics[width=.3\textwidth]{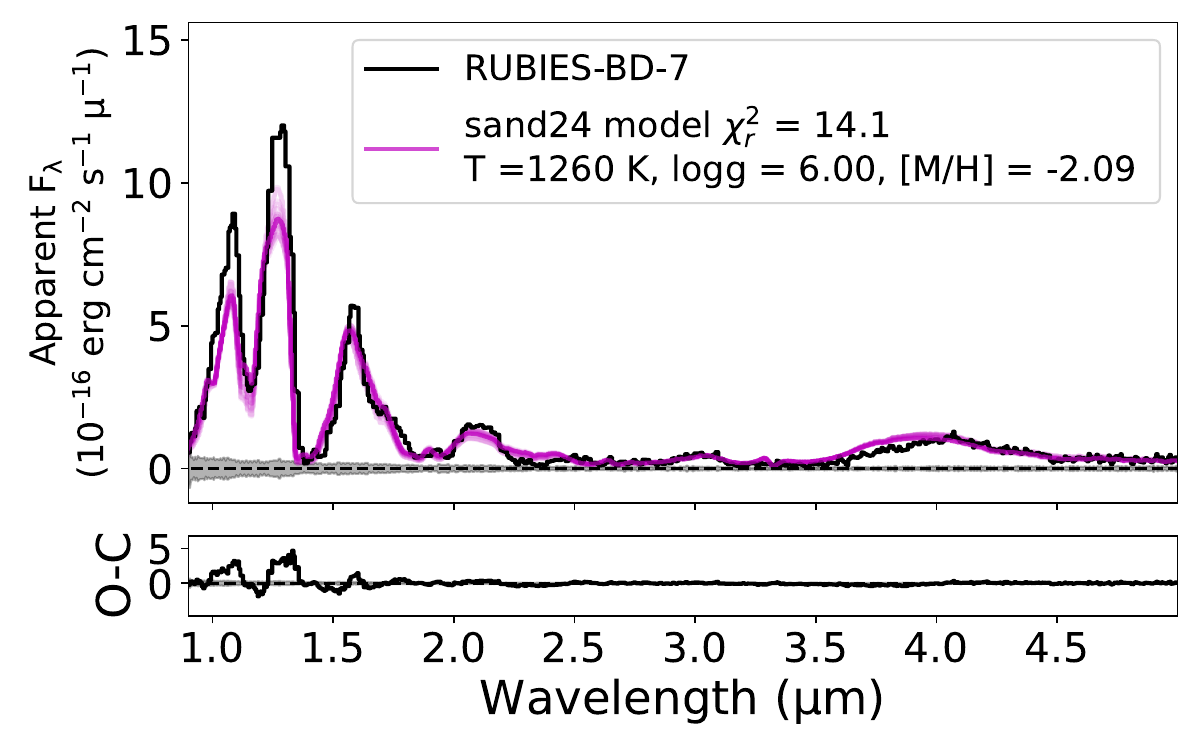}\hfill
    \caption{Continued.}
\end{figure}

% Restore figure numbering
\setcounter{figure}{\value{savedfigure}}
\renewcommand{\thefigure}{\arabic{figure}}

\begin{figure}[h]
    \centering
    \includegraphics[width=0.9\linewidth]{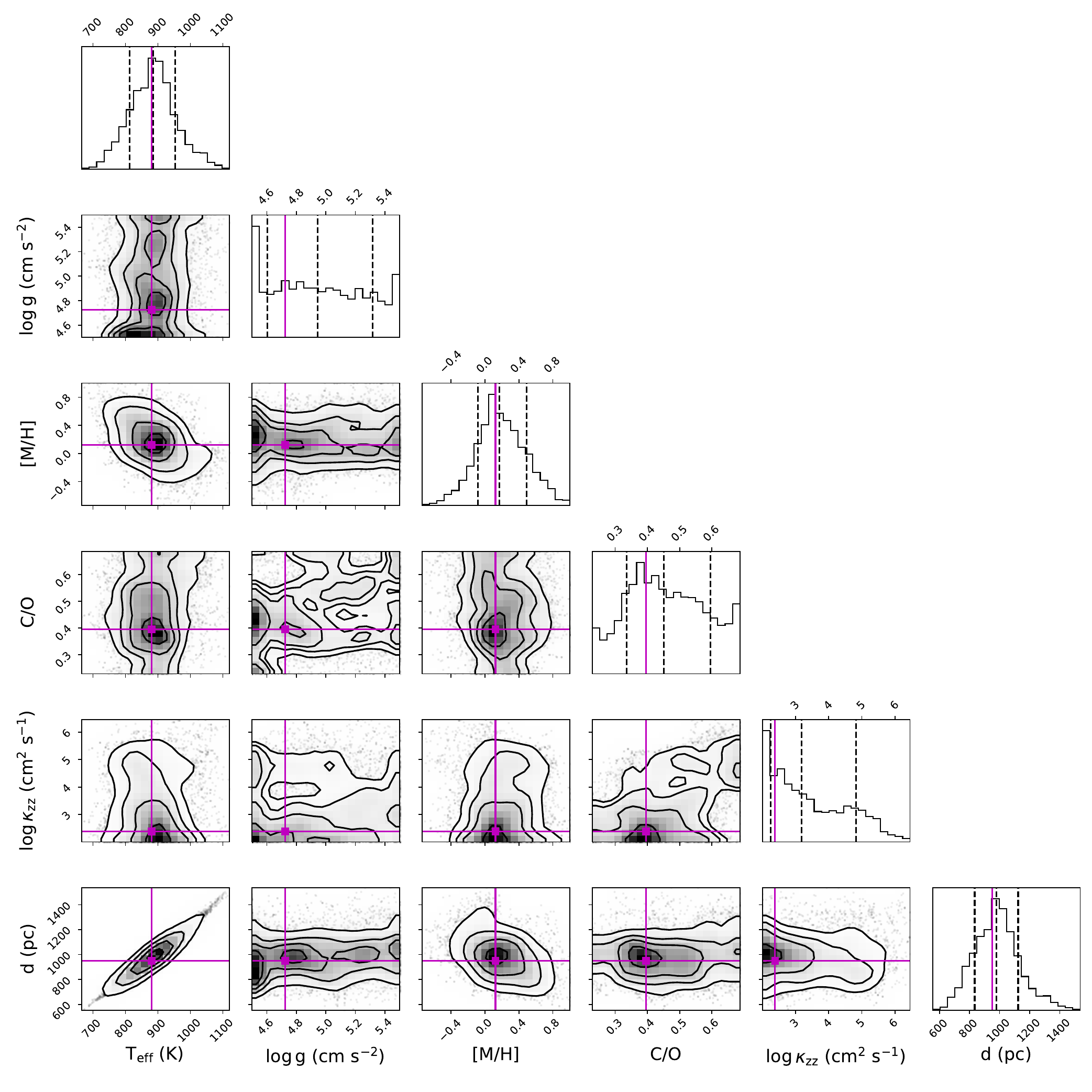}
    \caption{\added{Example} corner plot of Elf Owl fits to NIRSpec/Prism data for RUBIES-BD-1. Diagonal panels display the marginalized posterior distributions for {\teff}, {\logg}, [M/H], C/O, {\kzz}, and distance, while the inner panels display contour plots of distributions marginalized over parameter pairs to illustrate correlations. Contour shading \added{indicates} likelihood \added{values}, and lines \added{trace} 25\%, 50\%, and 75\% likelihood \added{thresholds}.
    Magenta lines indicate the parameter values for the best-fitting individual model, while the vertical dashed lines in the marginalized distributions indicate the 16\%, 50\%, and 84\% quantiles.}
    \label{fig:corner}
\end{figure}

\begin{deluxetable*}{lcccccccc}[h]
    \tabletypesize{\footnotesize}
    \tablecaption{Properties of RUBIES Brown Dwarfs \label{tab:summary}}
    \tablehead{
    Parameter & Unit &  \colhead{BD-1} & \colhead{BD-2} & \colhead{BD-3} & \colhead{BD-4} & \colhead{BD-5} & \colhead{BD-6} & \colhead{BD-7}}
    \startdata 
    SpT & \nodata & T7$\pm$0.5 & L1$\pm$0.5 & sd:L6$\pm$1 & T2$\pm$3 & T1$\pm$2 & T8$\pm$0.5 & T5$\pm$0.5\\
%    \hline
%    NIR Best Model(s) & \nodata & E(D) & D & EDS & EDS & S & EDS & E \\
    Best Model(s) & \nodata & E(D) & DS & (E)DS & EDS & S & E(DS) & E(D) \\
    {\teff} & (K) & $930_{-80}^{+100}$ & $2040_{-80}^{+60}$ & 2170$\pm$130 & $1800_{-700}^{+300}$ & $1270_{-60}^{+80}$ & 830$\pm$130 & 1180$\pm$80  \\
    {\logg} & (cm~s$^{-2}$) & 4.9$\pm$0.4 & $4.7_{-0.6}^{+0.7}$\tablenotemark{a} & $4.6_{-0.4}^{+0.5}$ & $5.0_{-0.5}^{+0.4}$ & $4.9_{-0.4}^{+0.5}$ & 5.2$\pm$0.5 & 4.9$\pm$0.4  \\
    {[M/H]} & (dex) & $+0.1_{-0.4}^{+0.3}$ & +0.1$\pm$0.3 & +0.2$\pm$0.4 & $-0.1_{-0.5}^{+0.6}$ & $-0.7_{-0.3}^{+0.4}$ & $-0.1_{-0.4}^{+0.3}$ & $-$0.2$\pm$0.3  \\
    $d_\mathrm{model}$ & (pc) & $1070_{-160}^{+220}$ & $860_{-60}^{+50}$ & 2500$\pm$300 & $5400_{-3400}^{+2200}$ & $2050_{-200}^{+260}$ & $1300_{-350}^{+390}$ & $780_{-90}^{+110}$  \\
    % {\teff} & (K) & 940$^{+80}_{-90}$ & 2040$^{+70}_{-80}$ & 2170$\pm$130 & 1700$^{+400}_{-700}$ & 1270$^{+80}_{-70}$ & 860$^{+120}_{-160}$ & 1170$^{+90}_{-80}$ \\
    % {\logg} & (cm~s$^{-2}$) & 5.0$^{+0.4}_{-0.6}$ & 4.8$^{+0.6}_{-0.7}$\tablenotemark{a} & 4.7$\pm$0.4 & 5.0$^{+0.4}_{-0.6}$ & 5.0$^{+0.5}_{-0.4}$ & 5.2$^{+0.6}_{-0.4}$ & 4.9$\pm$0.4 \\
    % {[M/H]} & (dex) & +0.1$^{+0.3}_{-0.4}$ & +0.1$\pm$0.3 & +0.3$^{+0.5}_{-0.4}$ & +0.0$^{+0.5}_{-0.7}$ & $-$0.7$\pm$0.4 & $-$0.1$\pm$0.4 & $-$0.2$\pm$0.3 \\
    % $d_\mathrm{mod}$ & (pc) & 370$\pm$70 & 890$\pm$50 & 2240$\pm$250 & 5800$^{+2800}_{-3700}$ & 2140$^{+290}_{-220}$ & 1210$\pm$360 & 790$\pm$110 \\
    $d_\mathrm{spt}$  & (pc) & 1000$\pm$170 & 1320$\pm$100 & 1870$\pm$180 & 2990$\pm$230 & 2500$\pm$250 & 940$\pm$140 & 830$\pm$90  \\
    $d_\mathrm{adopted}$ & (pc) & 1030$\pm$140 & 980$\pm$210\tablenotemark{b} & 2000$\pm$300 & 3000$\pm$400 & 2300$\pm$300 & 980$\pm$170 & 810$\pm$70 \\
    $z$ & (pc) & 890$\pm$120 & 850$\pm$180 & 1700$\pm$300 & 2990$\pm$200 & 2200$\pm$200 & 810$\pm$210 & 720$\pm$80 \\
%    Population & \nodata &thin  & thin  & thick  & thick  & thick  & thin  & thin  \\
    $P$(thin disk) & \nodata & 50$\pm$7\%  & 52$\pm$10\%  & 13$\pm$7\%  & 2.3$\pm$2.2\%  & 9$\pm$5\%  & 56$\pm$8\%  & 64$\pm$3\%  \\
    $P$(thick disk) & \nodata & 45$\pm$5\%  & 43$\pm$8\%  & 68$\pm$3\%  & 62$\pm$6\%  & 67.9$\pm$1.7\%  & 40$\pm$7\%  & 33$\pm$3\%  \\
    $P$(halo disk) & \nodata & 5.1$\pm$1.2\%  & 4.8$\pm$1.2\%  & 18$\pm$5\%  & 35$\pm$7\%  &23$\pm$5\%  & 4.4$\pm$1.4\%  & 3.0$\pm$0.4\%  \\
    \enddata
    \tablecomments{Best model(s) are for NIRSpec fits, and are indicated by labels 
    E = Elf Owl \citep{2024ApJ...963...73M}, D = Diamondback \citep{Morley_2024}, and 
    S = SAND \citep{2024ApJ...971...65G, 2024RNAAS...8..134A}.
    When more than one model is listed, 
    models outside of parentheses have indistinguishable best-fit $\chi_r^2$, while
    models in parentheses have slightly higher (more poorly fit) best-fit $\chi_r^2$. 
    Model parameters are determined from an uncertainty-weighted mean of the posterior distributions of all the best-fit models listed (see Eqns.\ref{eqn:mcmcweight} and~\ref{eqn:uncweighted}).
%    {\teff}, {\logg}, [M/H], and $d_\mathrm{model}$ are uncertainty-weighted averages for NIRSpec fits for the 
    The adopted distance $d_\mathrm{adopted}$ is an uncertainty-weighted average of the model fit distance ($d_\mathrm{model}$) and spectrophotometric distance ($d_\mathrm{spt}$) assuming the larger of the error bounds.
    }
    \tablenotetext{a}{The SAND NIRSpec fit included in this average has a significantly discrepant value of {\logg}.}
    \tablenotetext{b}{Model and spectrophotometric distance estimates differ significantly for this source.}
\end{deluxetable*}

%% NEED DISCUSSION
% 
% metallicity offset for Dback and Elfowl
% R-3 metal rich?
% R-2 cloudy and metal-rich
% let's do a closeup of R-1 in 4-5 µm range - ph3?
% importance of MCMC for parameter uncertainties, but also other approaches are possible: RFR, bayesian, retrieval
%  maybe plot of 
% 
% 
% 

\section{Discussion} \label{sec:discussion}

\subsection{Population Assignments} \label{sec:population}

% to do:
% - CONFIM DISTANCES
% - calculate scaleheights for which thin --> thick --> halo

% old text

The large distances and high Galactic latitudes of these sources 
%Both the EGS and UDS fields are inclined approximately 60$^o$ relative to the Galactic plane, such that the large estimated distances of our sources 
translate into large vertical scaleheights from the plane, giving them
%. At these displacements, stars have 
a high probability of belonging to the Milky Way's thick disk or halo populations \citep{2024ApJ...962..177B}. To quantify population membership, we used the star density model of \citet{2008ApJ...673..864J} to estimate the relative space densities of thin disk, thick disk, and halo sources in the directions and at the distances for each of our targets, sampling distance uncertainties using Monte Carlo methods. The resulting probabilities are listed in Table~\ref{tab:summary}. Not surprisingly, the most distant sources in our sample, RUBIES-BD-3, -4, and -5, have high probabilities of thick disk membership, with RUBIES-BD-4 \added{having} a significant probability (35$\pm$7\%) of halo membership. This subsample includes both the classified L subdwarf RUBIES-BD-3 and the only source in the sample whose modeled parameters indicate subsolar metallicity, RUBIES-BD-5, \added{although the metallicity values of all three of these sources are more consistent with thin/thick disk membership than with the halo.} The mid- to late-T dwarfs in our sample---RUBIES-BD-1, -6, and -7---are intrinsically fainter and detected at vertical scaleheights $\lesssim$1~kpc, and are thus more likely to be thin disk dwarfs, as supported by their spectral similarity to local spectral standards.  
The mix of thin disk, thick disk, and halo brown dwarfs mimics the composition of distant stars and brown dwarfs detected in other deep JWST survey samples \citep{2024ApJ...962..177B,2024ApJ...964...66H,2024ApJ...975...31H,2025arXiv251000111H}.

\subsection{Spectral Variations Among Distant T Dwarfs} \label{sec:diagnostics}

\begin{figure*}
    \centering
    \includegraphics[width=.35\textwidth]{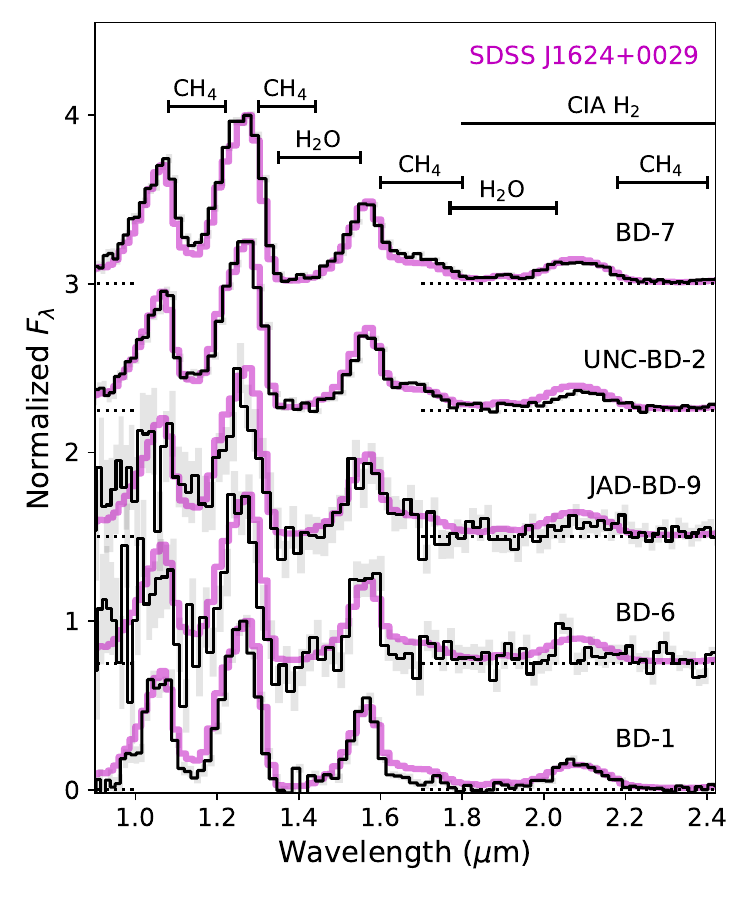}
    \includegraphics[width=.35\textwidth]{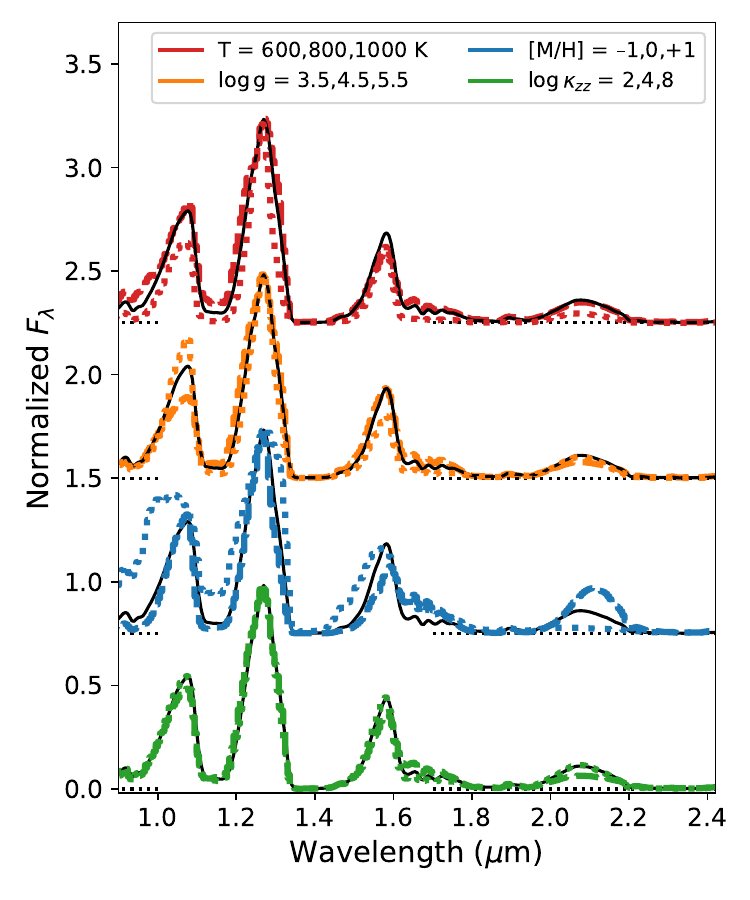} \\
    \includegraphics[width=.35\textwidth]{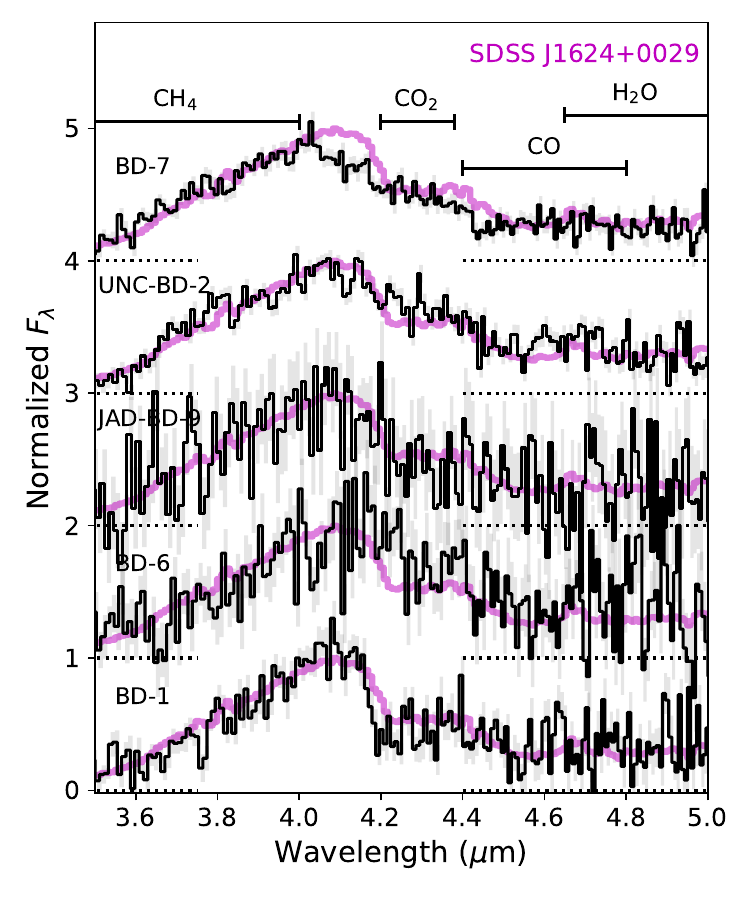}
    \includegraphics[width=.35\textwidth]{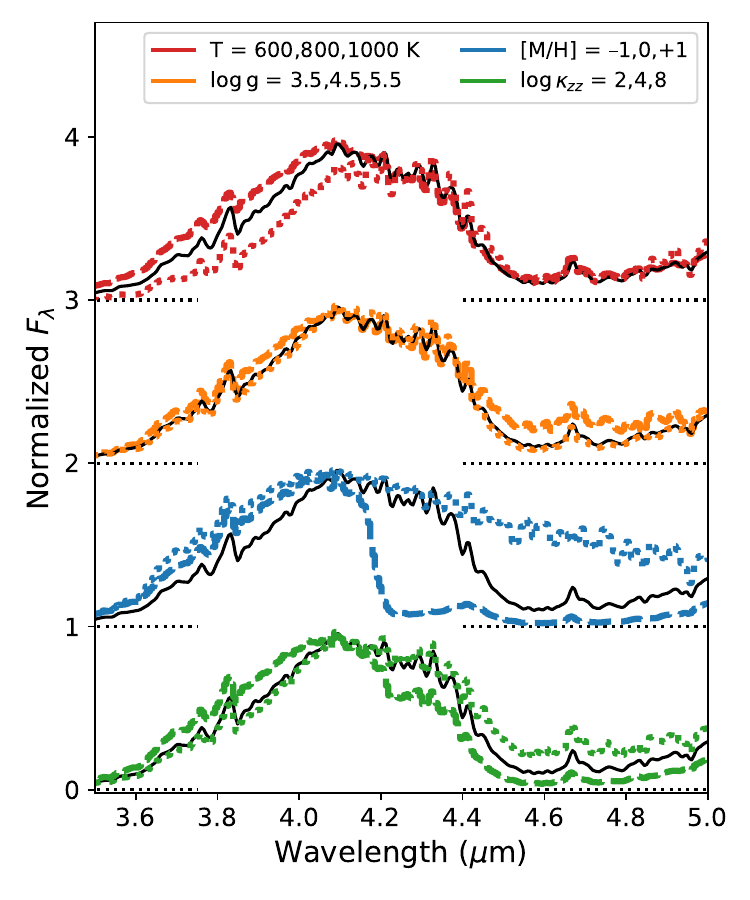} \\
    \caption{ Spectral variations among distant mid- and late-T dwarfs identified by JWST.
    (Left panels): Comparison of NIRSpec prism spectra of (from top to bottom): RUBIES-BD-7, 
    UNCOVER-BD-2 (UNC-BD-2; data from \citealt{2024ApJ...962..177B}),
    JADES-GS-BD-9 (JAD-BD-9; data from \citealt{2024ApJ...975...31H}),
    RUBIES-BD-6, and RUBIES-BD-1 (solid black lines).
    These \added{spectra} are compared to 
    equivalent data for the T6 spectral standard SDSS~J1624+0029 (repeated magenta line, data from \citealt{2024ApJ...973..107B})
    in the 0.9--2.4~{\micron} (top) and 3.5--5.0~{\micron} (bottom) ranges. 
    All spectra are normalized at \added{the} 1.25~{\micron} and 4~{\micron} peaks, and offset for clarity (dotted lines).
    (Right panels): Comparison of parameter variations in Elf Owl atmosphere models for T dwarf spectra in the same wavelength ranges. The black line is a reference spectrum with {\teff} = 800~K, {\logg} = 4.5, [M/H] = 0, and {\kzz} = 4. The colored dashed and dotted lines show the changes in spectral morphologies for
    $\Delta${\teff} = $\pm$200~K (red lines), 
    $\Delta${\logg} = $\pm$0.5~dex (orange lines),
    $\Delta$[M/H] = $\pm$0.5~dex (blue lines),
    \added{and} {\kzz} = 2 and 8 (green lines). 
    Model spectra are normalized and offset in the same manner as the observed spectra.}
    \label{fig:tdwarfcomp}
\end{figure*}

The mid- and late-type T dwarfs that have been identified in deep JWST/NIRSpec spectral surveys extend out to 1--2~kpc, a distance that encompasses a diversity of populations, ages, and compositions. As such, these early discoveries provide the opportunity to examine the influence of secondary parameters such as surface gravity and metallicity on cool brown dwarf spectra. To illustrate this, Figure~\ref{fig:tdwarfcomp} compares the NIRSpec/Prism spectra of five distant mid- and late-type T dwarfs, including RUBIES-BD-1 (T7), -6 (T8), and -7 (T5), 
the T6 dwarf UNCOVER-BD-2 (estimated distance of 2.1--2.6~kpc; \citealt{2024ApJ...962..177B}),
and the T6 dwarf JADES-GS-BD-9  (estimated distance of 1.8--2.3~kpc; \citealt{2024ApJ...975...31H}).
All of these sources are similarly normalized to focus on spectral morphology, and are
compared to the NIRSpec/Prism spectrum of the T6 dwarf standard SDSS~J1624+0029 \citep{2024ApJ...973..107B}.
We separately evaluate the 0.9--2.4~{\micron} NIR and 3.6--5.0~{\micron} MIR regions to focus on key spectral features.
In the NIR, we observe subtle variations between these spectra in several key features, including 
the depth of the 1.15~{\micron} and 1.6~{\micron} {\chhhh} bands (shallowest in RUBIES-BD-7, deepest in RUBIES-BD-1),
the width of the 1.25~{\micron} peak (widest in RUBIES-BD-7, narrowest in JADES-GS-BD-9 and RUBIES-BD-6),
and the shape and brightness of the 2.1~{\micron} peak (broad in RUBIES-BD-7, shifted redward in UNCOVER-BD-2, shifted blueward in RUBIES-BD-1, flat in JADES-GS-BD-9 and RUBIES-BD-6).
In the MIR, the broad 4~{\micron} peak shows a variety of shapes among these sources, from a relatively featureless and blueshifted peak in the spectra of RUBIES-BD-7 and UNCOVER-BD-2, to the more structured peaks in the spectra of JADES-GS-BD-9 and RUBIES-BD-7, the latter showing strong {\coo} and CO absorption bands at 4.3~{\micron} and 4.6~{\micron}, also observed in the spectrum of SDSS~J1624+0029. 

To contextualize these variations, we compare in the right panels of Figure~\ref{fig:tdwarfcomp} \added{to} Elf Owl atmosphere models with variations in temperature, surface gravity, metallicity, and vertical mixing efficiency relative to a baseline model  appropriate for a thin disk T6 dwarf ({\teff} = 800~K, {\logg} = 4.5, [M/H] = 0, and {\kzz} = 4).
NIR variations sensitive to these features have been discussed extensively in the literature (e.g., \citealt{2002ApJ...564..421B,2010ApJ...710.1627L,2012MNRAS.422.1922P,2017MNRAS.468..261Z,2024ApJ...963...73M,2025ApJ...982...79B}).
Here, we highlight that variations in the depths of {\chhhh} bands, the width of the 1.25~{\micron} peak, and \added{the} shape of the 2.1~{\micron} peak are all driven primarily by temperature and metallicity variations in the models. Surface gravity plays a greater role in setting the relative brightness of the 1.05~{\micron} and 1.6~{\micron} peaks, while vertical mixing plays only a minor role in shaping the 2.1~{\micron} peak.
In the MIR, the models show that metallicity plays \added{the} greatest role in shaping the 4~{\micron} peak, with low metallicity models missing the {\coo} and CO bands and producing a relatively featureless broad plateau. Vertical mixing also modulates the {\coo} feature. Temperature primarily shifts the location of the 4~{\micron} peak (blueward for warmer temperatures), while surface gravity only modestly influences the depth of the CO band.

This analysis suggests that all of these parameters can drive the variations observed among these similarly-classified distant T dwarfs. As such, a comparison of multiple features is necessary to isolate metallicity variations from other secondary parameters. There is an important caveat to this analysis, in that current models do not fully capture the complex atmosphere chemistry among the observed molecular constituents, Most importantly, the 4.3~{\micron} {\coo} band is known to be far stronger than can be explained even with efficient vertical mixing \citep{2010ApJ...722..682Y,2024ApJ...973...60B}.
We also note that none of these spectra show clear evidence of {\phhh} absorption that was initially claimed in the distant (0.6--1.2~kpc) late T dwarf UNCOVER-BD-3 \citep[][see \citealt{2024ApJ...973...60B} for further discussion]{2024ApJ...962..177B} and more recently detected in the metal-poor T dwarf companion Wolf~1130C \citep{burgasser-ph3}.
The absence or weakness of this feature in most brown dwarf and exoplanet spectra \added{remains} a challenge to chemical models \citep{2018ApJ...858...97M,2024ApJ...977L..49R}.

% \subsection{Improvements to Model Fits} \label{sec:discussfits}

% \subsection{Anticipated Yield for Future JWST Spectral Surveys} \label{sec:surveys}

\section{Summary} \label{sec:summary}

The key results of this study are as follows:
\begin{itemize}
\item We have identified seven brown dwarfs observed with JWST/NIRSpec as part of the RUBIES survey. All seven sources show spectral features characteristic of L- and T-type brown dwarfs in the 1--5~{\micron} range, including strong H$_2$O, CH$_4$, CO, and CO$_2$ molecular features.
\item We determined spectral types for these sources by comparing their near-infrared (0.9--2.4~{\micron}) spectra to dwarf and subdwarf standards, inferring classifications spanning L1 to T8. One source, RUBIES-BD-3, is best matched to mild L subdwarf (d/sdL) and L subdwarf (sdL) standards, suggesting that is has a metal-poor atmosphere.
\item We compared flux-calibrated spectra to three sets of atmosphere models: Sonora Elf Owl, Sonora Diamondback, and SAND. We find the L dwarf spectra are best matched to the Diamondback and SAND models, likely due to the presence of condensates in their atmospheres; the early-T dwarf spectra are best matched to the SAND models; and the mid- and late-T dwarf spectra are best matched to the Elf Owl and Diamondback models, the latter with efficient sedimentation or no clouds. Spectra of the two L dwarfs, RUBIES-BD-2 and -3, show the most significant deviations from the models, possibly due to incorrect treatment of condensates or the failure to incorporate reddening corrections, as previously noted by \citet{2025ApJ...980..230T}. Only one source, RUBIES-BD-5, shows significant evidence of being metal-poor based on the model fits.
\item Three of our sources---RUBIES-BD-3, -4, and -5---have estimated vertical scaleheights exceeding 1~kpc, and are mostly likely members of the Galactic thick disk population. RUBIES-BD-4, our faintest, most distant, and most poorly characterized source, also has a significantly probability of being a member of the Galactic halo population.
\item Examination of the present sample of mid- and late-type T dwarfs found in deep JWST spectral surveys at distances of 1--2~kpc show subtle but distinct variations in the strengths of major molecular bands and the shapes of inter-band peaks, most notably the 2.1~{\micron} flux peak and the 4.3~{\micron} {\coo} band. Examination of comparable Elf Owl models point to multiple factors that may be responsible for these variations, requiring further examination to isolate the \added{subsolar metallicities expected for} these distant objects.
\end{itemize}

This study adds to the growing sample of deep brown dwarf spectra from JWST/NIRSpec Prism data collected in deep survey programs such as 
CEERS \citep{2023ApJ...946L..13F},
UNCOVER \citep{2024ApJ...974...92B},
JADES \citep{2025ApJS..277....4D}, 
CANUCS \citep{2022PASP..134b5002W},
BoRG \citep{2025ApJ...983...18R},
NEXUS \citep{2024arXiv240812713S}, and
CAPERS \citep{2024jwst.prop.6368D},
among others (cf.~\citealt{2025arXiv251002026T}).
\added{Collectively,} these discoveries
%demonstrated the capacity to identify and characterize several dozen distant cool brown dwarfs in high galactic latitude fields, 
increase the diversity of physical parameters sampled among known brown dwarfs,
and provide the opportunity to improve
% complementary deep JWST/NIRCam imaging is also proving critical for characterizing these distant brown dwarfs, 
% through their identification by color and morphological selection (e.g., \citealt{2023ApJ...957L..27L,2024ApJ...964...39G,2025ApJ...980..230T}),
% the calibration of spectral observations for 
atmosphere and evolutionary models of low-temperature atmospheres more broadly, not only for thick disk and halo brown dwarf but also candidate brown dwarfs identified in
globular clusters \citep{2024ApJ...971...65G}
and stellar stream populations \citep{2025ApJ...982...79B},
as well as gas giant exoplanets orbiting metal-poor stars \citep{2019A&A...621A.110B,2022MNRAS.517.4447A}.
% and secondary indicators of physical properties and population membership 
% (e.g., proper motions; \citealt{2024ApJ...975...31H}).
Building larger and more diverse samples of cool brown dwarfs from these surveys
can also facilitate their use to address broader questions in Milky Way science, such as the evolution and distribution of chemical abundances, the Milky Way's star formation and accretion history, and the dynamical history of various Galactic populations \citep{2004ApJS..155..191B,2017ApJ...847...53R,2022ApJ...934...73A,2024ApJ...967..115B,2024ApJ...974..222R,2025ApJ...985...48H}.
% However, 
% %while the broad spectral range covered by the NIRSpec prism mode samples multiple diagnostic features, 
% improvements in modeling, most notably condensate and {\coo} chemistry, will be required to accurately reproduce the observed spectra, and to disentangle and constrain the physical properties these ubiquitous low-temperature sources.

\begin{acknowledgments}
\added{The authors thank our anonymous referee for their prompy and thorough review of the original manuscript.}
S.M. acknowledges the guidance and financial support received from the UCSD Summer Triton Research \& Experiential Learning Scholars (TRELS) program.
A.J.B. acknowledges funding support from NASA/STScI through JWST general observer program GO-4668, under NASA contract NAS 5-03127.
Some of the data products presented herein were retrieved from the Dawn JWST Archive (DJA). DJA is an
initiative of the Cosmic Dawn Center (DAWN), which
is funded by the Danish National Research Foundation
under grant DNRF140.
The Cosmic Dawn Center is funded by the Danish National Research Foundation (DNRF) under grant DNRF140.
\added{JWST/NIRSpec spectral data reported in this program are available on the Mikulski Archive for Space Telescopes (MAST) and are associated with dataset DOI 10.17909/qk5z-7p30.}
%
%[OTHERS?]
%
\end{acknowledgments}


\vspace{5mm}
\facilities{JWST(NIRCam), JWST(NIRSpec)}


\software{
    {\tt astropy} \citep{2013A&A...558A..33A,2018AJ....156..123A,2022ApJ...935..167A},  
    {\tt msaexp} \citep{brammer_2023_8319596}, 
    JWST Calibration pipeline \citep{2024ApJ...962..195B}, 
    {\tt SPLAT} \citep{2017ASInC..14....7B}, and
    {\tt ucdmcmc} \citep{ucdmcmc}
}

\appendix

\section{Near-Infrared Spectral Standards}

Table~\ref{tab:standards} lists the near-infrared spectral standards used for the classification of our sources in Section~\ref{sec:classify}.
Dwarf standards are defined in \citet{Burgasser_2006} and \citet{2010ApJS..190..100K}; 
subdwarf standards are defined in \citet{2017MNRAS.464.3040Z,2018MNRAS.479.1383Z,2019AJ....158..182G}; and \citet{2025ApJ...982...79B}.
We drew from previously-published data acquired with several different instruments, which were resampled to the wavelength-dependent 
resolution of the JWST/NIRSpec prism mode \citep{Birkmann_2022}.
The resampled spectra are included as part of the online materials for the article.

\startlongtable
\begin{deluxetable}{cccc}
\tabletypesize{\footnotesize}
\tablecaption{Near-Infrared Spectral Standards \label{tab:standards}}
\tablehead{
\colhead{SpT} & 
\colhead{Source} & 
\colhead{Instrument} & 
\colhead{Data Ref.}
}
\startdata 
L0 &  2MASP J0345432+254023 & IRTF/SpeX & [1] \\
L1 &  2MASSW J2130446$-$084520  & IRTF/SpeX &  [2] \\
L2 &  Kelu$-$1AB & IRTF/SpeX &  [3] \\
L3 &  2MASSW J1506544+132106 & IRTF/SpeX &  [4] \\
L4 &  2MASS J21580457$-$1550098 & IRTF/SpeX &  [2] \\
L5 &  SDSS J083506.16+195304.4  & IRTF/SpeX &  [5] \\
L6 &  2MASSI J1010148$-$040649 & IRTF/SpeX &  [6] \\
L7 &  2MASSI J0103320+193536 & IRTF/SpeX &  [2] \\
L8 &  2MASSW J1632291+190441 & IRTF/SpeX &  [4] \\
L9 &  DENIS-P J0255$-$4700  & IRTF/SpeX &  [7] \\
T0 &  SDSS J120747.17+024424.8  & IRTF/SpeX &  [8] \\
T1 &  SDSSp J083717.22$-$000018.3  & IRTF/SpeX &  [9] \\
T2 &  SDSSp J125453.90$-$012247.4 & IRTF/SpeX &  [10] \\
T3 &  2MASS J12095613$-$1004008AB & IRTF/SpeX &  [10] \\
T4 &  2MASSI J2254188+312349  & IRTF/SpeX &  [10] \\
T5 &  2MASS J15031961+2525196  & IRTF/SpeX &  [10] \\
T6 &  SDSSp J162414.37+002915.6 & IRTF/SpeX &  [11] \\
T7 &  2MASSI J0727182+171001  & IRTF/SpeX &  [11] \\
T8 &  2MASSI J0415195$-$093506  & IRTF/SpeX &  [10] \\
T9 &  UGPS J072227.51$-$054031.2 & IRTF/SpeX &  [9] \\
\hline
d/sdL0 & 2MASS J00412179+3547133 & IRTF/SpeX &  [10] \\ % Zhang 2017 sdL0.5
d/sdL1 &  2MASS J17561080+2815238 & IRTF/SpeX?? &  [12] \\ % Zhang 2017 sdL1
d/sdL6 &  SDSS J133148.92$-$011651.4 & IRTF/SpeX &  [2] \\ % Burgasser 2025
d/sdL8 &  2MASS J11582077+0435014 & IRTF/SpeX &  [12] \\ % Burgasser 2025
d/sdL9	&	CWISE J202130.11+152418.3	& Keck/NIRES &  [13] \\ % Burgasser 2025
d/sdT0	&	2MASS J06453153$-$6646120	& VLT/Xshooter &  [14] \\ % Burgasser 2025
d/sdT1	&	 WISE J030119.47$-$231921.8	& IRTF/SpeX &  [15] \\ % Burgasser 2025
d/sdT2	&	 WISE J000458.59$-$133655.2	& IRTF/SpeX &  [15]	\\ % Burgasser 2025
d/sdT2.5	&	CWISE J211255.59+303037.6	& Keck/NIRES &  [13] \\ % Burgasser 2025
d/sdT4	&	 ULAS J002135.97+155226.8	& VLT/Xshooter &  [16] \\ % Burgasser 2025
d/sdT5.5	&	CWISE J113010.07+313944.7	& Keck/NIRES &  [13] \\ % Burgasser 2025
d/sdT5.5	&	GJ 576B	& Keck/NIRES &  [13] \\ % Burgasser 2025
d/sdT6	&	CWISE J201342.30$-$032643.4	& Keck/NIRES &  [13]	\\ % Burgasser 2025
d/sdT7.5	&	LHS 6176B	& Keck/NIRES &  [13] \\ % Burgasser 2025
d/sdT8	&	2MASS J09393548$-$2448279	& IRTF/SpeX &  [7] \\ % Burgasser 2025
d/sdT9	&	ULAS J083338.11+005206.2	& Gemini/GNIRS &  [17] \\ % Burgasser 2025
\hline
sdL0 & WISE J04592121+1540592 & IRTF/SpeX &   [12] \\
sdL1 & ULAS J124947.04+095019.8 & VLT/Xshooter &  [18] \\  % Zhang 2017 sdL1
sdL4 & ULAS J021642.97+004005.6 & VLT/Xshooter &  [18] \\  % Zhang 2017 sdL1
sdL5 &  SDSS J141624.08+134826.7 & IRTF/SpeX &  [19] \\ % Zhang 2017 sdL7
sdT0	&	 WISEA J152443.14$-$262001.8	& Keck/NIRES &  [13] \\ % Burgasser 2025
sdT1	&	CWISE J211250.11$-$052925.2	& Keck/NIRES &  [13] \\ % Burgasser 2025
sdT3	&	 CWISE J062316.19+071505.6	& Keck/NIRES &  [13] \\ % Burgasser 2025
sdT4	&	 CWISE J155349.96+693355.2	& Keck/NIRES &  [20] \\ % Burgasser 2025
sdT5.5	&	CWISE J113019.19$-$115811.3	& Keck/NIRES &  [21] \\ % Burgasser 2025
sdT6	&	2MASS J09373487+2931409	& IRTF/SpeX & 	[11] \\ % Burgasser 2025
sdT7	&	ULAS J141623.94+134836.3	& IRTF/SpeX &  [22]	\\ % Burgasser 2025
sdT7.5	&	ULAS J001354.85+063445.4	& Gemini/GNIRS &  [17] \\ % Burgasser 2025
\hline
esdL0 & SSSPM J10130734$-$1356204 & IRTF/SpeX &  [23] \\ % Zhang 2017 usdL0
esdL1 & SSSPM J144420.67$-$201922.2 & IRTF/SpeX &  [2] \\ % Zhang 2017 esdL1
esdL3 &  SDSS J125637.16$-$022452.2 & IRTF/SpeX &  [24] \\ % Zhang 2017 usdL3
esdL4 & 2MASS J16262034+3925190  & IRTF/SpeX &  [23] \\ % Zhang 2017 usdL4
esdL6	&	 2MASS J061643.51$-$640719.8	& VLT/Xshooter &  [18]	\\ % Zhang 2017 esdL6, Burgasser 2025 esdT0
esdL7	&	2MASS J053253.46+824646.5	& Keck/LRIS+NIRES &  [13]	\\ % Zhang 2017 esdL7
esdT3	&	 WISEA J181006.18$-$101000.5	& GTC/OSIRIS+Palomar/TSpec &  [25,26]	\\ % Burgasser 2025
esdT6	&	WISEA J041451.67$-$585456.7	& Magellan/FIRE & 	[25] \\ % Burgasser 2025
\hline
\enddata
\tablerefs{[1] \citet{2006AJ....131.1007B}; [2] \citet{2014ApJ...794..143B}; [3] \citet{2007ApJ...658..557B}; [4] \citet{2007ApJ...659..655B}; [5] \citet{2006AJ....131.2722C}; [6] \citet{2006AJ....132..891R}; [7] \citet{2006ApJ...637.1067B}; [8] \citet{2007AJ....134.1162L}; [9] \citet{2017ASInC..14....7B}; [10] \citet{2004AJ....127.2856B}; [11] \citet{2006ApJ...639.1095B}; [12] \citet{2010ApJS..190..100K}; [13] \citet{2025ApJ...982...79B}; [14] \citet{2018MNRAS.480.5447Z}; [15] \citet{2019AJ....158..182G}; [16] \citet{2019MNRAS.486.1260Z}; [17] \citet{2014MNRAS.437.1009P}; [18] \citet{2017MNRAS.464.3040Z}; [19] \citet{2010AJ....139.1045S}; [20] \citet{2021ApJ...915..120M}; [21] \citet{2021ApJS..253....7K}; [22] \citet{2010AJ....139.2448B}; [23] \citet{2004ApJ...614L..73B}; [24] \citet{2009ApJ...697..148B}; [25] \citet{2020ApJ...898...77S}; [26] \citet{2022AA...663A..84L}.}
\end{deluxetable}

\bibliography{bibtex}
\bibliographystyle{aasjournal}
\end{document}